\documentclass[12pt,preprint]{aastex}

\usepackage{multirow}
\usepackage{graphicx}
\usepackage{epsfig}
\usepackage{txfonts}
\usepackage{psfrag}
\usepackage{units}
\DeclareMathAlphabet{\mathpzc}{OT1}{pzc}{m}{it}

\def\sun{\hbox{$\odot$}}
\def\nh{{n_{\rm H}}}

\def\nh2{{n(\rm H_2)}}
\def\h2{${\rm H_2}$}

\def\3cm{\rm {cm^{-3}}}
\def\2cm{\rm {cm^{-2}}}
\def\s-1{\rm {s^{-1}}}
\def\etal {et al.}

\def\Msun{$M_{\sun}$}
\def\mum {\hbox{$\mu\rm m$}}

\def\Kkms {\hbox{${\rm K}\,{\rm km}\,{\rm s}^{-1}$}}

\def\ergs{{\rm {erg}~{s^{-1}}}}
\def\ergcms{{\rm {erg}~{{cm^{-2}}~s^{-1}}}}
\def\ergscm{{\rm {erg}~s^{-1}~{{cm^{-2}}}}}
\def\ergscmsr{{\rm {erg}~s^{-1}~{{cm^{-2}}}~sr^{-1}}}

\def\hcop{{{HCO$^+$}}}
\def\hcn{{{HCN}}}
\def\hnc{{{HNC}}}
\def\cn{{{CN}}}

\def\h2s1{{\rm {H$_2$}S(1)}}
\def\h2s2{{\rm {H$_2$}S(2)}}
\def\twco{{$^{12}$CO}}
\def\thco{{$^{13}$CO}}

\def\oi{{[O {\scriptsize I}]}}

\def\ci{{[C {\scriptsize I}]}}
\def\cii{{[C {\scriptsize II}]}}

\def\nii{{[N {\scriptsize II}]}}
\def\neii{{[Ne~{\scriptsize II}]}}

\def\nev{{[Ne~{\scriptsize V}]}}

\def\hi{{H~{\scriptsize I}}}

\def\c18o{{C$^{18}$O}}

\defcitealias{wada02}{WN02}
\defcitealias{wada05}{WT05}
\defcitealias{yamada07}{YWT07}
\defcitealias{wada09}{WPS09}

\newcommand{\pb}{P\'erez-Beaupuits}
\newcommand{\myemail}{jp@mpifr.de}

\slugcomment{Received 2010 November 29;}

\shorttitle{An AGN torus}
\shortauthors{\pb\ \etal}

\begin{document}


\title{The Structure and Dynamics of an AGN torus:\\
       CO line predictions for ALMA from 3D hydrodynamical simulations with X-ray driven chemistry}


%\altaffilmark{1,3}
%\thanks{Most of the work was done as PhD student at Kapteyn institute, but finished at MPIfR as Humboldt Fellow.}
\author{J.P. \pb }
\affil{Max-Planck-Institut f\"ur Radioastronomie, Auf dem H\"ugel 69, 53121 Bonn, Germany}
\affil{Kapteyn Astronomical Institute, Rijksuniversiteit Groningen, 9747 AD Groningen, The Netherlands}
\email{\myemail}

\and

%\altaffilmark{2}
\author{K. Wada}
\affil{Graduate School of Science and Engineering, Kagoshima University, Kagoshima 890-0065, Japan}
\affil{Visiting researcher, Research Center for Space and Cosmic Evolution, Ehime University}
\email{wada@sci.kagoshima-u.ac.jp}

\and

%\altaffilmark{5}
\author{M. Spaans}
\affil{Kapteyn Astronomical Institute, Rijksuniversiteit Groningen, 9747 AD Groningen, The Netherlands}
\email{spaans@astro.rug.nl}


%\altaffiltext{1}{Visiting Astronomer, Cerro Tololo Inter-American Observatory.
%CTIO is operated by AURA, Inc.\ under contract to the National Science
%\altaffiltext{2}{Society of Fellows, Harvard University.}
%\altaffiltext{3}{present address: Center for Astrophysics,
%    60 Garden Street, Cambridge, MA 02138}


\begin{abstract}
Many efforts have been made to model the mass distribution and dynamical evolution of the circumnuclear gas in active galactic nuclei (AGNs). However, chemical evolution is not included in detail in three-dimensional (3-D) hydrodynamic simulations. The X-ray radiation from the AGN can drive the gas chemistry and affect the thermodynamics, as well as the excitation of the interstellar medium (ISM).
Therefore, we estimate the effects (on chemical abundances and excitation) of X-ray irradiation by the AGN, for atomic and molecular gas in a 3-D hydrodynamic model of an AGN torus.
We obtain the abundances of various species from an X-ray chemical model. A 3-D radiative transfer code estimates the level populations, which result in line intensity maps.
Predictions for the CO $J=1\rightarrow0$ to $J=9\rightarrow8$ lines indicate
that mid-$J$ CO lines are excellent probes of density and dynamics in the central ($\lesssim60~\rm pc$) region of the AGN, in contrast to the low-$J$ CO lines.
Analysis of the $X_{\rm CO}/\alpha$ conversion factors shows that only the higher-$J$ CO lines can be used for gas mass determination in AGN tori.
The \cii\ $158~\mum$ emission traces mostly the hot ($T_k>1000~\rm K$) central region of the AGN torus. The \cii\ $158~\mum$ line will be useful for ALMA observations of high redshift ($z\gtrsim1$) AGNs. The spatial scales ($\ge0.25~\rm pc$) probed with our simulations match the size of the structures that ALMA will resolve in nearby ($\le45~\rm Mpc$ at $0.01''$) galaxies.
\end{abstract}





\section{Introduction}

The formation and growth of a central black hole and its interaction with intense star-forming regions is one of the topics most debated in the context of galaxy evolution. There is observational evidence for a common physical process from which most active galactic nuclei (AGNs) and starbursts originate \citep[e.g.,][]{soltan82, magorrian98, ferrarese00, graham01, haring04}.
A plausible scenario considers that starbursts, super-massive black hole growth, and the formation of red elliptical and submillimeter galaxies, are connected through an evolutionary sequence caused by mergers between gas-rich galaxies \citep{hopkins06, hopkins08, tacconi08, narayanan09, narayanan10}.
In this scenario, the starbursts and (X-ray producing) AGNs seem to be co-eval, and the interaction processes between them (phase d and e in Figure~1 of \citealt{hopkins08}), that dominate the formation and emission of molecular gas, is one of the long-standing issues concerning active galaxies.

Numerous molecules tracing different (AGN and starburst driven) gas chemistry have been detected in Galactic and (active) extragalactic environments. Studies have shown that chemical differentiation observed within Galactic molecular clouds is also seen at larger ($\sim$100 pc) scales in nearby galaxies \citep[e.g.,][]{henkel87, nguyen91, martin03, usero04, tacconi08, pb07, pb09, pb10, baan10, vdwerf10}.

The evolution of the ISM in the inner $100~\rm pc$ region around a $10^8~M_{\sun}$ supermassive black hole (SMBH) was investigated by \citet{wada02} (hereafter \citetalias{wada02}) using three-dimensional (3-D) Euler-grid hydrodynamic simulations. They took into account self-gravity of the gas, radiative cooling and heating due to supernovae (SNe). A clumpy and filamentary torus-like structure was found to be reproduced on a scale of tens of pc around the SMBH, with highly inhomogeneous ambient density and temperature, and turbulent velocity field.
Their results indicated that AGNs could be obscured by the circumnuclear material. This represents theoretical support for observational evidence showing that some AGNs are obscured by nuclear starbursts \citep[e.g.,][and references therein]{levenson01, levenson07, ballantyne08}.

Several efforts have been made to estimate the molecular line emission from the nuclear region in these 3-D hydrodynamic simulations, and to compare the results with observational data. For instance, 
% From Wada & Tomisaka 2005
\citet{wada05} (hereafter \citetalias{wada05}) derived molecular line intensities emitted from the nuclear starburst region around a SMBH in an AGN. They used the 3-D hydrodynamic simulations (density, temperature, and velocity field data) of the multi-phase gas modeled by \citetalias{wada02} as input for 3-D non-LTE radiative transfer calculations of \twco\ and \thco\ lines. They found that the CO-to-H$_2$ conversion factor ($X$-factor) is not uniformly distributed in the central 100 pc and the $X$-factor for \twco\ $J=1\rightarrow0$ is not constant with density, in contrast with the \twco\ $J=3\rightarrow2$ line. Similarly, the role of the \hcn\ and \hcop\ high-density tracers in the inhomogeneous molecular torus of \citetalias{wada02} was studied by \citet{yamada07} (hereafter \citetalias{yamada07}). These non-LTE radiative transfer calculations suggested a complicated excitation state of the rotational lines of \hcn\ (with maser action) and \hcop, regardless of the spatially uniform chemical abundance assumed.

However, all these previous efforts to estimate the molecular line emissions from the central 100 pc of an AGN leave room for improvements. First of all, the radiative cooling in the simulations by \citetalias{wada02} are not consistent with the chemical abundances in the cold and dense gas because (collisional) formation and (radiative) destruction of H$_2$ by far ultraviolet radiation (FUV) was not included. Therefore, the cold and dense gas in the simulations by \citetalias{wada02} does not necessarily represent the dusty molecular gas phase around an AGN.

% New 3-D HD model...
Hence, in order to study the distribution and structures of the various density regimes of the H$_2$ gas, the 3-D hydrodynamic simulations of \citetalias{wada02} were extended by \citet*{wada09} (hereafter \citetalias{wada09}) to solve the nonequilibrium chemistry of hydrogen molecules along with the hydrodynamics. 
The formation of H$_2$ on dust and its radiative destruction by far ultraviolet radiation (FUV) from massive stars are also included in the model by \citetalias{wada09}. This allows to track the evolution of molecular hydrogen and its interplay with the \hi\ phase in the central $64 \times 64 \times 32~\rm pc$ region. 
Thus, the radiative cooling in the model by \citetalias{wada09} is more consistent with the chemical abundances expected in the cold ISM, in comparison with the models by \citetalias{wada02}.
Different SN rates and strengths of the uniform FUV field were also explored in order to study their effects on the structures of molecular gas.

On the other hand, the inhomogeneous density and temperature structures observed in the 3-D hydrodynamic models are not the only factors that drive molecular abundances and excitation conditions of molecular lines. There is observational and theoretical evidence in the literature that supports different chemical evolution scenarios due to X-ray and UV radiation from the central AGN and circumnuclear starburst, as well as mechanical heating produced by turbulence and supernovae \citep[e.g.,][]{kohno01, kohno07, kohno05, imanishi04, imanishi06a, imanishi06b, imanishi06, aalto07a, meijerink07, garcia07, loenen08, garcia08, pb09}. The strong UV and X-ray radiation from the AGN and accretion disk could affect both the dynamics and excitation of the molecular gas \citep[e.g.,][]{ohsuga01a, ohsuga01b, meijerink07, vdwerf10}. However, the radiation field from the AGN itself was not taken into account in the earlier estimates of molecular line emissions from hydrodynamical simulations. 

%While a uniform UV radiation field was assumed in the molecular line emission estimates by WT05 and YWT07, the radiation from the massive stars distributed in the clumpy nuclear region is actually expected to affect, in a non-uniform way, the dissociation and excitation of the molecular gas, as well as the heating of the ISM.

A preliminary estimate of the potential effects of hard X-rays ($E>1$ keV) on the molecular gas was done by \citetalias{wada09} using the X-ray Dissociated Region (XDR) models of \citet{meijerink05}. It was found that XDR chemistry may change the distribution of H$_2$ around an AGN, if X-ray effects are explicitly included in the hydrodynamic model. The X-ray chemistry depends mainly on $H_X/n$, where $H_X$ is the X-ray energy deposition rate and $n$ is the number density of the gas \citep{maloney96}.
Although the H$_2$ abundance is robust in a clumpy medium like the one found in the hydrodynamical models, the temperature of the gas affected by an X-ray flux is expected to be a factor of $\sim5$ higher than that found in standard models of a Photon Dominated Region (PDR; e.g., \citealt{hollenbach99}), for $log(H_X)/n > −26$ \citep{meijerink07}. This is because the ionization heating by X-rays is more efficient than photo-electric emission by dust grains. 
%The higher temperatures of the molecular gas will therefore produce stronger emission in the pure rotational H$_2$ lines (e.g., S(0) and S(1) at 28 \mum\ and 17 \mum, respectively). 

Other molecular and atomic lines have also been suggested as tracers of the AGN and starburst activity in nearby galaxies ($z<1$) as well as in high ($z\ge1$) redshift galaxies. \citet{spaans08} studied the possibility of using \twco\ and H$_2$ emission lines to trace a young population of accreting massive ($\ge10^6~M_{\sun}$) black holes at redshifts $z=5-20$ and radiating close to the Eddington limit. 
An enhancement in the intensities of various \twco\ transitions up the rotational ladder, as well as other molecular and atomic lines like \thco, \hcn, \hcop, \ci, \cii, \oi\ and \nii, is also expected to be observed when X-ray irradiation dominates the local gas chemistry \citep{meijerink07, spaans08}.
Simulations of quasars at $z\sim6$ with massive ($10^{12}-10^{13}~M_{\sun}$) halos and different merging histories showed that mid-$J$ \twco\ lines are highly excited by a starburst, while high velocity peaks are expected to be produced by AGN-driven winds \citep{narayanan08a, narayanan08b}. It was further found by \citet{narayanan09} that the compact \twco\ spatial extents, broad linewidths and high excitation conditions observed in Submillimetre Galaxies (SMGs) at $z\sim2$ can be explained if SMGs are a transition phase of major merging events. 

%Also objects with black hole, bulge and H$_2$ masses similar to those observed in SMGs are obtained from a mass sequence of merger models constrained by observational data (Narayanan \etal\ 2010).

% This work....
In this work we use the XDR/PDR chemical model by \citet{meijerink05} to estimate the abundances of more than 100 species (atoms and molecules) at each grid point in the computational box of the extended 3-D hydrodynamical models of an AGN torus by \citetalias{wada09}. We also estimate the actual X-ray flux emerging from the AGN, derived from the central black hole mass. Flux attenuation by photo-absorption of X-rays along the ray path and the distance from the central black hole is included. Thus, we estimate non-homogeneous abundances at each grid point that depend on the local density and impinging X-ray flux. An extended version of the non-LTE 3-D radiative transfer code $\beta$3D by \citet{poelman05} is used to compute the level populations of any molecule or atom for which collision data exist in the LAMDA\footnote{http://www.strw.leidenuniv.nl/$\sim$moldata/} database \citep{schoier05}. Molecular and atomic line intensities and profiles are calculated with a line tracing approach for an arbitrary viewing angle. Model predictions for future ALMA observations of CO lines and the \cii\ $158\mum$ fine structure line are presented.
The organization of this article is as follows. In Sec.~\ref{sec:3Dmodel} we describe the numerical method. The results and analysis are presented in Sec.~\ref{sec:results}. The final remarks are presented in Sec.~\ref{sec:remarks}.
%Implications of the results and future observational applictions are discussed in Sec.~\ref{sec}. 

%Shocks and turbulence can heat as well, therefore observational data on line profile kinematics would be useful in the future.

%Including the molecular gas in numerical models will eventually yield a much richer interface with current and future observations. This is because unlike the dust continuum, the effectively optically thin molecular line emission (even large line-center optical depths allow full view of turbulent media), is a direct probe of the velocity fields and thus of the gaseous disk dynamics, unobscured by the concomitant dust. The upcoming commissioning of ALMA will drastically enlarge such an interface by enabling the imaging of a whole suite of molecular lines probing different density and temperature regimes at high angular resolution. 

\section{Numerical Method}\label{sec:3Dmodel}

The three-dimensional hydrodynamic model of the AGN torus used in this work includes inhomogeneous density fields and mechanical
heating effects due to turbulence and supernova explosions, with a resolution (pixel size) of $0.25~\rm pc$ in diameter. Detailed descriptions of the hydrodynamic equations and simulations can be found in \citetalias{wada09}. 
The 3-D hydrodynamic model considers the formation and destruction of H$_2$ in a self-consistent way, including formation of H$_2$ on grains, and the destruction of it by FUV radiation. This allows the code to compute the total density, local temperature (and velocity field) as well as the fraction of H$_2$ at each grid element. The hydrodynamical code runs for an equivalent time of $\gtrsim3.5~\rm Myr$ until it reaches a quasi-steady state \citepalias{wada02}, and the gas forms a highly inhomogeneous and clumpy torus with some spiral structures. It comprises a flared disk of H$_2$ gas $\sim50~\rm pc$ in diameter, and about $10~\rm pc$ in height \citepalias[their Fig.2c]{wada09}.

One of the main criticism that hydrodynamical models receive in general, is that the density (and temperature) distribution that they show does not mimic closely actual observations of the gas structure and distribution in galaxies and galaxy nuclei (like in \citetalias[their Fig.2a\&b]{wada09}). However, observational data do not show the actual density or temperature of the gas either. The information we get from observations is the intensity and distribution of particular atomic and molecular emission (or absorption) lines, from which the ambient conditions (density, temperature and radiation field) can be estimated. Therefore, hydrodynamical models need to be complemented with atomic and gas chemistry that allow us to infer how the emission of different species would look like given the density and temperature structure obtained from the hydrodynamic simulations.

\begin{figure}[tp]
%\centering

%\hfill\includegraphics[angle=0,height=5.5cm]{nH2_GE_100cm3_xy_cinv}%
%\includegraphics[angle=0,height=5.5cm]{nH2_GE_100cm3_yz_cinv}\hspace*{\fill}\\

\hfill\includegraphics[angle=0,height=5.5cm]{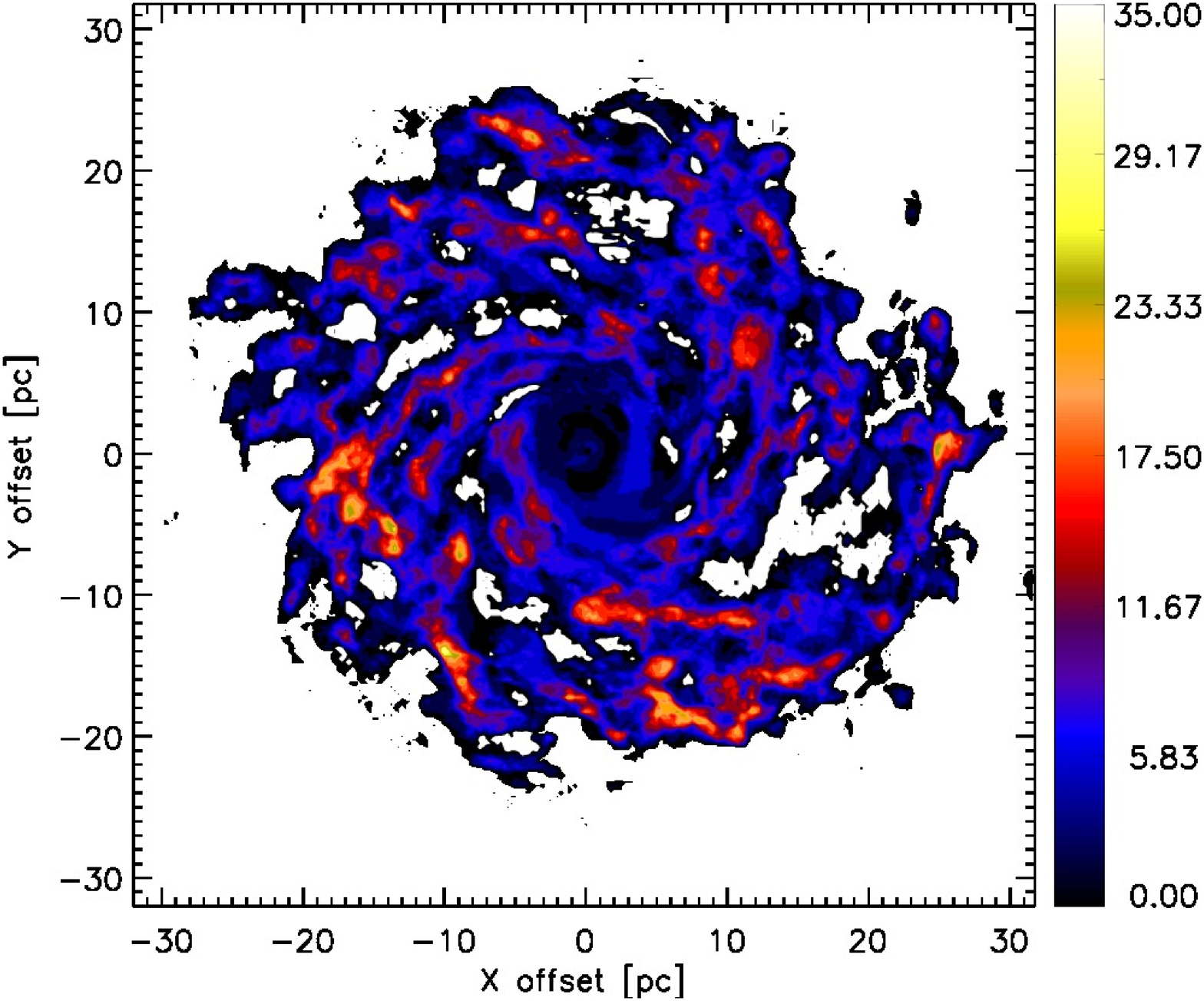}%
\hfill\includegraphics[angle=0,height=5.5cm]{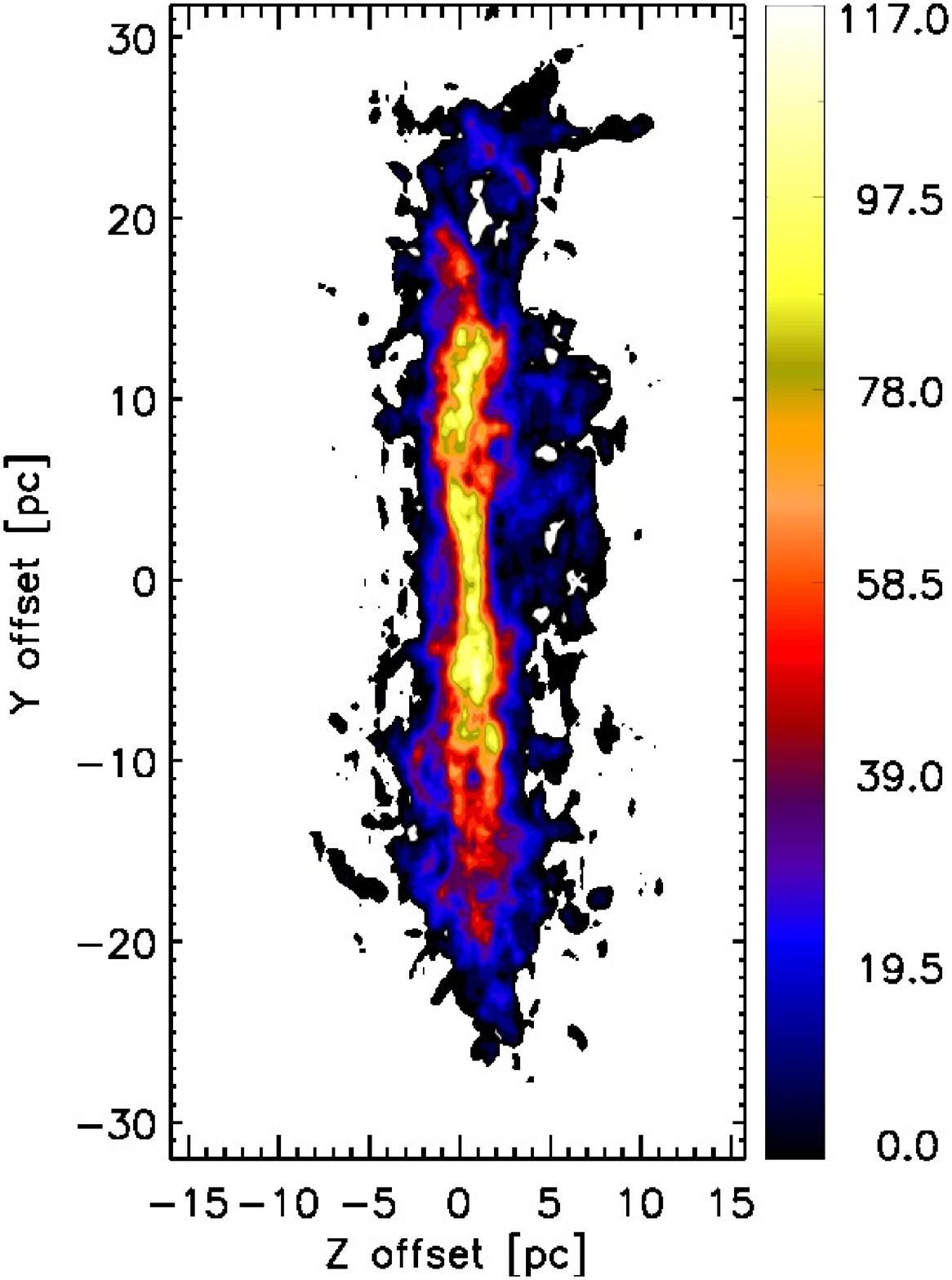}\hspace*{\fill}\\

\vspace{-0.5cm}

%\hfill\includegraphics[angle=0,height=5.5cm]{Tk_LE_100K_xy_cinv}%
%\includegraphics[angle=0,height=5.5cm]{Tk_LE_100K_yz_cinv}\hspace*{\fill}

\caption{{\footnotesize Number of grid elements in the 3-D hydrodynamical model that has both density $n(\rm H_2)$ higher than $10^2~\3cm$ and temperature $T_k$ lower than $10^4~\rm K$, along the line-of-sight of the X-Y plane (\textit{left}) and the Z-Y plane (\textit{right}). These are the grid elements that would contribute the most to the emission of molecular gas irradiated by the X-ray flux emitted from the central SMBH.}}
\label{fig:contributing-cells}
\end{figure}

One aspect that needs to be kept in mind is that not all the grid elements shown in density and temperature distribution maps of the 3-D code by \citetalias{wada09} will contribute to the emission of, particularly, molecular lines.
Figure~\ref{fig:contributing-cells} shows the number of grid elements with relatively cold temperature ($T_K<10^4~\rm K$) and moderate density ($n(\rm H_2)>10^2~\3cm$) that can contribute to the molecular emission emerging along the line-of-sight of the face-on (X-Y plane) and edge-on (Z-Y plane) viewing angles. At higher temperatures and lower densities, the fractional abundances of molecular species would be very low ($<10^{-10}$) and their contribution to the molecular emission lines would be negligible due to the low collisional excitation. The structure observed then is quite different than when considering the full density and temperature distribution at any cross section of the computational box.
Although the maps of the main contributing grid elements represent a close estimate of the structure that we would expect to observe in molecular emission, they do not take into account the optical depth effects nor the (sub-)thermal excitation of the molecular energy levels that are treated in the radiative transfer calculations described in Sec.~\ref{sec:beta3D}. The actual structure of emission lines then depends on the local density, temperature, and the radiation flux impinging at each grid element.

%-----------------------------------------------------------------------------
%-----------------------------------------------------------------------------

The heating of the gas and dust by X-rays emanating from the AGN, as well as the chemical abundances in the cold ($<500~\rm K$) gas, are not computed in hydrodynamical models. Including the (time dependent) chemical evolution at each step in the hydrodynamical simulations would take too long with the current computational resources. Therefore, after a realization of the state-of-art 3-D hydrodynamical simulation, the XDR/PDR code by \citet{meijerink05} is used to estimate the chemical abundances, based on the local density and impinging X-ray flux at each grid cell of the computational box. This code is depth dependent up to large columns ($N_{\rm H}\sim10^{25}~\2cm$), and considers a large (over 100 species) chemical network.
In the following sections we describe the calculation of the X-ray flux, and the impact that it has on the chemistry and heating of the atomic and molecular gas around the AGN.

\subsection{The X-ray flux model}\label{sec:x-ray-flux}

For our 3-D hydrodynamical model we have a $M_{BH}=1.3\times 10^7 M_{\sun}$ SMBH \citepalias{wada09}, so we can estimate the monochromatic luminosity of the AGN model at the rest frame wavelength $\lambda=510$ nm as follows

\begin{equation}\label{eq:luminosity}
 \lambda L_{\lambda}(510~{\rm nm})= 10^{44}\times \left[ \frac{10^{-7}}{a} \left( \frac{M_{BH}}{M_{\sun}}\right)\right]^{1/b}~~~~~~~~{\rm erg~s^{-1}}.
\end{equation}

\noindent
With $a=5.71^{+0.46}_{-0.37}$ and $b=0.545\pm0.036$ \citep[e.g.,][]{kaspi00} we have $\lambda L_{\lambda}(510~{\rm nm})=6.62\times10^{42}~\ergs$. Using the same bolometric to monochromatic luminosity factor as in \citet{kaspi00} we can determine the bolometric (total radiant energy) luminosity as $L_{bol}\approx 9\lambda L_{\lambda}(510~{\rm nm})~\ergs$.

The incident specific flux is assumed to have a spectral shape of the form
\begin{equation}\label{eq:specific-flux}
 F_i(E)=F_0 \left( \frac{E}{1~{\rm keV}} \right)^{\alpha} exp\left(-E/E_c \right)~~~~~~~~{\rm erg~s^{-1}~cm^{-2}~eV^{-1}},
\end{equation}

\noindent
where $E=h\nu~\rm eV$, $\alpha\approx0.9$ is the characteristic spectral index of the power-law components of Seyfert 1 galaxies \citep[e.g.,][]{pounds90, madejski95, zdziarski95}, $E_c$ is the {f high energy} cut-off which can be $\gtrsim100~\rm keV$, $200~\rm keV$ or $550~\rm keV$ depending on the sample of AGNs \citep[e.g.,][]{madejski95}, and $F_0$ {is a constant we estimate later to match the fraction of the total luminosity emitted in X-rays at the central grid point in the data cube}. {On the other hand, the lower energy cut-off would depend on the shielding column density seen by the X-ray flux at each grid point. Soft X-rays (photons with energy $<1~\rm keV$) cannot be effectively attenuated by columns $<10^{22}~\2cm$. However, as it is shown later in Sec.3.1, the column densities typically found in the inner $\sim25~\rm pc$ of the torus are larger than $10^{22}~\2cm$. Therefore, we only consider the hard X-rays between $1~\rm keV$ and $100~\rm keV$ as relevant for our X-ray chemical model.} So we integrate eq.(\ref{eq:specific-flux}) over this energy range in order to obtain the hard X-ray flux ($F_{hard}$) as

\begin{equation}\label{eq:bolometric-flux}
 F_{hard}=\int_{1~\rm keV}^{100~\rm keV} F_0 \left( \frac{E}{1~{\rm keV}} \right)^{\alpha} e^{-E/E_c} dE~~~~~~~~{\rm erg~s^{-1}~cm^{-2}}.
\end{equation}

Considering that only $\sim$10\% \citep{schleicher10} of the total luminosity is emitted in X-rays, we have that $F_{hard}=0.1\times L_{bol}/4\pi r_0^2$, where $r_0$ is the distance from the central black hole which, for our purpose, is assumed to be the size of a grid cell ($r_0=\parallel \vec{r}_0(x_0,y_0,z_0) \parallel=0.25$ pc) for the central unresolved grid point. From this we find that $F_0\approx1.4\times 10^2~{\rm erg~s^{-1}~cm^{-2}~eV^{-1}}$.

For the rest of the cells, at position $\vec{r}(x,y,z)$ in the cube (a vector) the flux decreases not only with the square of the distance $r=\parallel \vec{r}(x,y,z) - \vec{r}_0(x_0,y_0,z_0) \parallel$ from the central black hole, but also because of the opacity $\tau(E,\vec{r})$ of each grid cell at position $\vec{r}(x,y,z)$ along the radial path. The opacity is defined as

\begin{equation}\label{eq:opacity}
\tau(E,\vec{r})=\sigma_{pa}(E)N_{\rm H}(\vec{r}), 
\end{equation}

\noindent
where $\sigma_{pa}(E)$ is the photoelectric absorption cross section per hydrogen nucleus, and $N_{\rm H}(\vec{r})$ is the total column density of hydrogen along the radial path from the central black hole to the position $\vec{r}$ in the computational box. The photoelectric absorption is calculated from all the species as
\begin{equation}
\sigma_{pa}(E)=\sum_i \mathpzc{A}_i^{total} \sigma_i(E),
\end{equation}

\noindent
with the total (gas and dust) elemental abundances, $\mathpzc{A}_i^{total}$, taken from \citet{meijerink05}, and the X-ray absorption cross sections, $\sigma_i(E)$, from \citet{verner95}.
The total {hard} X-ray flux $F_{hard}(\vec{r})$ ($\ergscm$) impinging on an arbitrary grid cell at position $\vec{r}$ is then calculated as

\begin{equation}\label{eq:X-ray-flux}
 F_{hard}(\vec{r})=\left( \frac{0.25~\rm pc}{r} \right)^2 \int_{1~\rm keV}^{100~\rm keV} F_0 \left( \frac{E}{1~{\rm keV}} \right)^{\alpha} e^{-E/E_c} e^{-\tau(E,\vec{r})} dE .
\end{equation}

%-----------------------------------------------------------------------------
\begin{figure}[htp]
%\centering

\hspace*{\fill}\includegraphics[angle=0,height=0.3\textwidth]{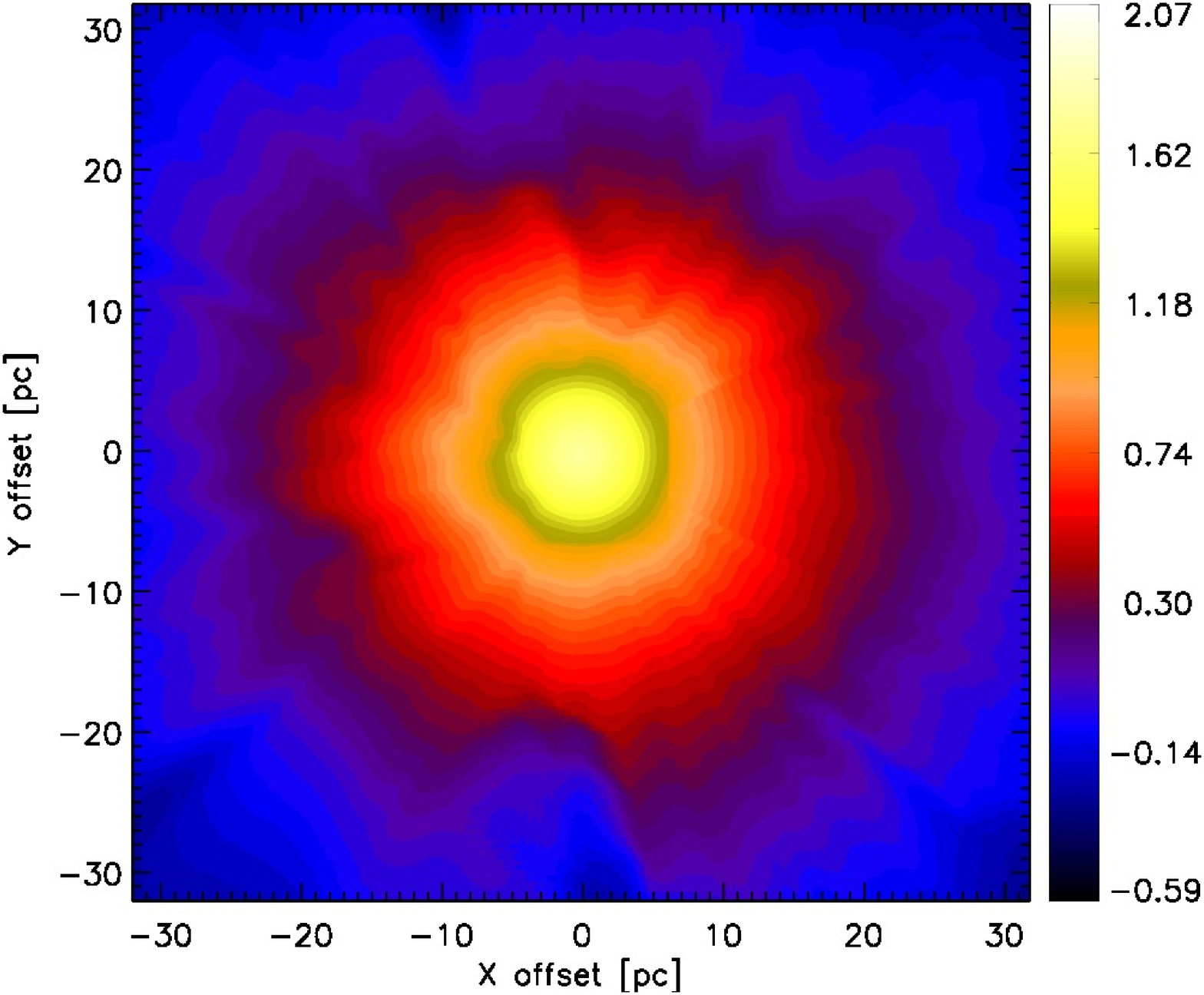}%
\hfill\includegraphics[angle=0,height=0.3\textwidth]{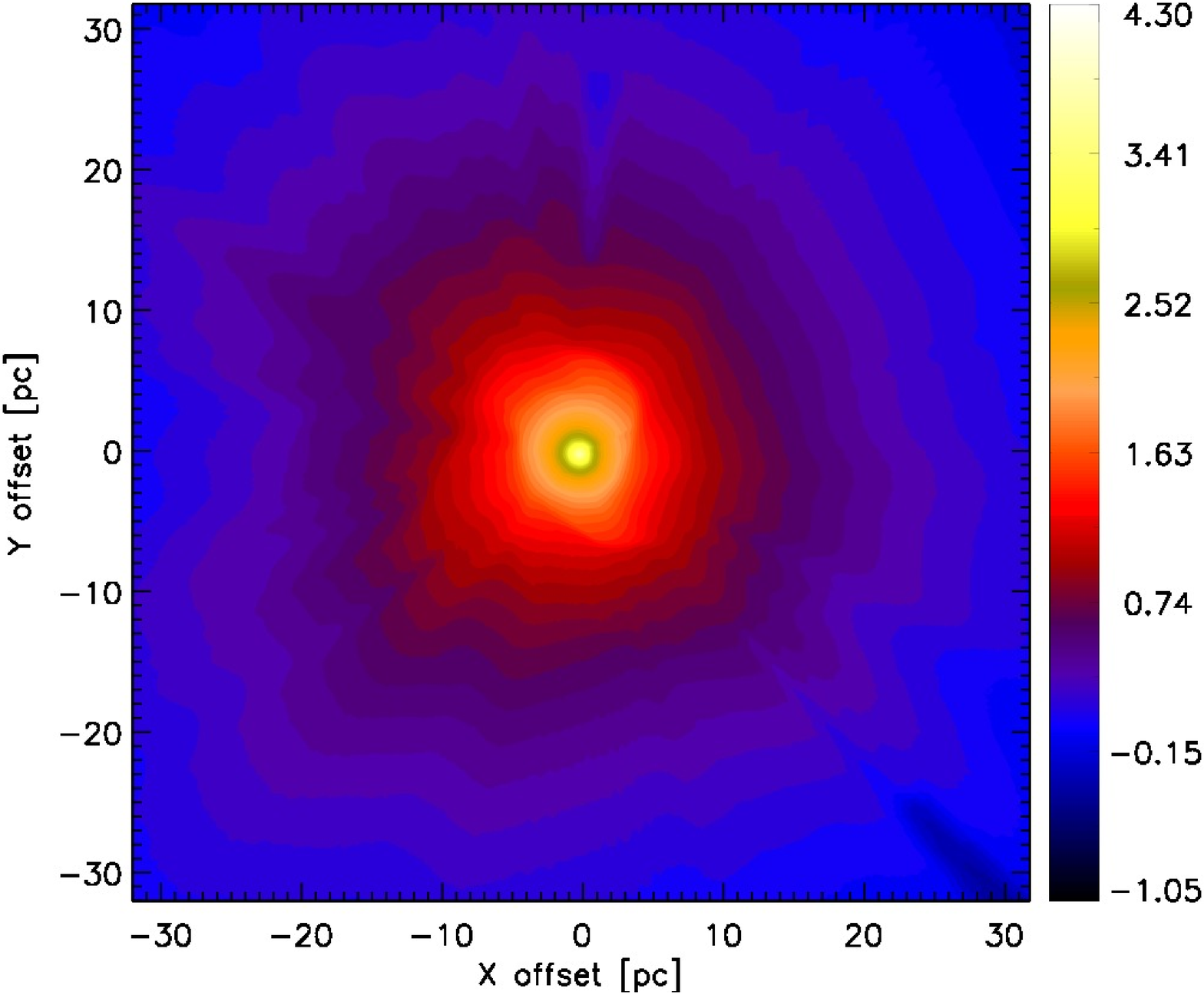}%
\hfill\includegraphics[angle=0,height=0.3\textwidth]{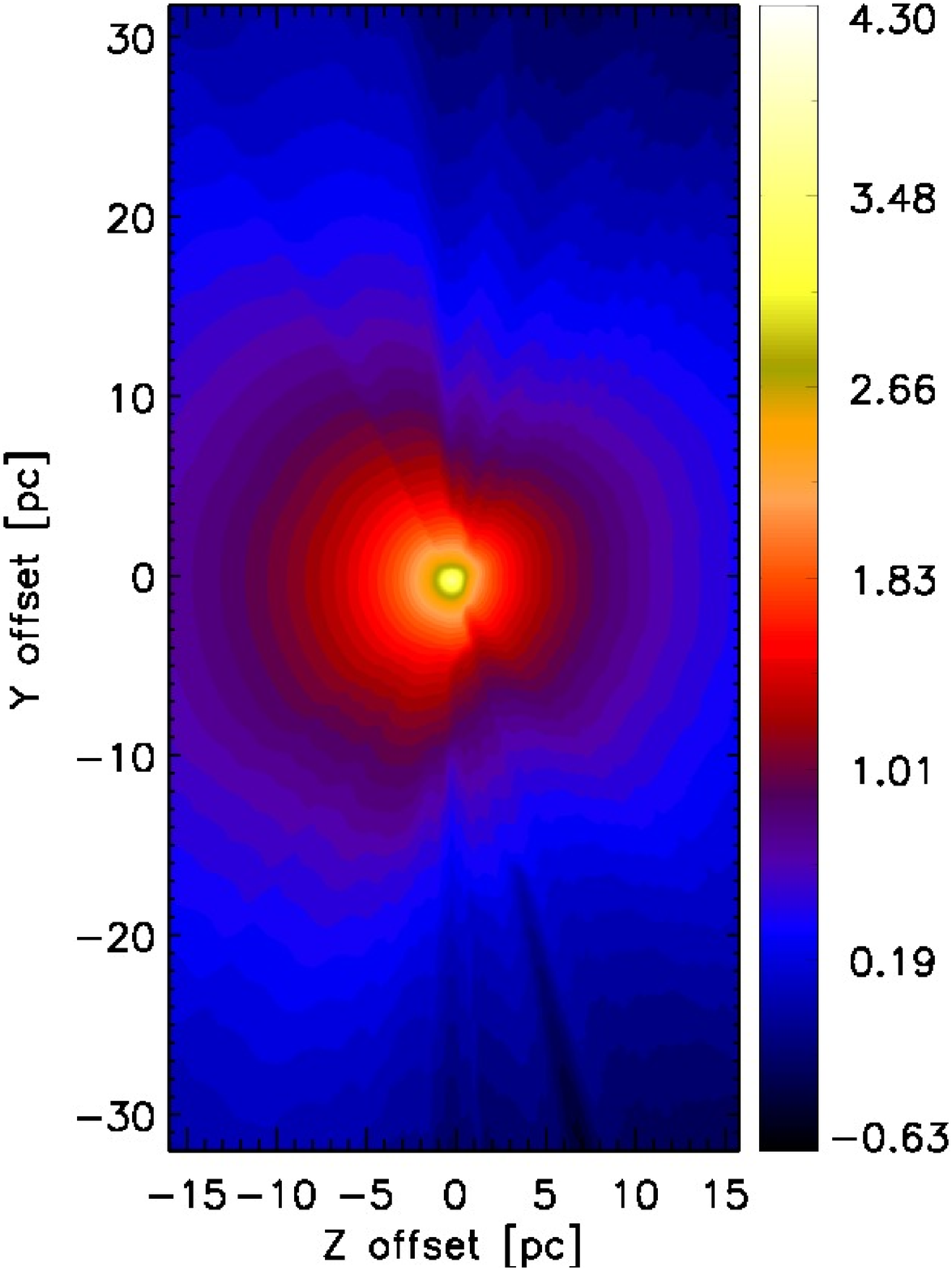}\hspace*{\fill}\\

\caption{{\footnotesize Impinging {hard} X-ray flux (in logarithmic scale and units of $\ergscm$) as seen in (\textit{left}) the X-Y plane $3.5~\rm pc$ below the mid-plane of the AGN torus; (\textit{middle}) the flux in the actual X-Y mid-plane; and (\textit{right}) the flux seen in the Y-Z plane. Note that the X-ray flux distribution is not homogeneous. The shadow-like shapes are due to the X-ray absorption by the grid cells with different densities found along the radial path from the AGN.}}
\label{fig:x-ray-flux}
\end{figure}
%-----------------------------------------------------------------------------

Figure~\ref{fig:x-ray-flux} shows the {hard} X-ray flux estimated for the 3-D hydrodynamical model by \citetalias{wada09} in the X-Y plane 3.5$~\rm pc$ below the mid-plane of the AGN torus (\textit{left panel}), as well as the flux in the actual X-Y mid-plane (\textit{middle panel}). The \textit{right panel} shows the flux in the Y-Z plane. The grid cells with different densities cause more or less absorption of the X-ray flux along the radial path from the central SMBH, producing shadow-like shapes and an inhomogeneous flux field.

This total bolometric X-ray flux (from now on $F_X$) is used along with the total gas density of a grid cell as input parameters of the XDR/PDR chemistry code to estimate the abundances of several species. The formalism is described in the next section.

\subsection{Chemical abundances and temperature}\label{sec:abundances}

For each grid point in the computational box we have the total gas density from the 3-D hydrodynamic model. Since each grid point represents a physical (unresolved) scale of $0.25~\rm pc$, we also know the total column density that the impinging radiation flux will go through. Thus, the total gas density, the radiation flux, and the column density of each grid point are used as input parameters for the XDR/PDR chemical model by \citet{meijerink05} to compute the densities and fractional abundances of different species at different column densities throughout the $0.25~\rm pc$ slab. In addition, we also get the temperature of the gas (as a function of the column density) derived self-consistently from the chemical and thermal balance computed in the XDR/PDR code. Hence, we can compare the temperatures and H$_2$ densities estimated from the X-ray free 3-D hydrodynamical model with those computed considering the X-ray effects (see Sec.~\ref{sec:temps-dens}).

%-----------------------------------------------------------------------------
\begin{figure}[tp!]
%\centering

\hspace*{\fill}\includegraphics[angle=0,width=0.46\textwidth]{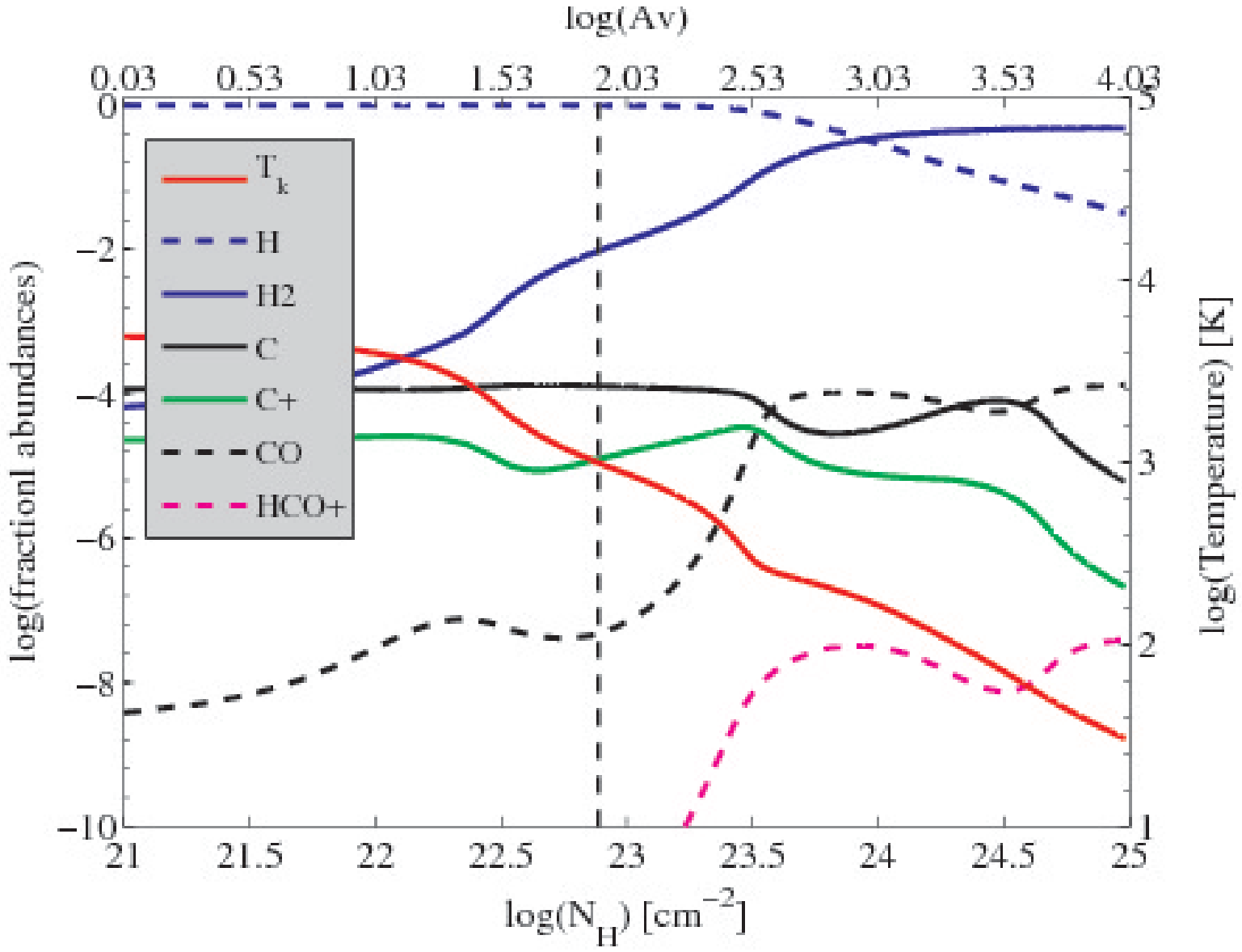}%
\hfill\includegraphics[angle=0,width=0.46\textwidth]{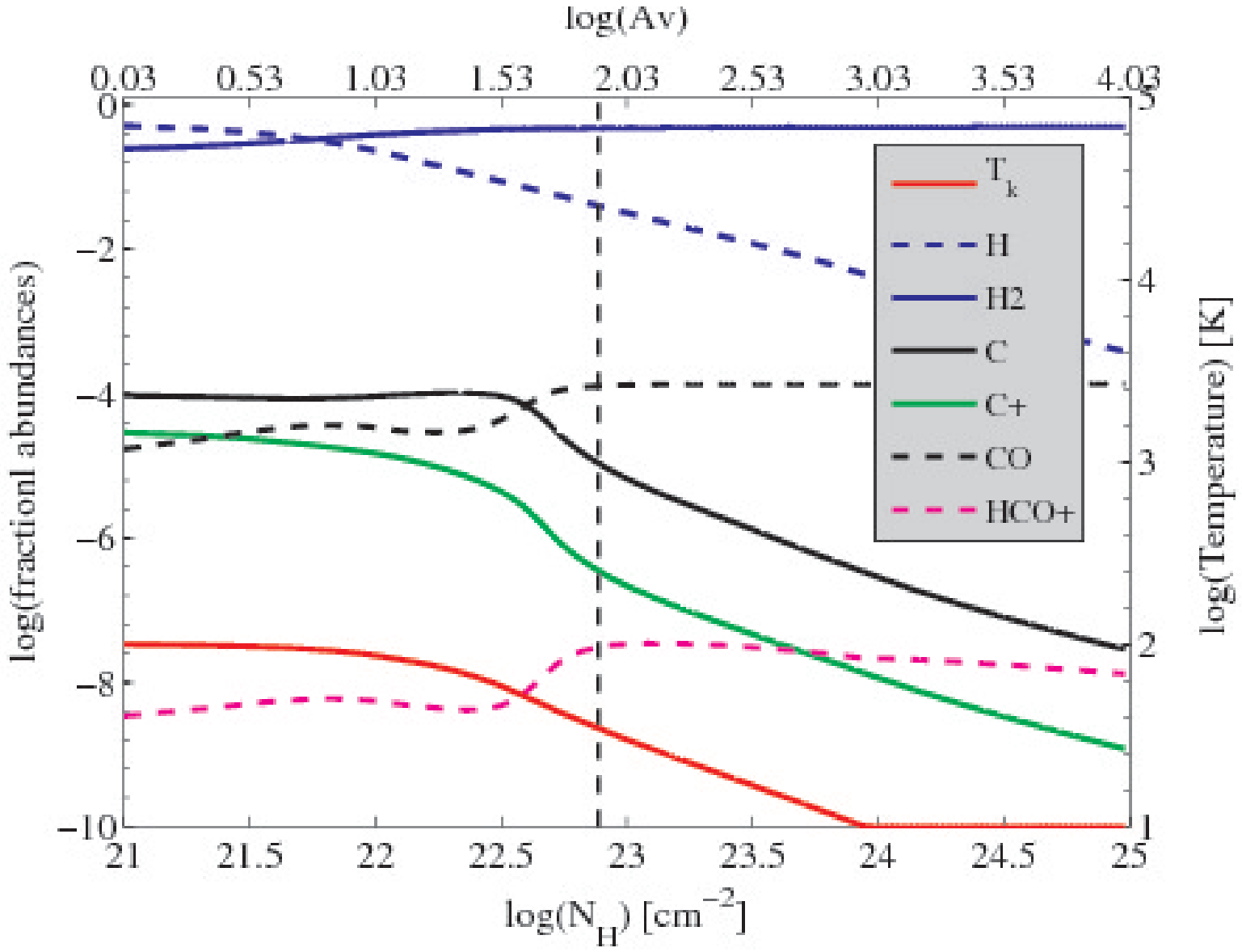}\hspace*{\fill}

\caption{{\footnotesize \textit{Left panel} -- Fractional abundances of various atomic and molecular species in a slab that represents one unresolved grid cell of $d=0.25$ pc and density of $n_{\rm H}=10^5~\3cm$, in the 3-D hydrodynamic model. The X-ray flux of $160~\ergcms$ penetrates on the left side of the slab and affects the chemistry as it is absorbed throughout the column of gas and dust. The vertical dashed-dotted line indicates the total column density $N_{\rm H}=n_{\rm H}\times d\approx8\times10^{22}~\2cm$ of the slab. For this particular X-ray flux, a denser slab would be required to observe a significant abundance of molecules like \hcop. \textit{Right panel} -- Fractional abundance for the same slab as above, but with an impinging radiation flux of $1.6~\ergcms$. Note the higher abundances of H$_2$, CO, and \hcop at lower column densities, while the abundance of H, C and C$^+$ decrease earlier and faster throughout the slab. The overall temperature is also lower in this case.
}}
\label{fig:XDR-model}
\end{figure}
%-----------------------------------------------------------------------------

Figure~\ref{fig:XDR-model} shows the fractional abundances of some species as a function of the column density for impinging radiation fluxes of $160~\ergcms$ (left panel) and $1.6~\ergcms$ (right panel). The slab represents a single unresolved grid point in the computational box, with a fixed scale of $d=0.25~\rm pc$ and total hydrogen density $n_{\rm H}=10^5~\3cm$. This gives a total column density of $N_{\rm H}=n_{\rm H}\times d\approx10^{23}~\2cm$, which is marked with a vertical slashed-dotted line. The XDR/PDR model, though, was executed to compute the abundances of the species up to a column of $10^{25}~\2cm$ in order to show how deep the X-rays can penetrate depending on the strength of the radiation field. For the strong X-ray flux (left panel) and the actual column density ($\sim10^{23}~\2cm$) of the slab, the abundances of molecules like \hcop\ will be negligible. Since the scale of the grid points is fixed, a slab denser than $10^5~\3cm$ would be required in order to observe a significant abundance of \hcop\ at larger ($N_{\rm H}>10^{23}~\2cm$) columns. Conversely, a weaker radiation flux of $1.6~\ergcms$ (right panel) impinging on the slab will produce a higher abundance of e.g., H$_2$, CO, \hcop\ at lower column densities, while H, C and C$^+$ would be less abundant, in a column averaged sense, than in the previous case. An overall lower temperature is also observed when the radiation flux is weaker, since X-ray photons are completely absorbed at a column of $\gtrsim10^{24}~\2cm$.

Since a grid point of the 3-D hydrodynamic model is unresolved, and the chemical abundances given by the XDR/PDR code depend on the column density, we estimate an abundance that is \textit{representative} for the particular slab and for each species in the chemical network from the abundances observed throughout the slab of $0.25~\rm pc$. We compute the total fractional abundance $<\mathpzc{A}_X>$ of the species $X$=\twco, \hcn, etc., as
\begin{equation}\label{eq:col-w-abun}
  <\mathpzc{A}_X>=\frac{\int n_X(l)~dl}{\int n_{\rm H}~dl},
\end{equation}

\noindent
where $n_X(l)$ ($\3cm$) is the density of the species $X$ at the layer $l$ ($\rm cm$) in the cloud, and $n_{\rm H}$ is the total density of the slab, so the denominator is actually the total column density $N_{\rm H}$ ($\2cm$) of a particular grid point. 

Because the variation of the abundance of a species through the slab is different for every species (e.g., \ci\ is more abundant at the edge of the slab where $T$ is high, while \twco\ is more abundant deep into the slab, where $T$ is low), we require a gas temperature that is representative of the layers in the slab where the abundance of the species is higher (since those layers contribute the most to the line emission). Therefore, we compute an abundance-weighted average temperature, throughout an unresolved grid point, as
\begin{equation}\label{eq:abun-wa-temp}
 <T_X>=\frac{\int \mathpzc{A}_X(l) T(l)~dl}{\int \mathpzc{A}_X(l)~dl},
\end{equation}

\noindent
which gives different temperatures for different species in the same grid point. For instance, with the lower ($1.6~\ergcms$) X-ray flux (\textit{right panel} in Fig.~\ref{fig:XDR-model}) the total fractional abundance of CO is $\sim10^{-4}$ with an abundance-weighted average temperature of $\sim37~\rm K$. While for C we have an abundance of $\sim3\times10^{-5}$ and a representative temperature of $\sim66~\rm K$.

Due to the large number of grid elements in the hydrodynamical model ($256\times256\times128\sim8.4\times10^6$ data points) we require an optimized process in order to run the XDR/PDR code for the essential grid elements. Our criterion was to process only the grid points with total density larger than $100~\3cm$ and temperature lower than $10^4~\rm K$. This is because lower gas densities will have little weight in the excitation of the molecular species, while higher temperatures would lead to mostly ionized gas.
Once the abundances and abundance-weighted average temperatures have been determined for the selected grid elements of the 3-D hydrodynamical model, we can proceed to perform the radiative transfer calculations for the entire cube.

\subsection{3-D radiative transfer and line tracing}\label{sec:beta3D}

The advanced 3-D radiative transfer code $\beta$3D \citep{poelman05} has been optimized for heavy memory usage, to be able to use the $256\times256\times128$ elements data cube of the low resolution ($0.25~\rm pc$ scale) 3-D hydrodynamical models by \citetalias{wada09}. In principle, the temperature, density and velocity field derived from the hydrodynamical simulations can be used as the ambient conditions for the radiative transfer formalism. However, the temperature derived from the XDR model is found to be significantly different than the temperature obtained in the hydrodynamical model, as discussed in Sec.~\ref{sec:temps-dens}.
A multi-zone radiative transfer approach is used, in which the calculation of the level populations in a grid cell depends on the level populations of all the other cells through the different escape probabilities connecting adjacent grid points \citep{poelman05, poelman06}.

The collisional rates available in the LAMDA database \citep{schoier05} are used in a similar way as in the 1-D and 2-D radiative transfer codes RADEX \citep{vdtak07} and RATRAN \citep{hogerheijde00}, to calculate the level populations of different atomic and molecular species (e.g. \ci, \cii, \oi, \twco, \thco, \hcn, \hcop, \hnc, \cn, etc.). For this we use the commonly adopted main collision partner H$_2$ for the radiative transfer calculations of all the molecules. Although the contribution of helium atoms to the total collision density for CO is just about $10^{-2}$, we also include (for completeness) He as an additional collision partner by extrapolating (see Appendix~\ref{sec:appendix}) the rate coefficients reported in \citet{cecchi02}. {For the case of \cii\ discussed in Sec.~\ref{sec:CII-to-CO}, we also use H and electrons as collision partners. With the exception of electrons, the density of all the collision partners $n(\rm H)$ and $n(\rm H_2)$ at each grid point is given by the hydrodynamical model (as described in \citepalias{wada09}), since our aim is to know how the line emissions would look like for this particular model of an AGN torus. The hydrodynamical model, however, does not yet trace the evolution of electron density. So we use $n(e^-)$ derived from the XDR model.}

The line intensities, including kinematic structures in the gas, and optical depth effects, are computed with a ray-tracing approach for arbitrary rotation (viewing) angles about any of the three axes of the computational box.
The emerging specific intensity is computed using the escape probability formalism described in \citet{poelman05}

%while the second one is the classical approach considering several layers of different systemic velocities along the line of sight (e.g., Rybicki \& Hummer 1978; Rybicki \& Hummer 1983).

\begin{equation}
dI_{\nu}^z=\frac{1}{4\pi} n_i A_{ij} h \nu_{ij} \beta\left( \tau_{ij} \right) \left( \frac{ S_{ij} - I_b^{loc}(\nu_{ij})}{S_{ij}}  \right) \phi(\nu_{ij}) dz~,
\end{equation}

%\begin{equation}
%dI_{\nu}^z=\left( S_{ij} - I_b^{loc}(\nu_{ij} \right) \left( 1 - exp(-\tau_{\nu}(z)\right) exp\left( \sum_{i=z}^N \tau_{\nu}(z) \right) 
%\end{equation}

\noindent
where $dI_{\nu}^z$ has units of ${\rm [erg~cm^{-2}~s^{-1}~sr^{-1}~Hz^{-1}]}$, $n_i$ is the population density in the $i^{\rm th}$ level, $A_{ij}$ the Einstein $A$ coefficient, $h\nu_{ij}$ the energy difference between the levels $i$ and $j$, $S_{ij}$ the source function for the corresponding transition, $I_b^{loc}(\nu_{ij})$ the local background radiation at the frequency of the transition, and $\tau_{ij}$ is the {\it cumulative optical depth} that increases away from the observer, from the edge of the ray path to the $z^{th}$ layer.

%$\tau_{\nu}(z)$ is the {\it monochromatic optical depth} of the $z^{th}$ layer. The cumulative optical depth increases from top to bottom.

\section{Analysis and results}\label{sec:results}

Once the level populations of particular transitions have been estimated with the radiative transfer code, and the line tracing at a particular inclination angle has been completed, the resulting two-dimensional emission map can be exported into a regular FITS data cube. %This data cube can in turn be processed by practically any single-dish or interferometric data reduction package and visualization tool. 

\subsection{CO maps}\label{sec:co-maps}

%\begin{figure}[tp]
%\centering

%\includegraphics[angle=-90,width=12cm]{CO1_small_chan_x-45_Vkms}\\

%\caption{{\footnotesize \twco~$J=1\rightarrow0$ emission line for the inner $21\times21\times21$ grid cells ($\sim5\times5\times5$ pc$^3$) of the original cube, at an adopted distance of $1~\rm Mpc$, and with an inclination angle of $45^{o}$ about the X-axis. The maps and spectra are produced using the {\it view} task of the GILDAS/CLASS data reduction package.}}
%\label{fig:CO1-class}
%\end{figure}

%The map of the $J=1\rightarrow0$ emission line of \twco, from the inner $\sim5\times5\times5$ pc$^3$ of the AGN torus, is shown in Fig.~\ref{fig:CO1-class} using the \textit{view} task of the GILDAS/CLASS\footnote[0]{http://iram.fr/IRAMFR/GILDAS/} data reduction package. The line intensity has been rescaled assuming a distance of $1~\rm Mpc$ for the source, in order to obtain the spatial scale in arcseconds. The effects of the velocity field are observed in the different velocity shifts across the region mapped. The spectral lines, produced by the different grid elements found along the line-of-sight through each pixel of the map, are distributed around the arbitrary systemic velocity chosen for this system. This distribution is consistent with gas rotating around the AGN. The spectral lines reveal the non-homogeneous distribution of the ambient conditions, and kinematics of the gas, at the particular inclination angle of $45^{o}$ about the X-axis. 

\begin{figure}[htp]
%\centering

\hspace*{\fill}\includegraphics[angle=0,width=0.33\textwidth]{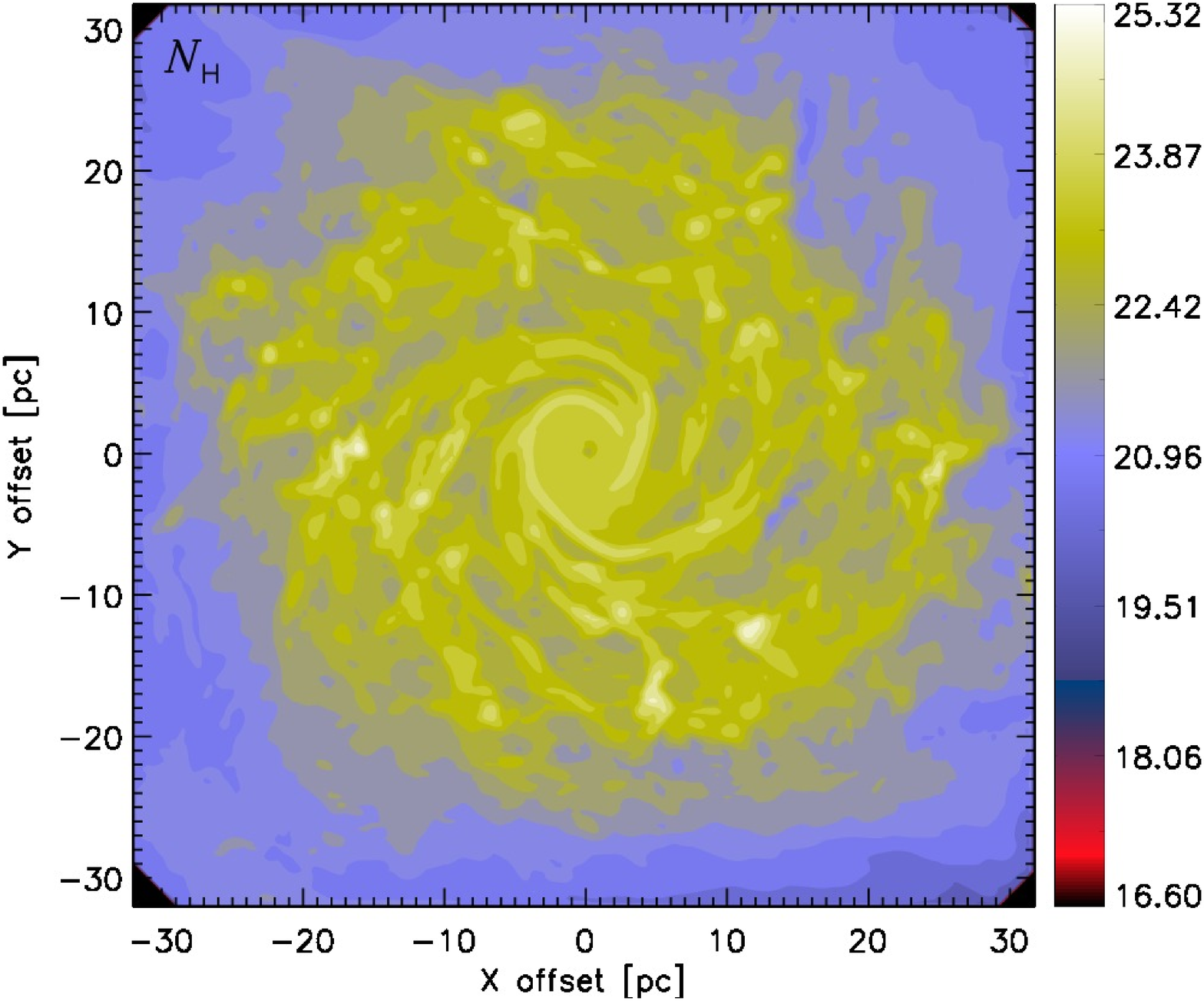}%
\hfill\includegraphics[angle=0,width=0.33\textwidth]{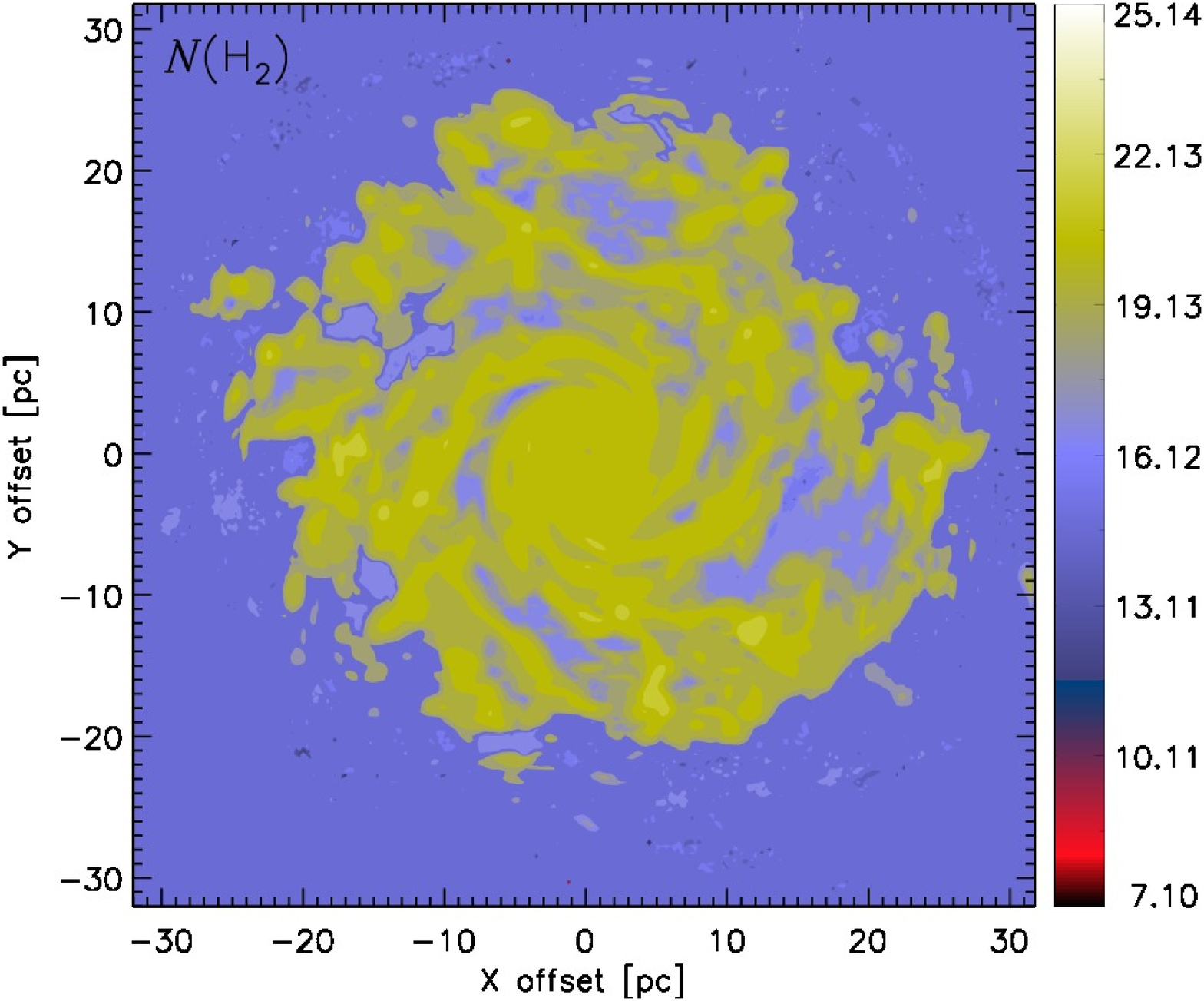}%
\hfill\includegraphics[angle=0,width=0.33\textwidth]{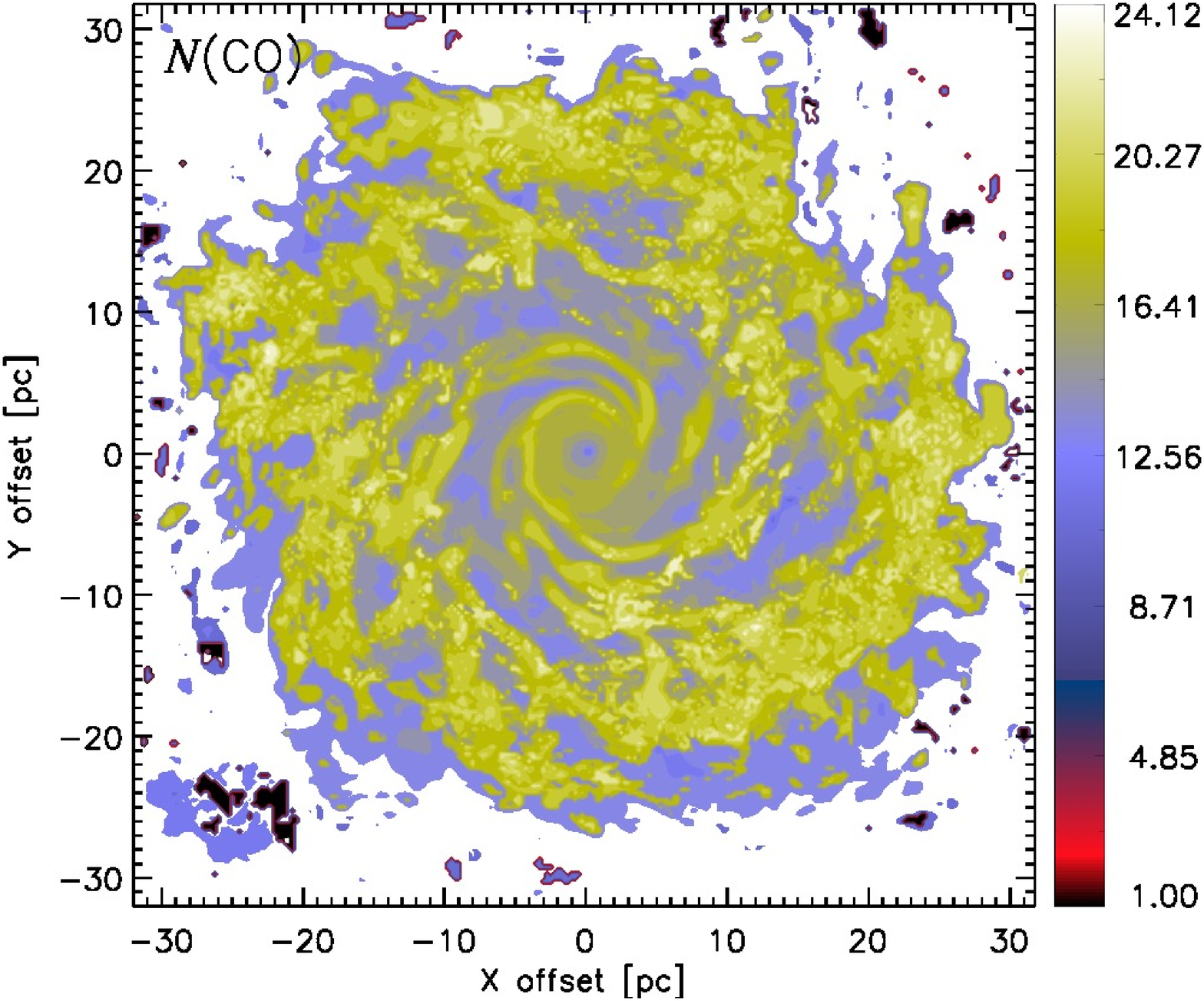}\hspace*{\fill}\\

\vspace{-0.6cm}

\hspace*{\fill}\includegraphics[angle=0,width=0.33\textwidth]{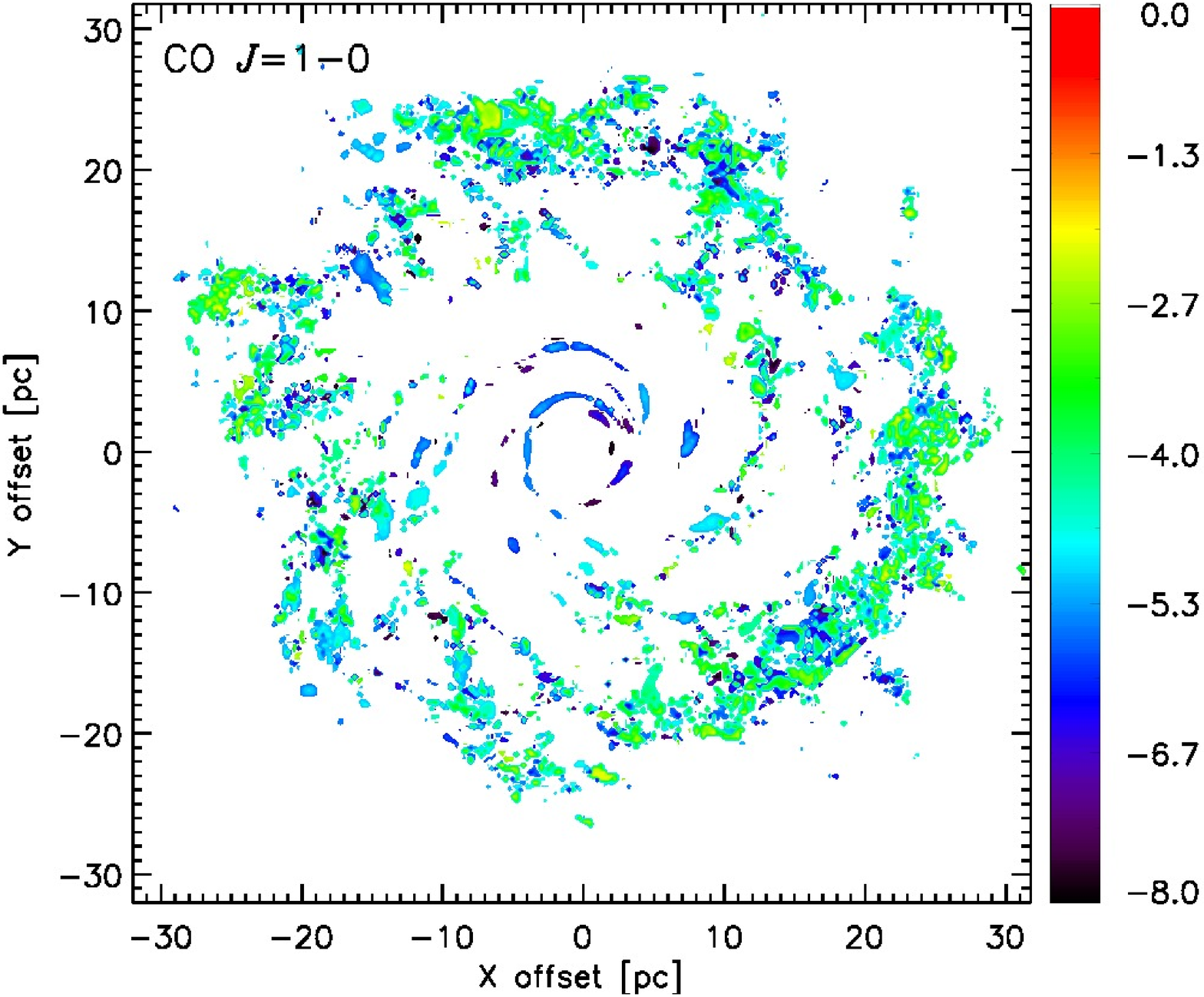}%
\hfill\includegraphics[angle=0,width=0.33\textwidth]{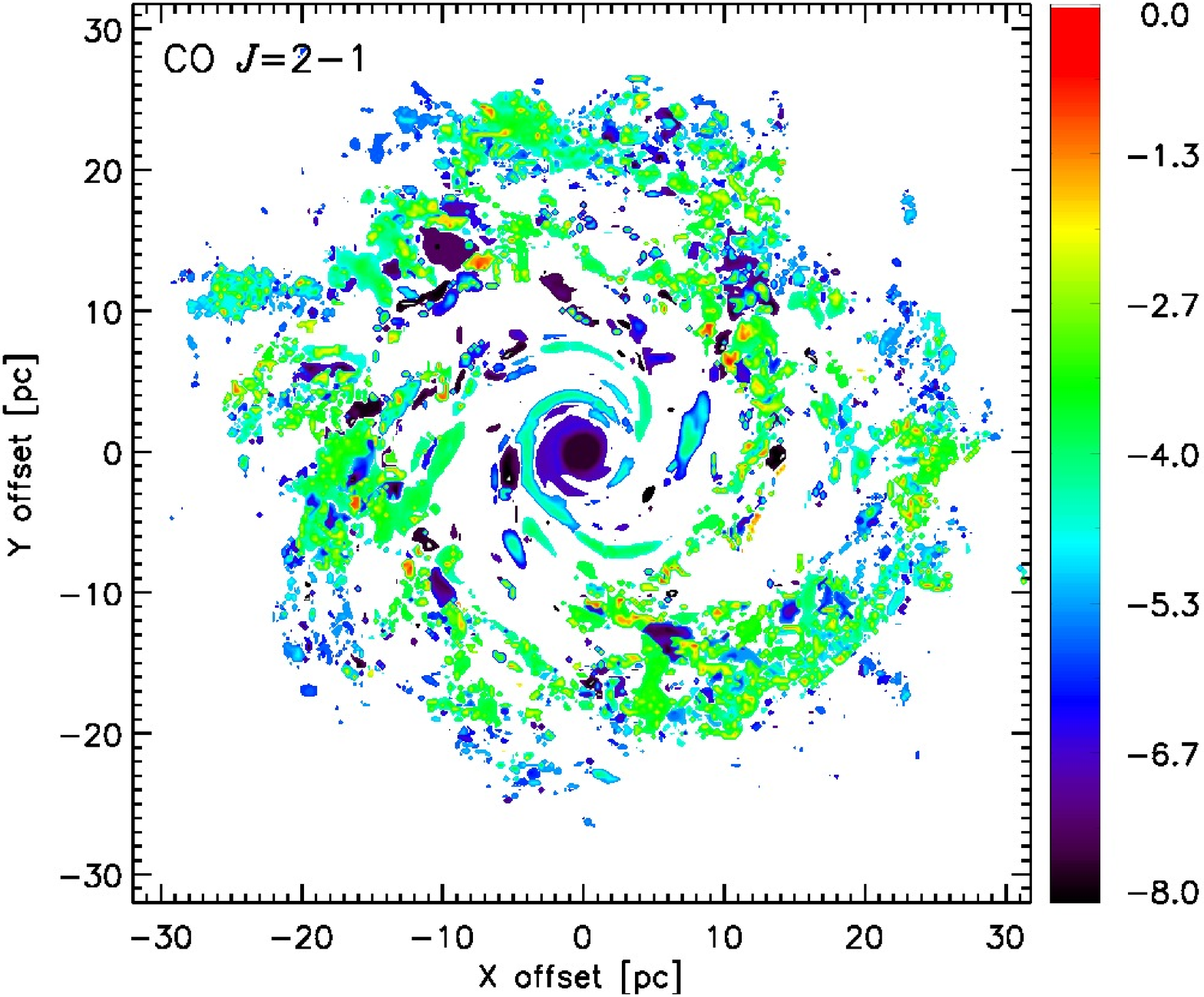}%
\hfill\includegraphics[angle=0,width=0.33\textwidth]{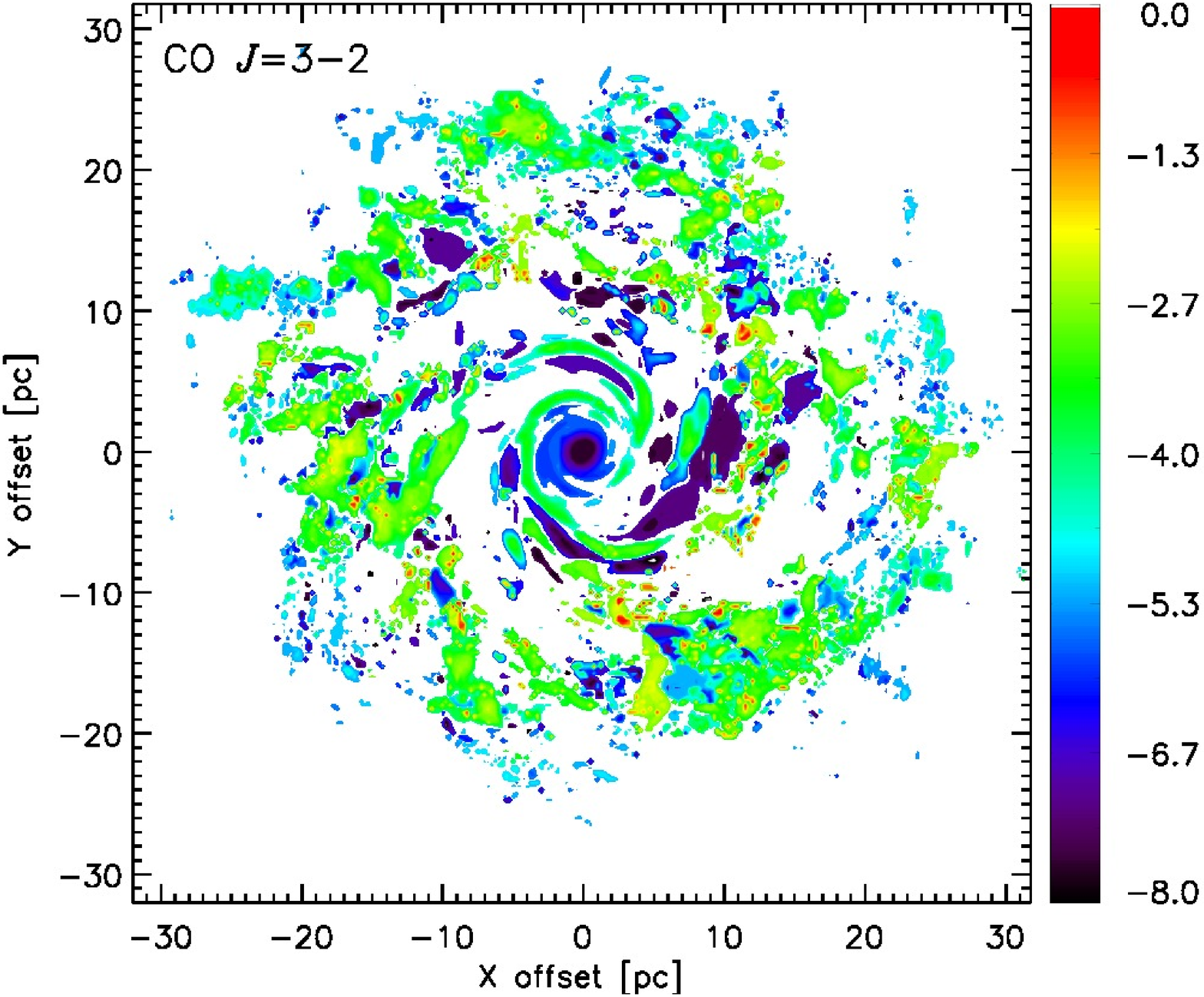}\hspace*{\fill}\\

\vspace{-0.6cm}

\hspace*{\fill}\includegraphics[angle=0,width=0.33\textwidth]{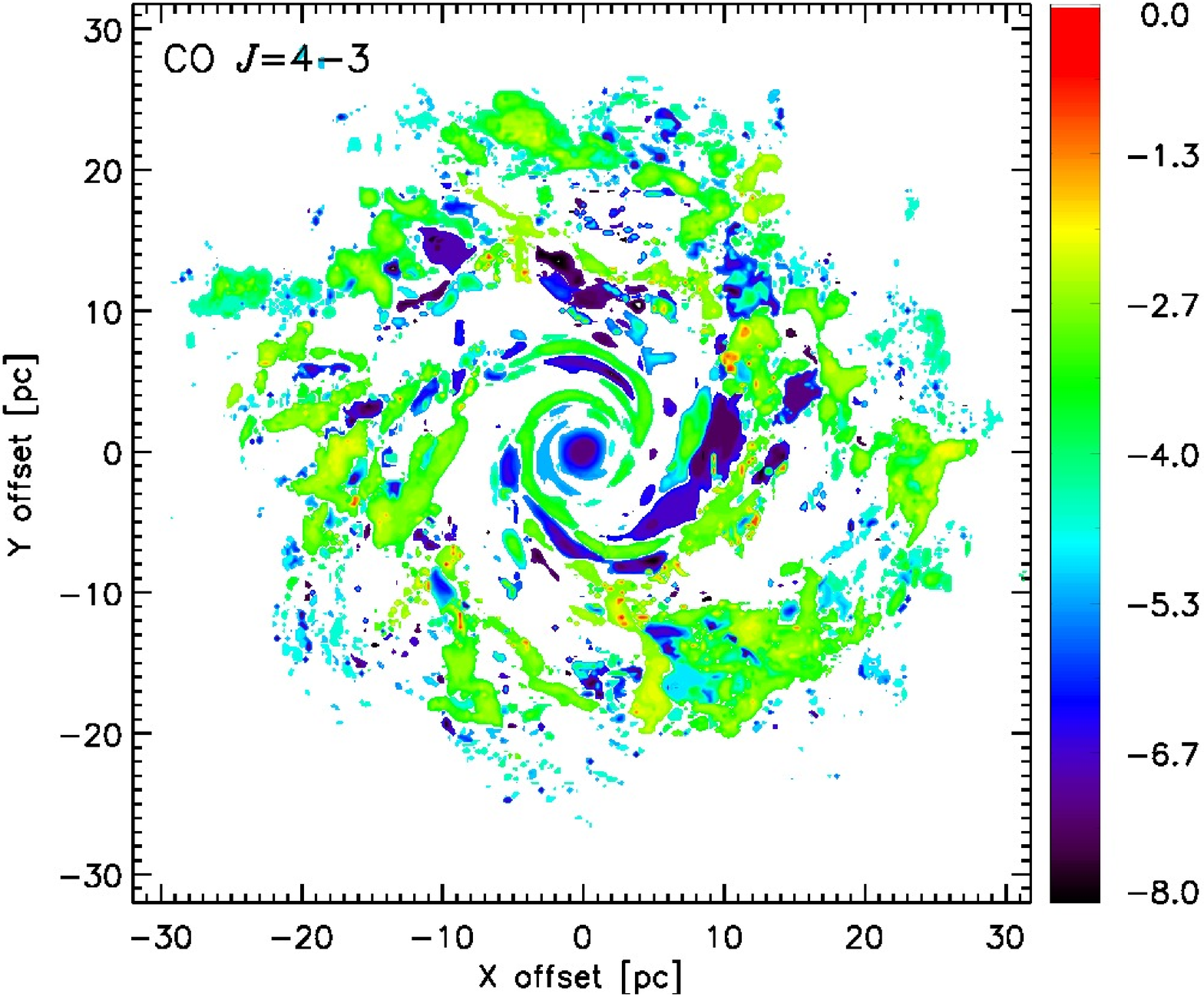}%
\hfill\includegraphics[angle=0,width=0.33\textwidth]{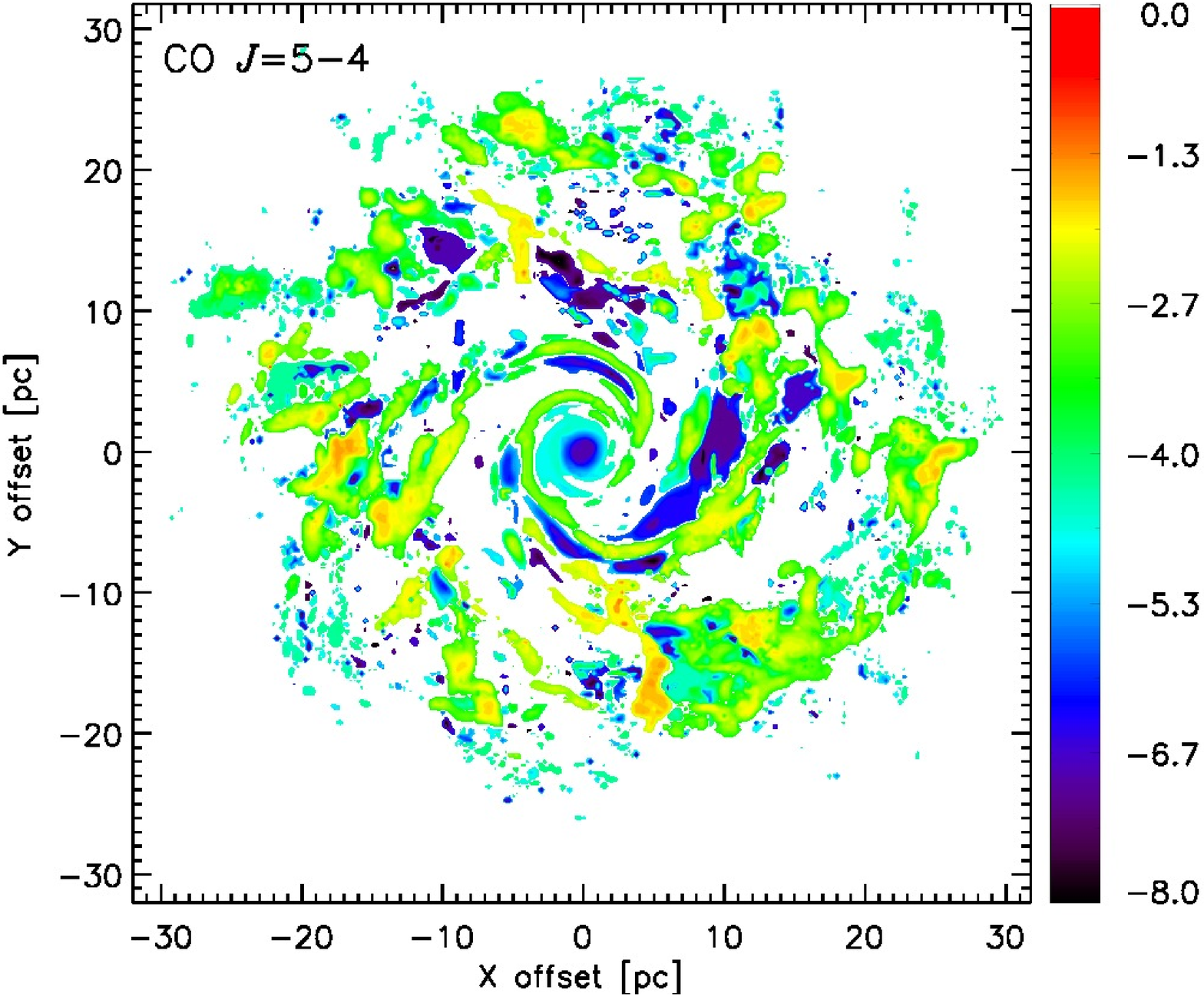}%
\hfill\includegraphics[angle=0,width=0.33\textwidth]{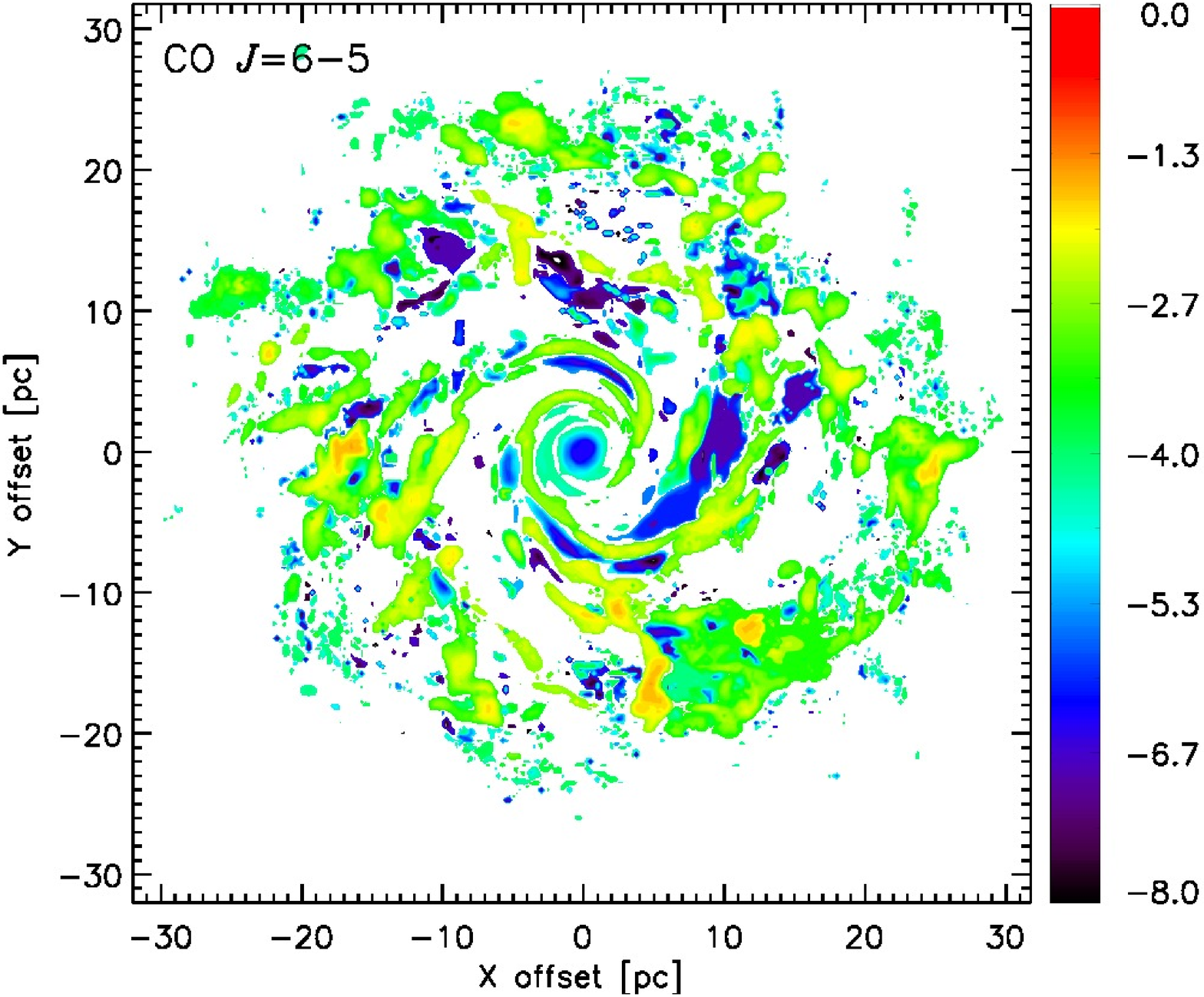}\hspace*{\fill}\\

\vspace{-0.6cm}

\hspace*{\fill}\includegraphics[angle=0,width=0.33\textwidth]{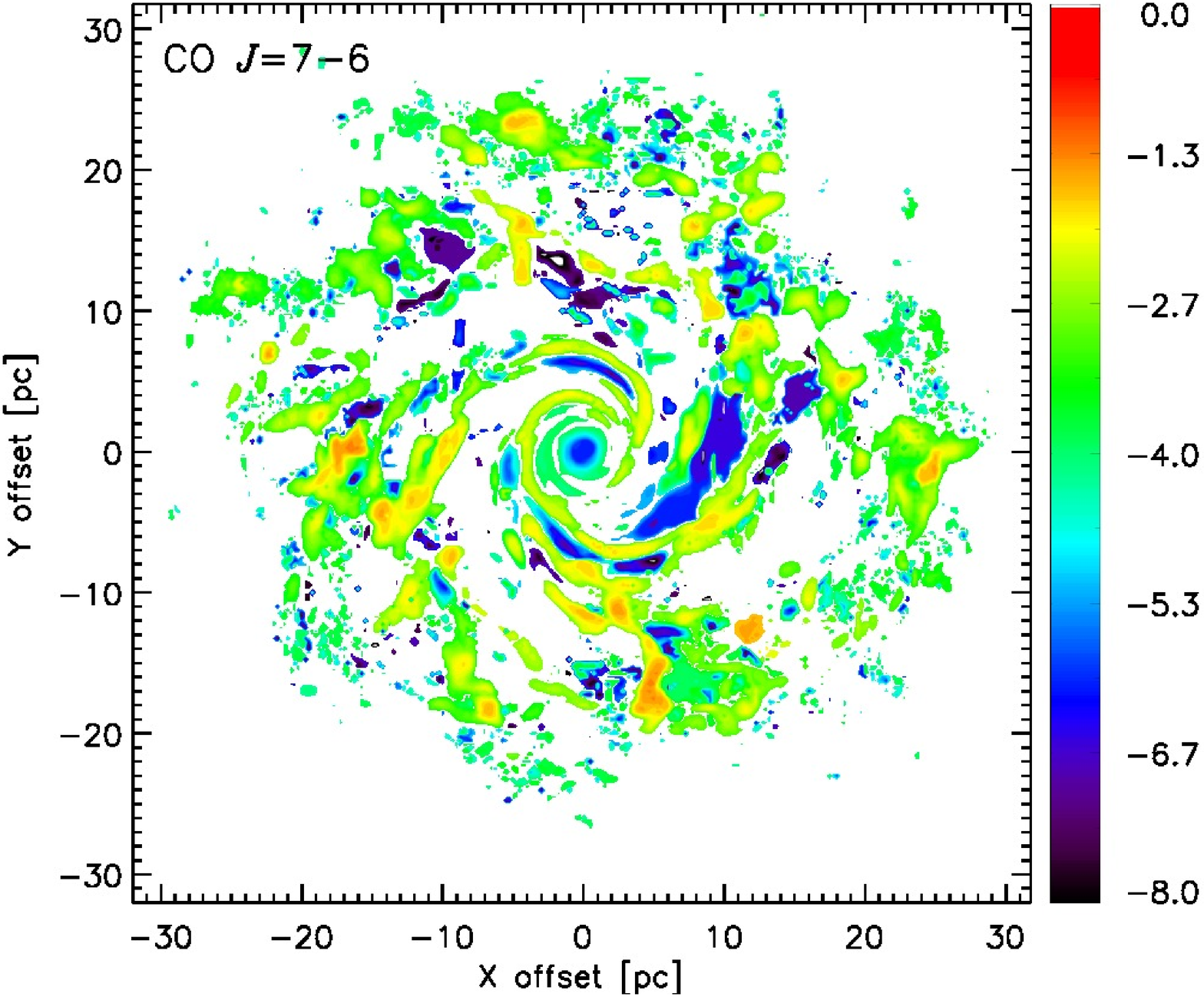}%
\hfill\includegraphics[angle=0,width=0.33\textwidth]{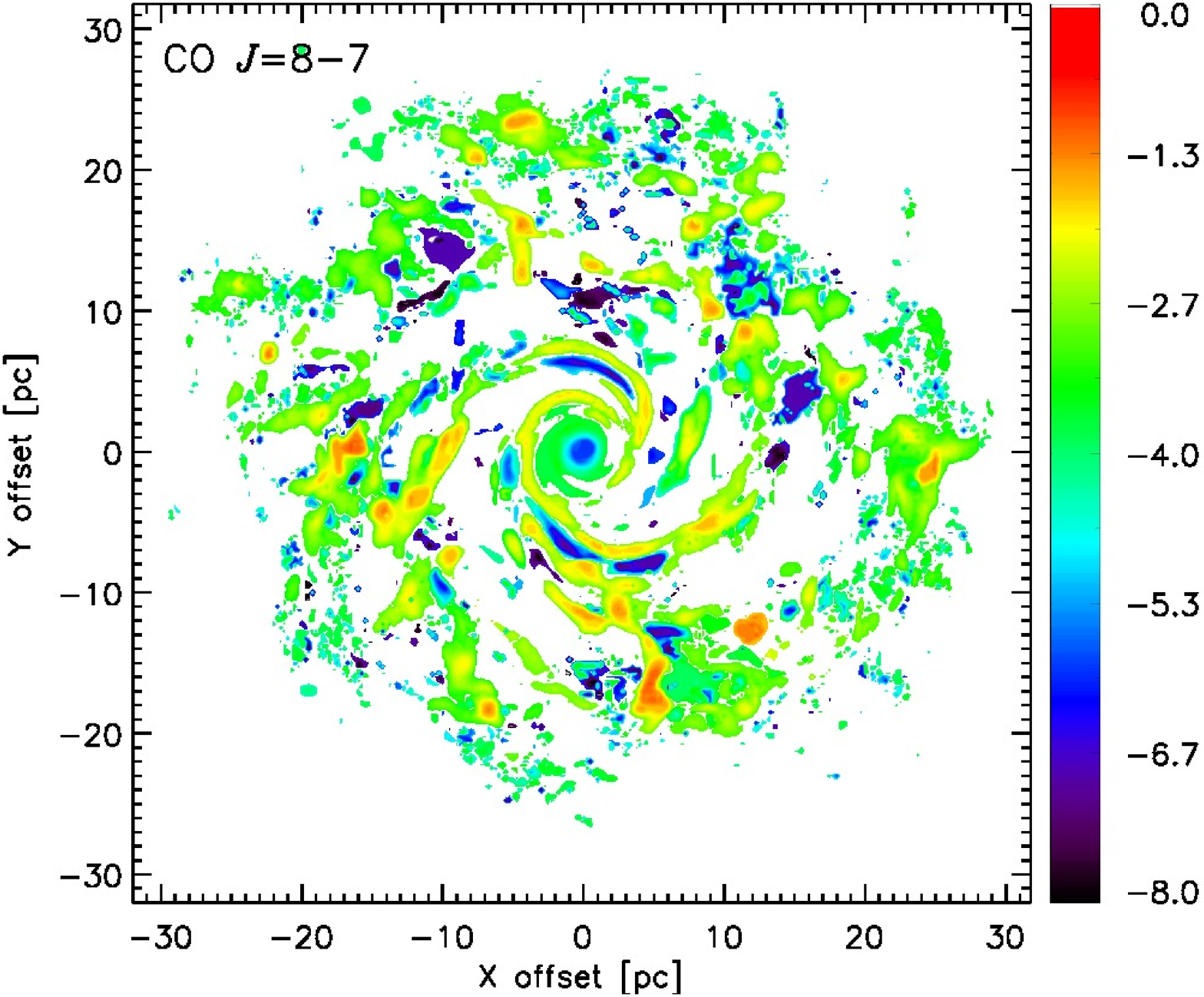}%
\hfill\includegraphics[angle=0,width=0.33\textwidth]{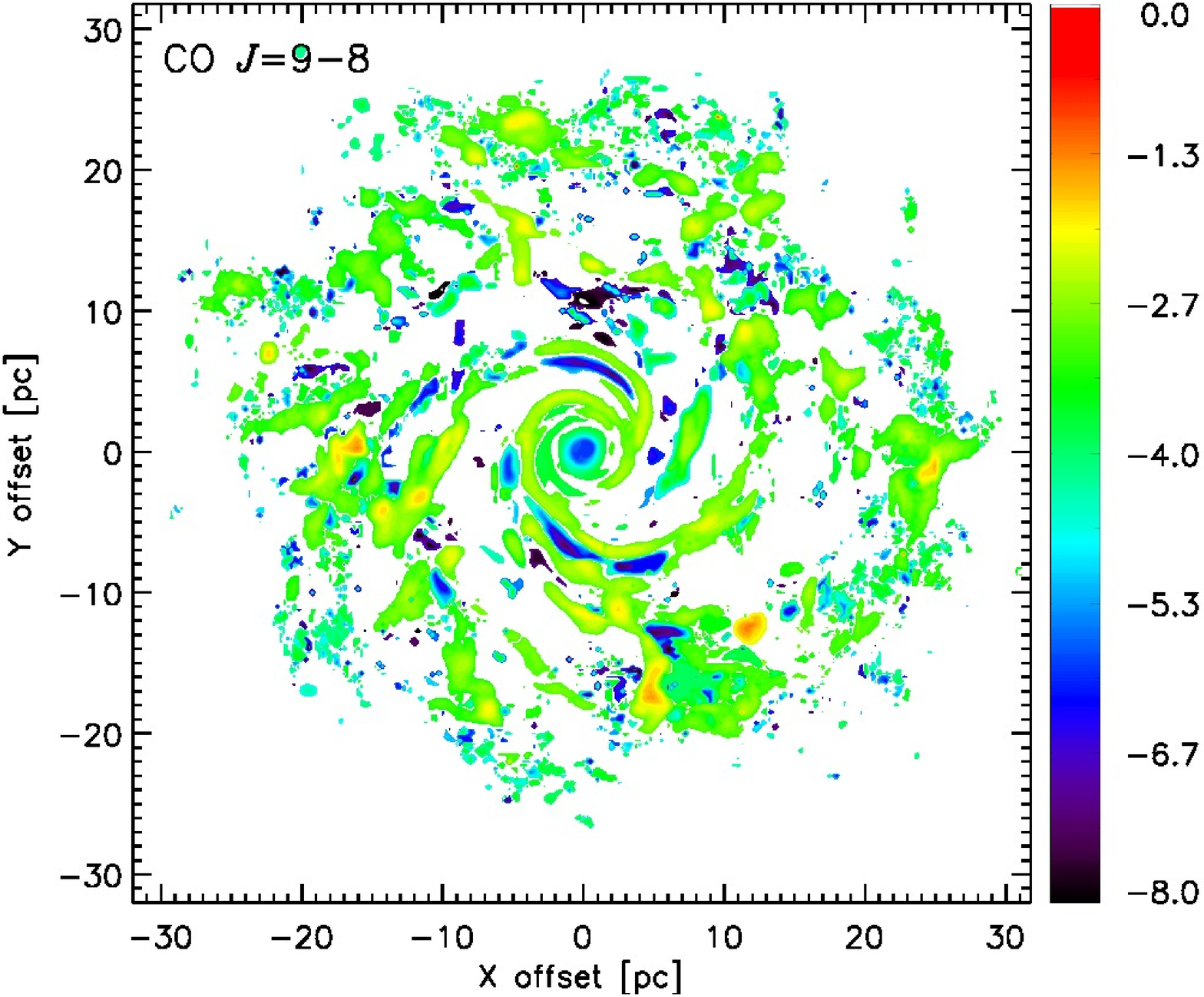}\hspace*{\fill}

\caption{{\footnotesize \textit{Top panels} - Face-on view of the total column density (units of $\2cm$) $N_{\rm H}$ (\textit{left}), the column density of molecular hydrogen $N({\rm H}_2)$ (\textit{middle}) and CO column $N({\rm CO})$ (\textit{right}) in logarithmic scale. 
\textit{Bottom panels} - Surface brightness maps of the $J=1\rightarrow0$ to $J=9\rightarrow8$ transitions of CO (in $log_{10}$ scale and units of $\ergscmsr$), as observed at the surface of the face-on data cube. The emission of the lower CO $J=1\rightarrow0$, $J=2\rightarrow1$, and $J=3\rightarrow2$ transitions do not trace (or just with relatively fainter emission) the inner region of the torus, while the higher CO lines (from $J=4\rightarrow3$ up to $J=9\rightarrow8$) do trace the inner spiral structures, including the {inner NLR}.}}
\label{fig:column-density}
\end{figure}

\begin{figure}[tp]
%\centering

\hspace*{\fill}\includegraphics[angle=0,width=0.35\textwidth]{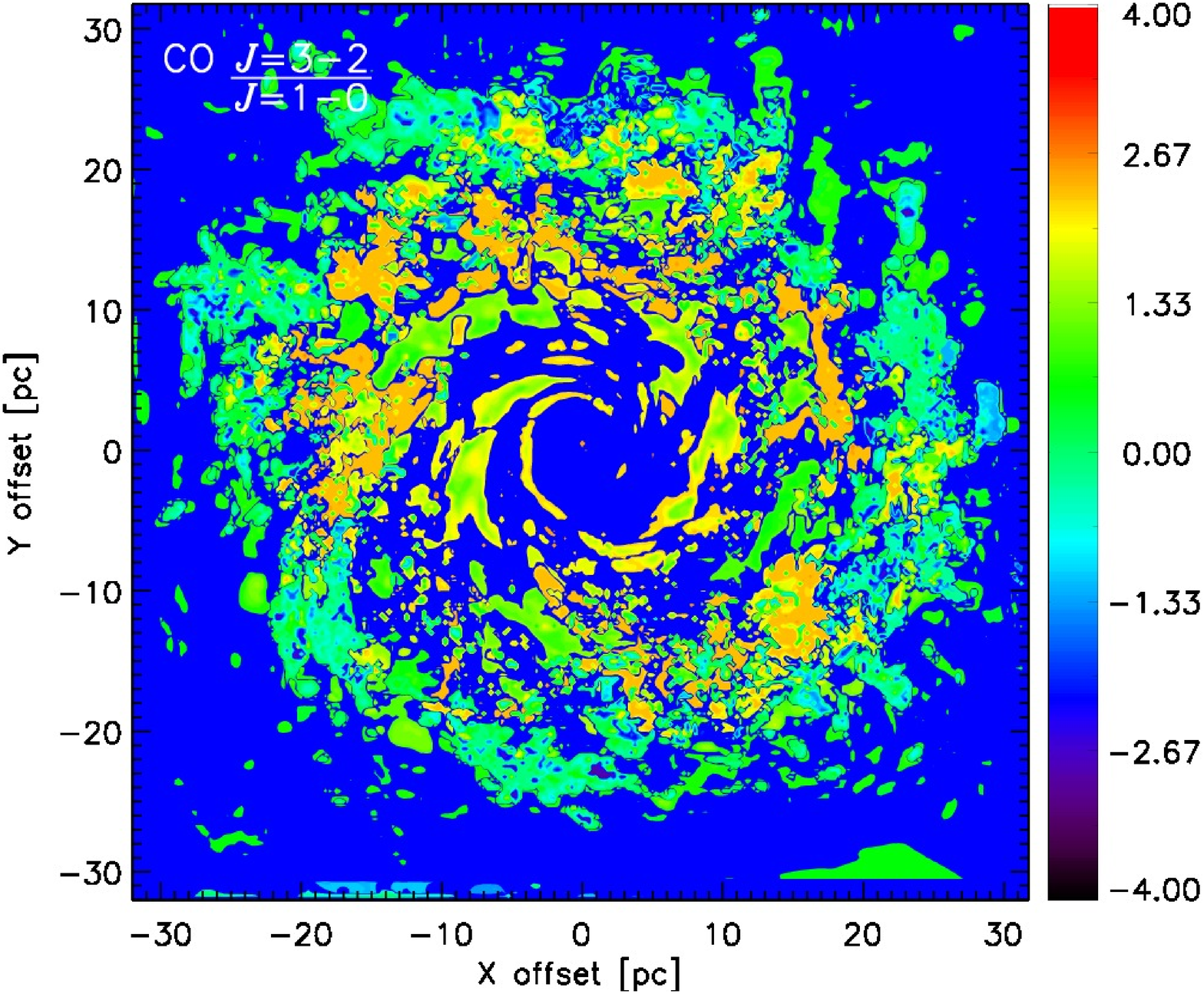}%
\hfill\includegraphics[angle=0,width=0.35\textwidth]{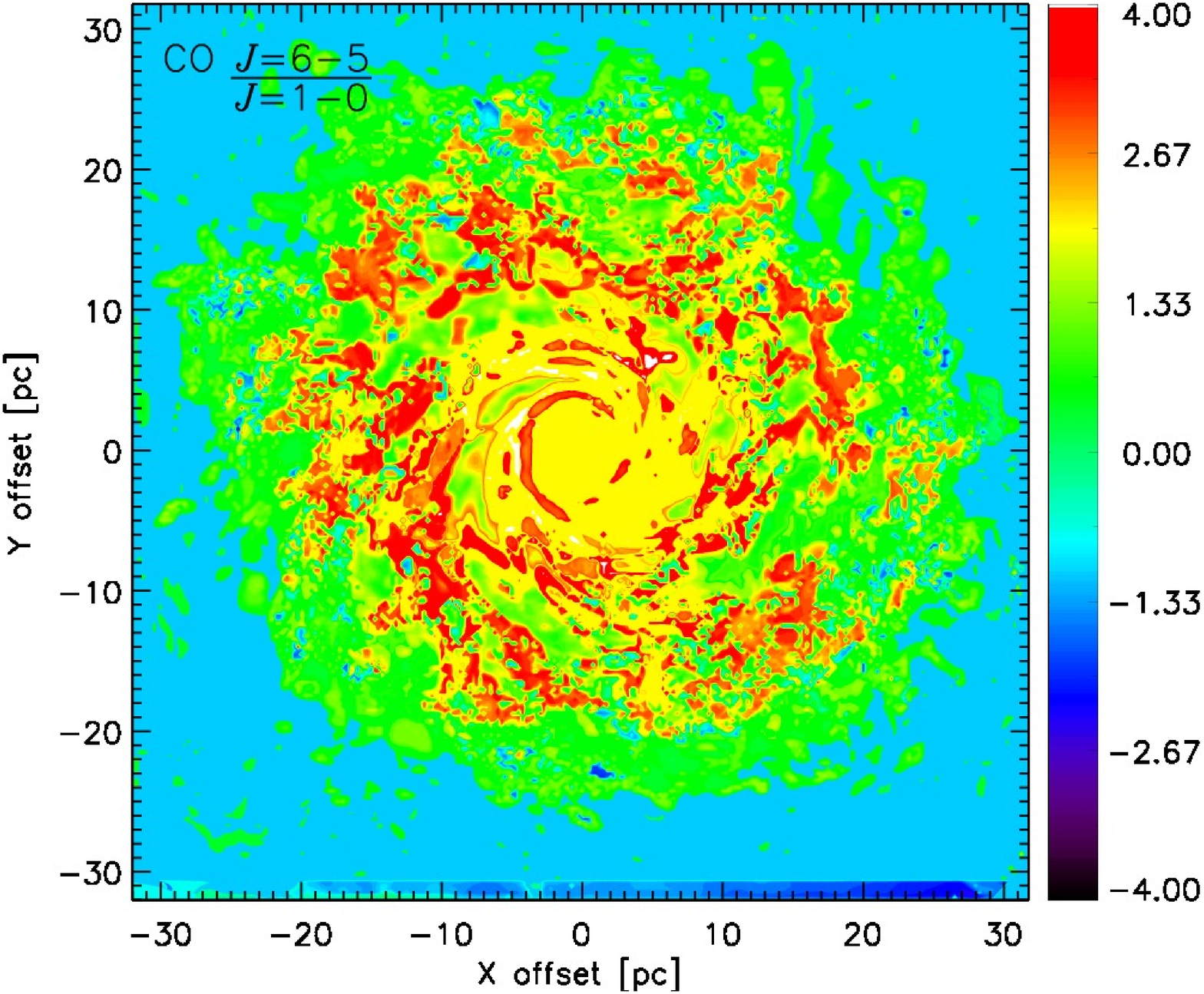}\hspace*{\fill}\\

\hspace*{\fill}\includegraphics[angle=0,width=0.35\textwidth]{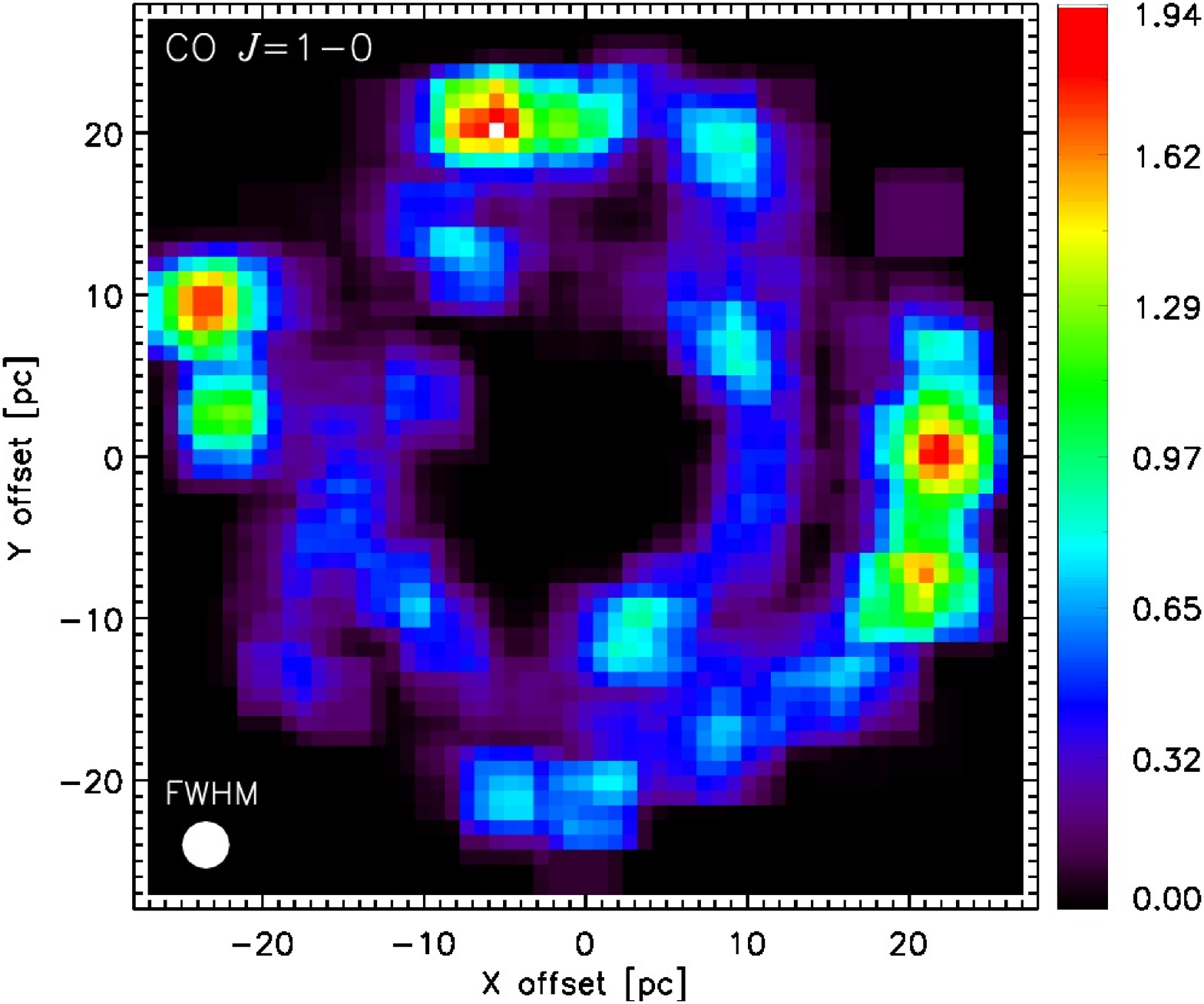}%
\hfill\includegraphics[angle=0,width=0.35\textwidth]{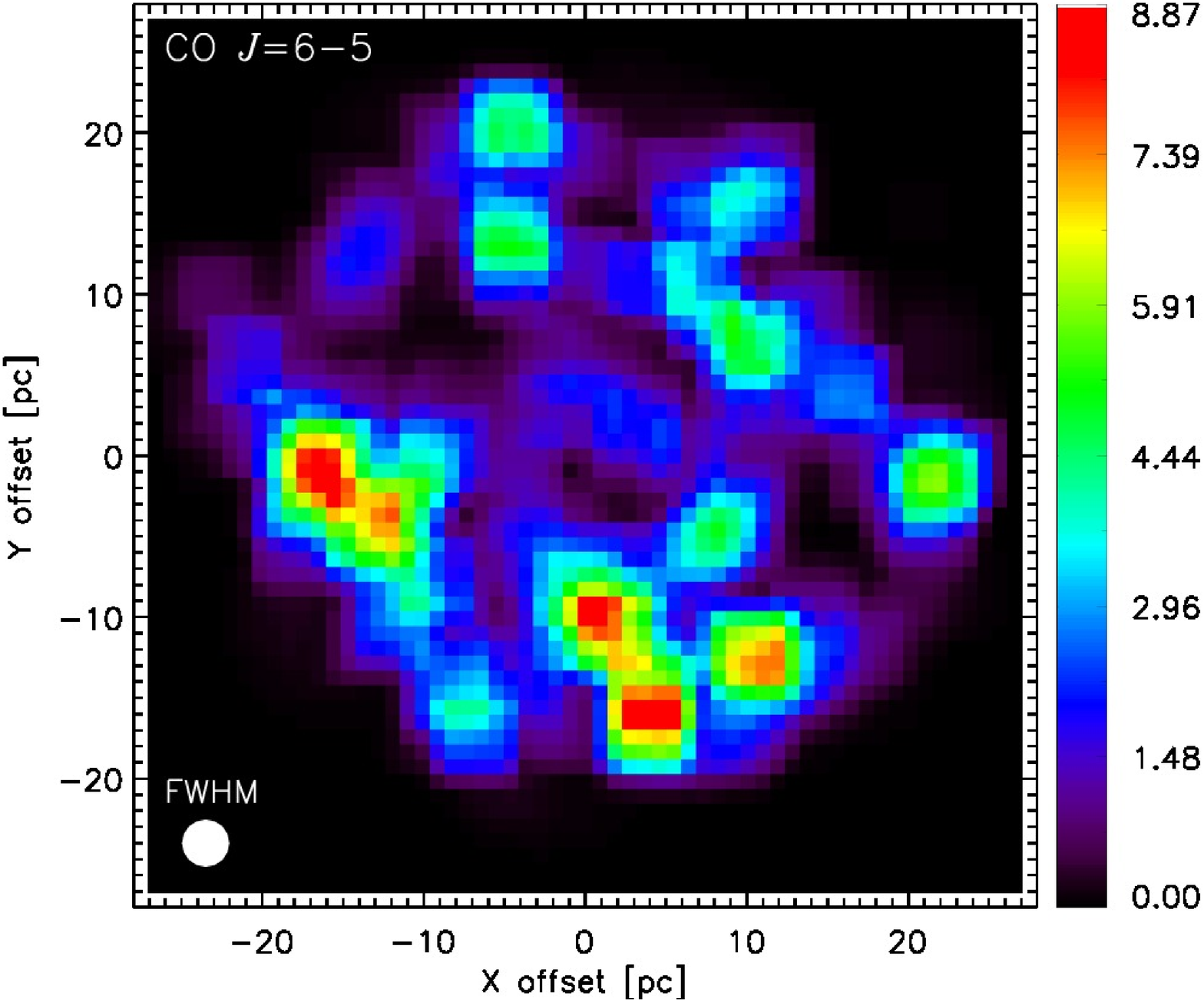}\hspace*{\fill}

\caption{{\footnotesize \textit{Top panels} - Maps of the CO $3-2/1-0$ (\textit{left}) and CO $6-5/1-0$ (\textit{right}) line intensity ratios.
\textit{Bottom panels} - Face-on view of the flux ($10^{-15} \ergscm$) of the CO $J=1\rightarrow0$ (\textit{left}) and  CO $J=6\rightarrow5$ (\textit{right}) lines, as mapped with a single dish beam of FWHM=$0.15''$ ($\sim2.8~\rm pc$) and adopting a distance $D=3.82~\rm Mpc$ to the source. Note how the relatively large beam smears out the intricate structure observed in the maps with the original resolution ($0.25~\rm pc$) shown in Fig.~\ref{fig:column-density}. These lower resolution maps make more evident the fact that the $J=1\rightarrow0$ line traces the {more diffuse} outer gas while the $J=6\rightarrow5$ traces the {denser} gas of the {inner NLR}.
}}
\label{fig:co-flux-fwhm}
\end{figure}

The total hydrogen column density $N_{\rm H}$, for a face-on viewing angle of the 3-D hydrodynamical model, is shown in the \textit{top left panel} of Fig.~\ref{fig:column-density}. As expected, the total column density of the CO molecule $N(\rm CO)$ (\textit{top right panel}) follows a similar distribution, although with about four orders of magnitude lower columns. The \textit{left} and \textit{right} \textit{bottom panels} of Fig.~\ref{fig:column-density} show the surface brightness of the CO $J=1\rightarrow0$ and $J=6\rightarrow5$ lines, respectively. These correspond to the brightness observed at the surface of the face-on data cube (i.e., not scaled for an arbitrary distance to the source).
{Given the relatively lower upper energy state ($\sim5.53~\rm K$) and critical density ($\sim2\times10^3~\3cm$ at 100 K) of the CO $J=1\rightarrow0$ line, most of the warm and dense gas and structure is traced by the CO $J=6\rightarrow5$ line instead.} 
The emission of the lower CO $J=1\rightarrow0$, $J=2\rightarrow1$, and $J=3\rightarrow2$ transitions trace with relatively fainter emission (or not trace at all) the inner region of the torus, while the higher CO lines (from $J=4\rightarrow3$ up to $J=9\rightarrow8$) strongly trace the inner structures, including the {inner Narrow Line Region (NLR)} of the torus. This is also shown by the CO $3-2/1-0$ and $6-5/1-0$ line intensity ratios of Fig.~\ref{fig:co-flux-fwhm} (top panels).
This can be explained by an optically thick $J=1\rightarrow0$ line ($N(\rm CO)>10^{18}~\2cm$) and by the modest presence of cold ($<100~\rm K$) gas, specially at the inner $\pm5~pc$ of the AGN torus.
This indicates that, by just considering the excitation of CO, the $J=1\rightarrow0$ line will not always be a good tracer of hydrogen column density $n_{\rm H}$ in the central region ($\lesssim60~\rm pc$) of an AGN. 
The same can be concluded from the emission maps obtained considering an inclination angle of $45^o$ about the X-axis. This is shown in Fig.~\ref{fig:maps-45x} of Appendix~\ref{sec:appendix-45x}.

As an exercise for comparison with future observations, we simulate a raster map of the AGN torus by adopting a distance $D=3.82~\rm Mpc$ (the distance to NGC~4945) to the source, and by convolving the surface brightness maps with a single dish beam of FWHM=$0.15''$ (about 11 pixels in the original map). This corresponds to a spatial scale of $\sim2.8~\rm pc$ at the distance chosen, and gives a flux with units of $10^{-15} \ergscm$ after multiplying the surface brightness by the solid angle $d\Omega=dR^2/D^2$ subtended by the original pixel scale ($dR=0.25~\rm pc$) at the adopted distance of the source.
A step size of one third the FWHM ($\sim0.92~\rm pc$, or about 4 pixels) degrades the original image from 256$\times$256 to a 61$\times$61 pixels image. Figure~\ref{fig:co-flux-fwhm} (bottom panels) shows the resulting flux maps of the CO $J=1\rightarrow0$ (\textit{left}) and  CO $J=6\rightarrow5$ (\textit{right}) lines. All the set of transitions from $J=1\rightarrow0$ to $J=9\rightarrow8$ is shown in Fig.~\ref{fig:maps-45x-fwhm} of the Appendix~\ref{sec:appendix-45x}. The smearing effect of the relatively large beam produces the loss of the intricate structure observed in the original maps with $0.25~\rm pc$ resolution shown in Fig.~\ref{fig:column-density}. While a torus like shape becomes more evident in the CO $J=1\rightarrow0$ line.

\subsection{The relation between CO and gas mass}\label{sec:CO-to-H2}

In the last few decades the integrated line fluxes of the lower \twco\ rotational lines (1--0, 2--1, and 3--2) have been used to estimate the gas masses of molecular clouds in the Milky Way \citep[e.g.][and references therein]{dickman86, solomon87, solomon91}. These estimates hold for the Milky Way and nearby normal galaxies where the CO emission emerges from moderately dense (volume-averaged densities of $n(\rm H_2)\sim500~\3cm$) giant molecular clouds (GMCs) in virial equilibrium (i.e., self-gravitating).
For spherical clouds supported by isotropic random motions (e.g., turbulence) in virial equilibrium, the resulting inferred theoretical conversion factors are $X\sim2\times10^{20}~[\2cm~(\Kkms)^{-1}]$ and $\alpha=4.3~[M_{\sun}~(\Kkms)^{-1}]$ \citep{strong96, dame01, dickman86, solomon87}.

The virial approach of the optically thick CO lines can be extended to other galaxies, considering an ensemble of virialized clouds, instead of a single one \citep{dickman86}. In galactic nuclei and starburst galaxies, however, the assumption of an ensemble of individual gas clouds in virial equilibrium does not hold. In these environments the gas motions are due to a combination of gas and stellar mass components, and the gas is expected to be in a smoother configuration along a disk. Nevertheless, \citet{downes93}, \citet{solomon97}, and \citet{downes98} have shown that a slightly modified version of the CO luminosity to gas mass conversion factors can be derived in these environments. For luminous or ultraluminous infrared galaxies the inferred theoretical conversion factors range between $\alpha=0.8$ and 1.6, or $X=3.7-7.3\times10^{19}$ \citep{solomon97, downes98}.

%-------------------------------------------------------------------------

\begin{figure}[htp]
%\centering

\hspace*{\fill}\includegraphics[angle=0,width=0.30\textwidth]{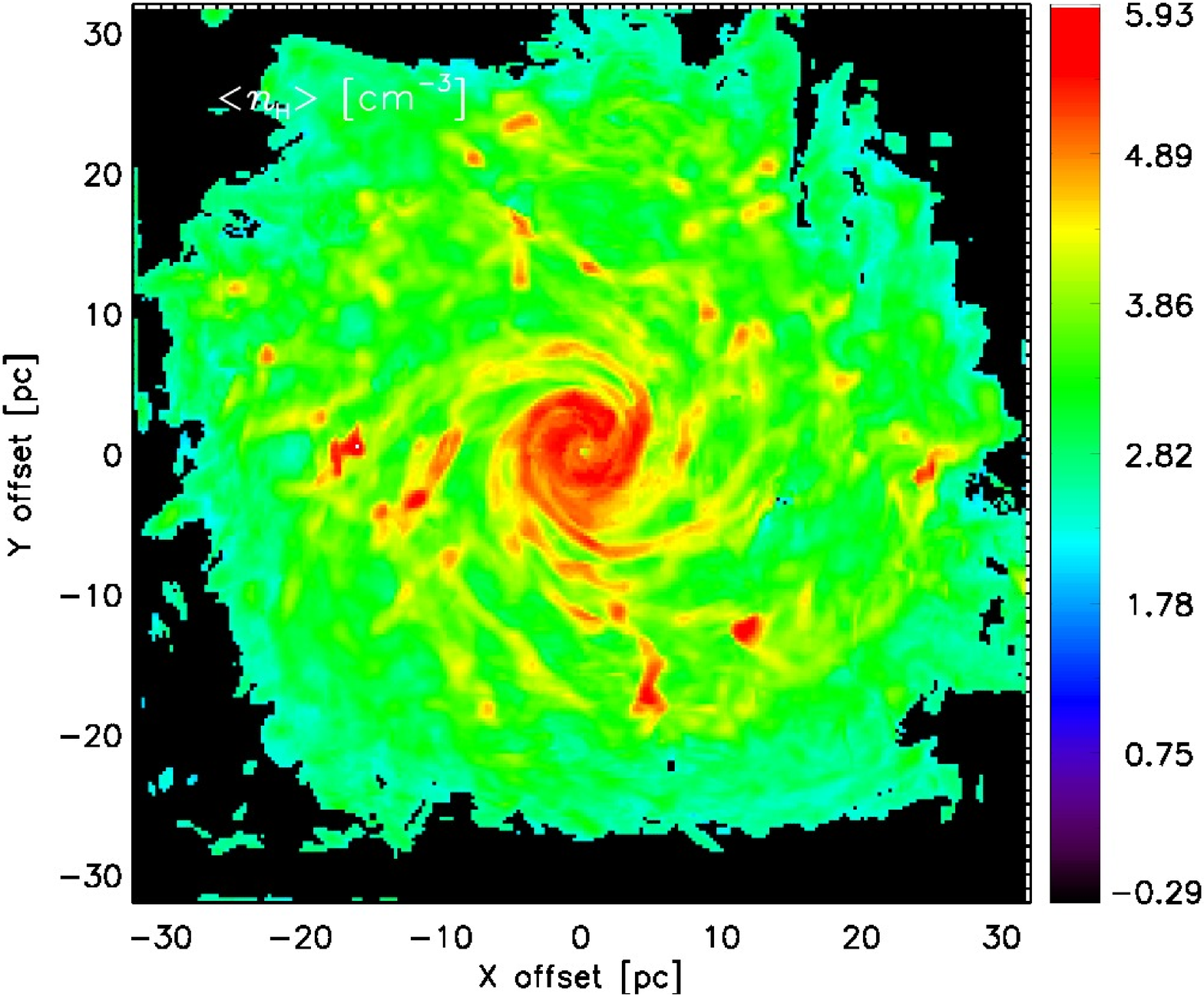}\hspace{-0.1cm}%
\hfill\includegraphics[angle=0,width=0.30\textwidth]{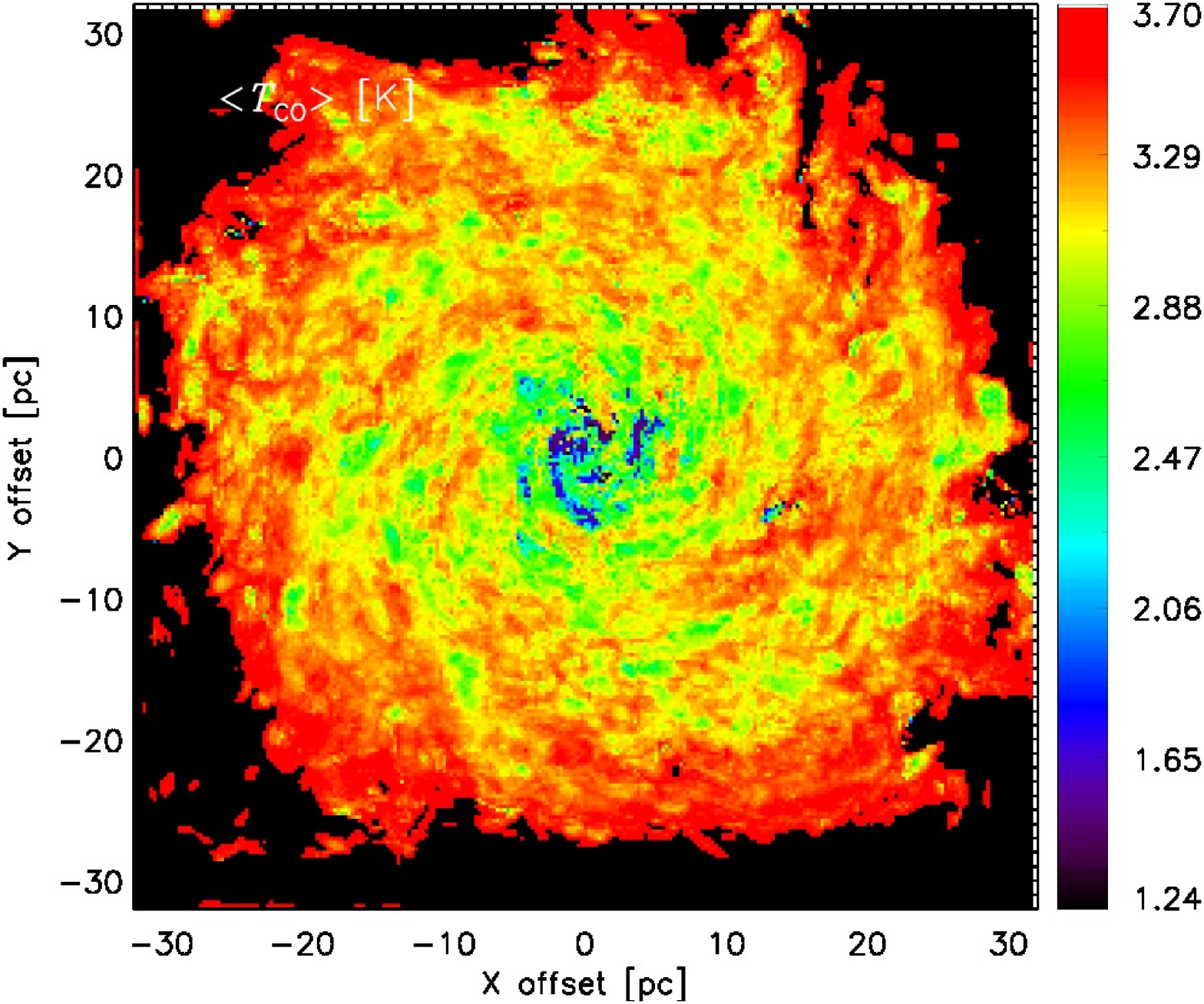}\hspace{-0.05cm}%
\hfill\includegraphics[angle=90,width=0.35\textwidth]{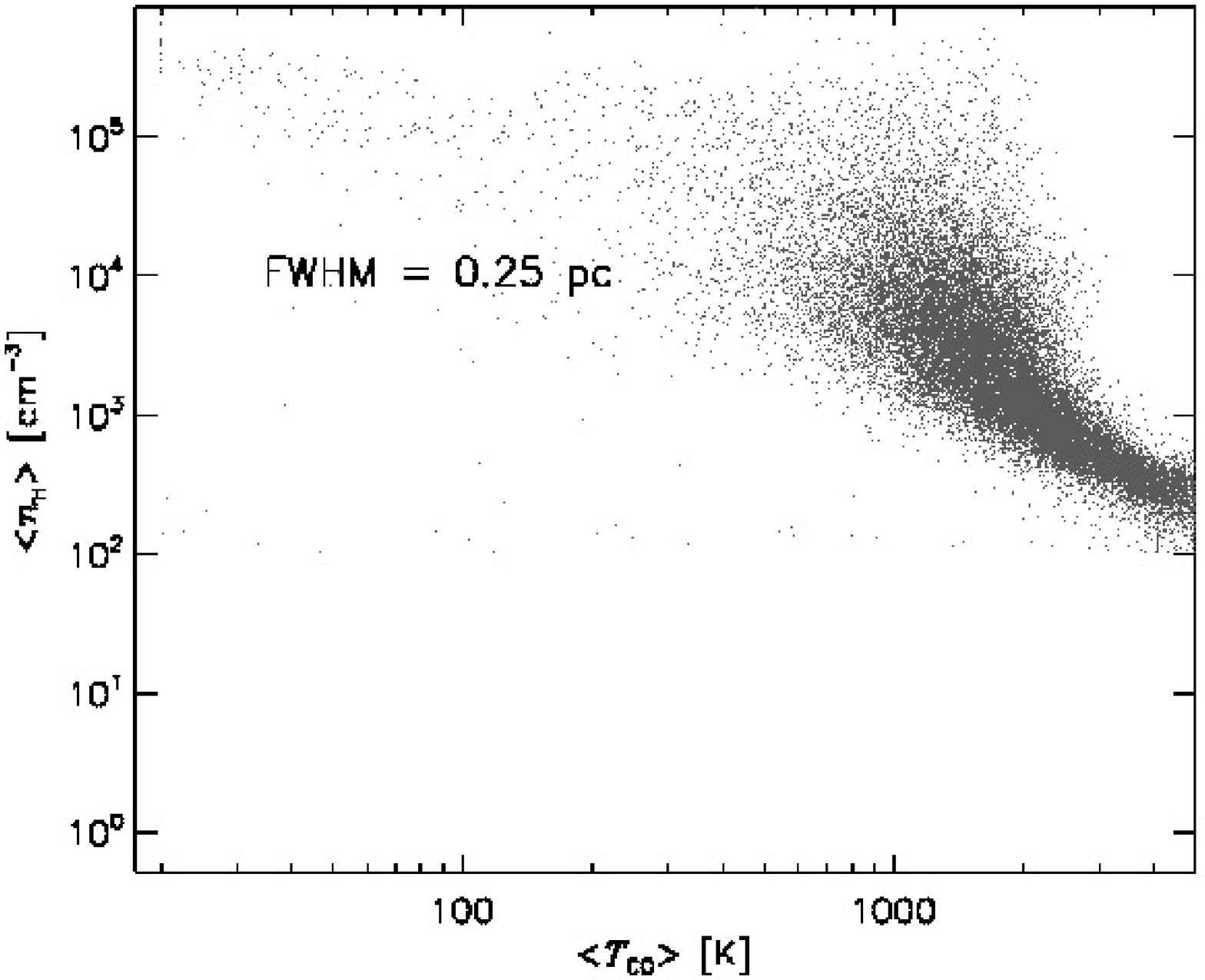}\hspace*{\fill}\\

\vspace{-0.5cm}

\hspace*{\fill}\includegraphics[angle=0,width=0.30\textwidth]{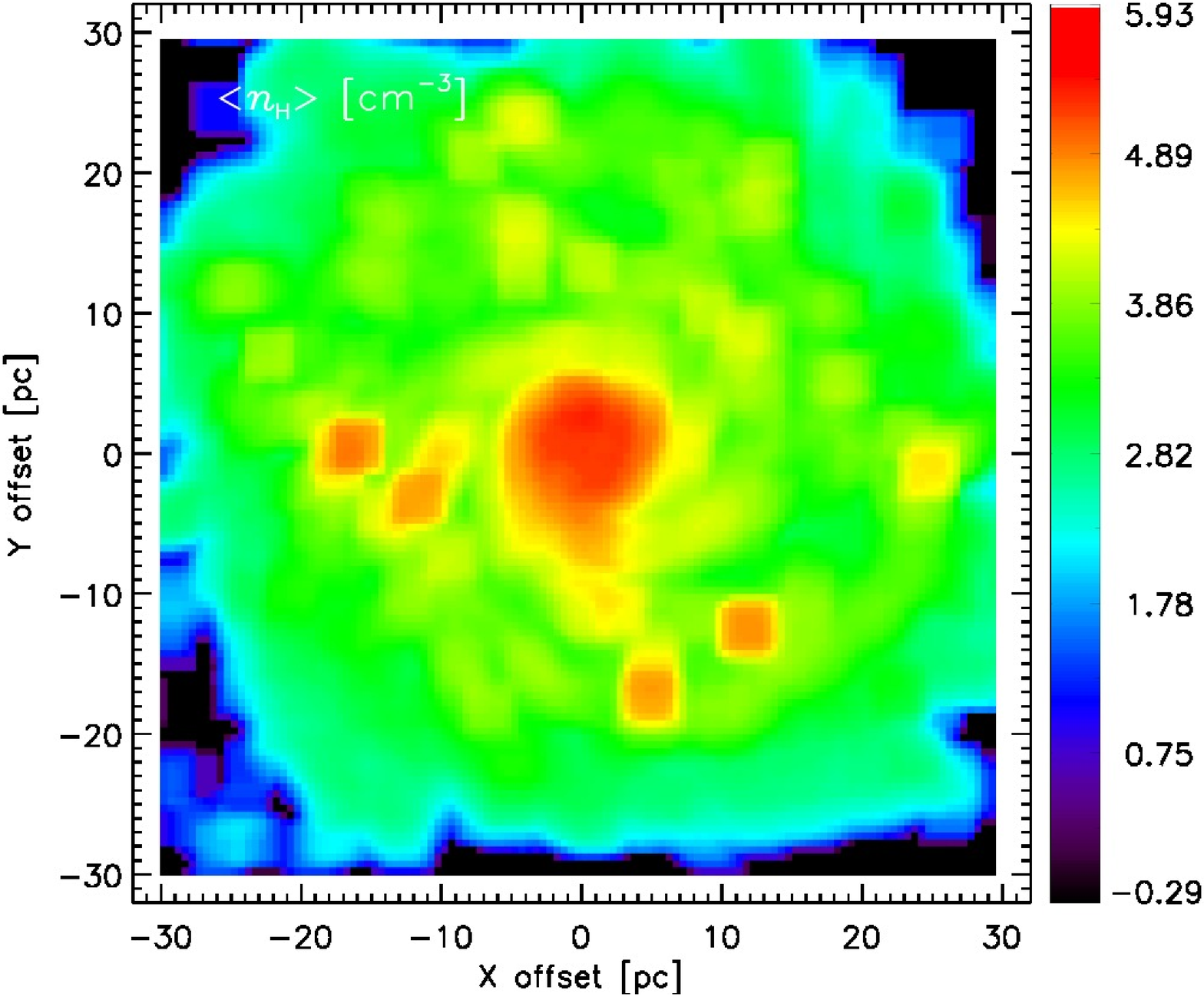}\hspace{-0.1cm}%
\hfill\includegraphics[angle=0,width=0.30\textwidth]{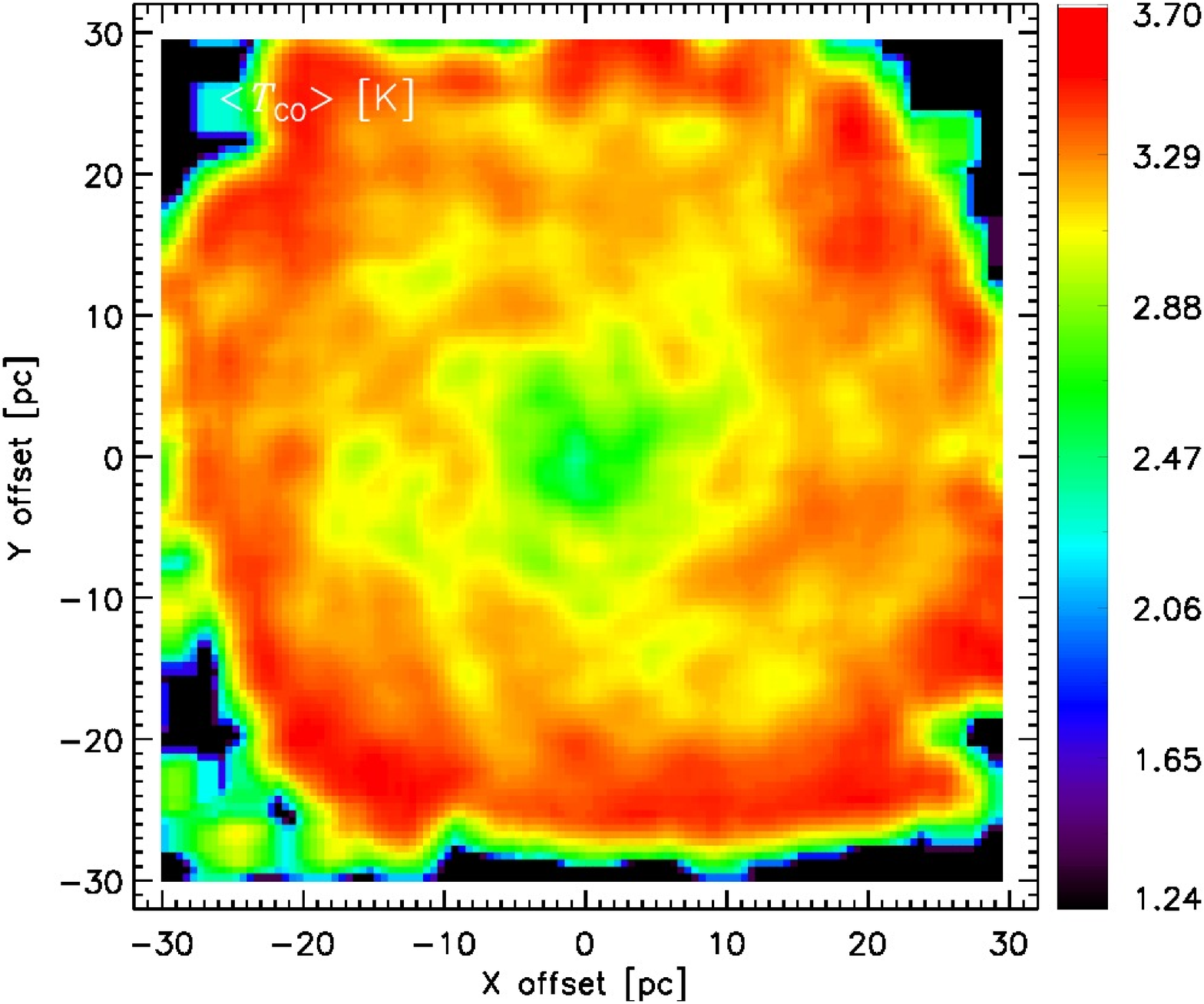}\hspace{-0.05cm}%
\hfill\includegraphics[angle=90,width=0.35\textwidth]{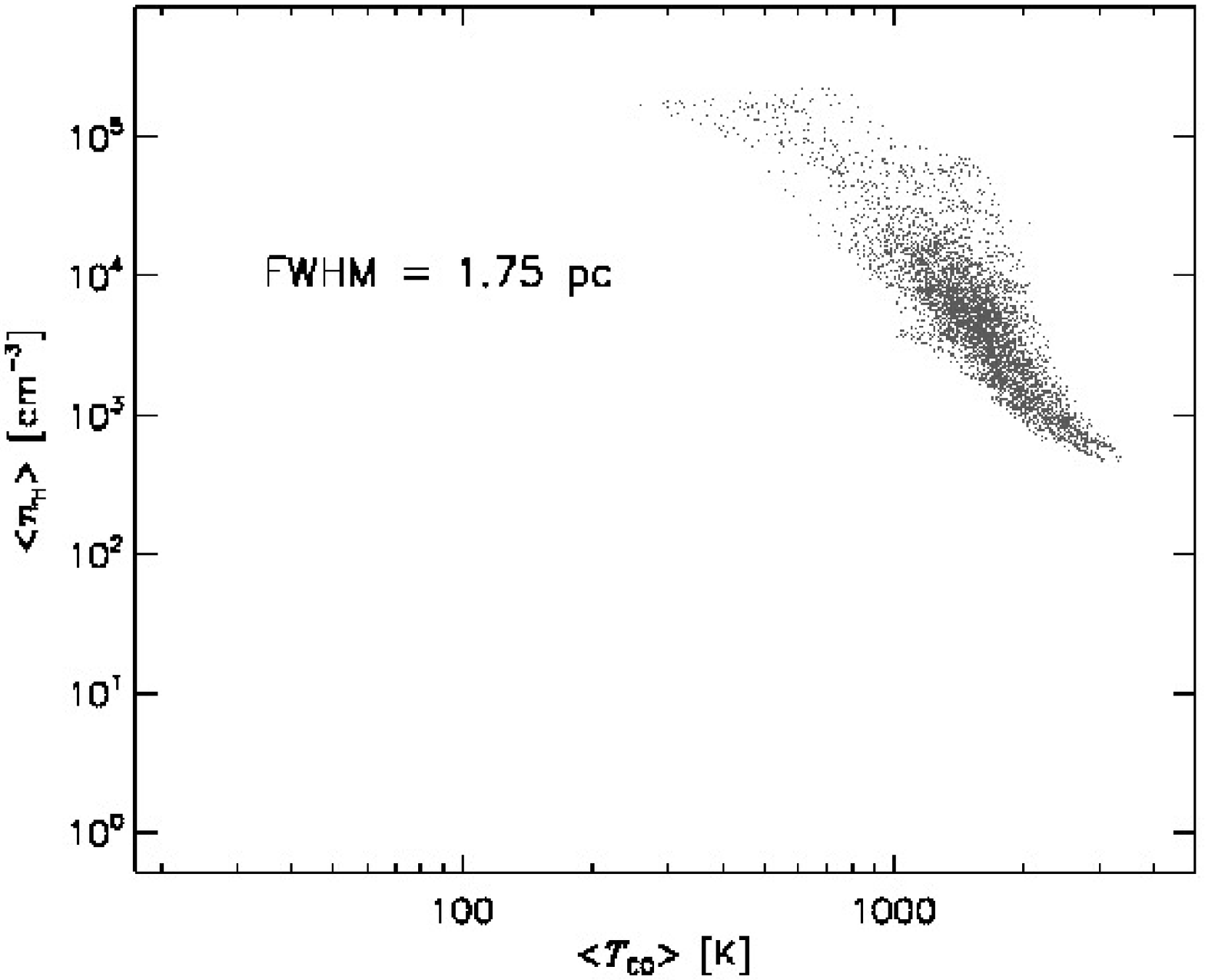}\hspace*{\fill}\\

\vspace{-0.5cm}

\hspace*{\fill}\includegraphics[angle=0,width=0.30\textwidth]{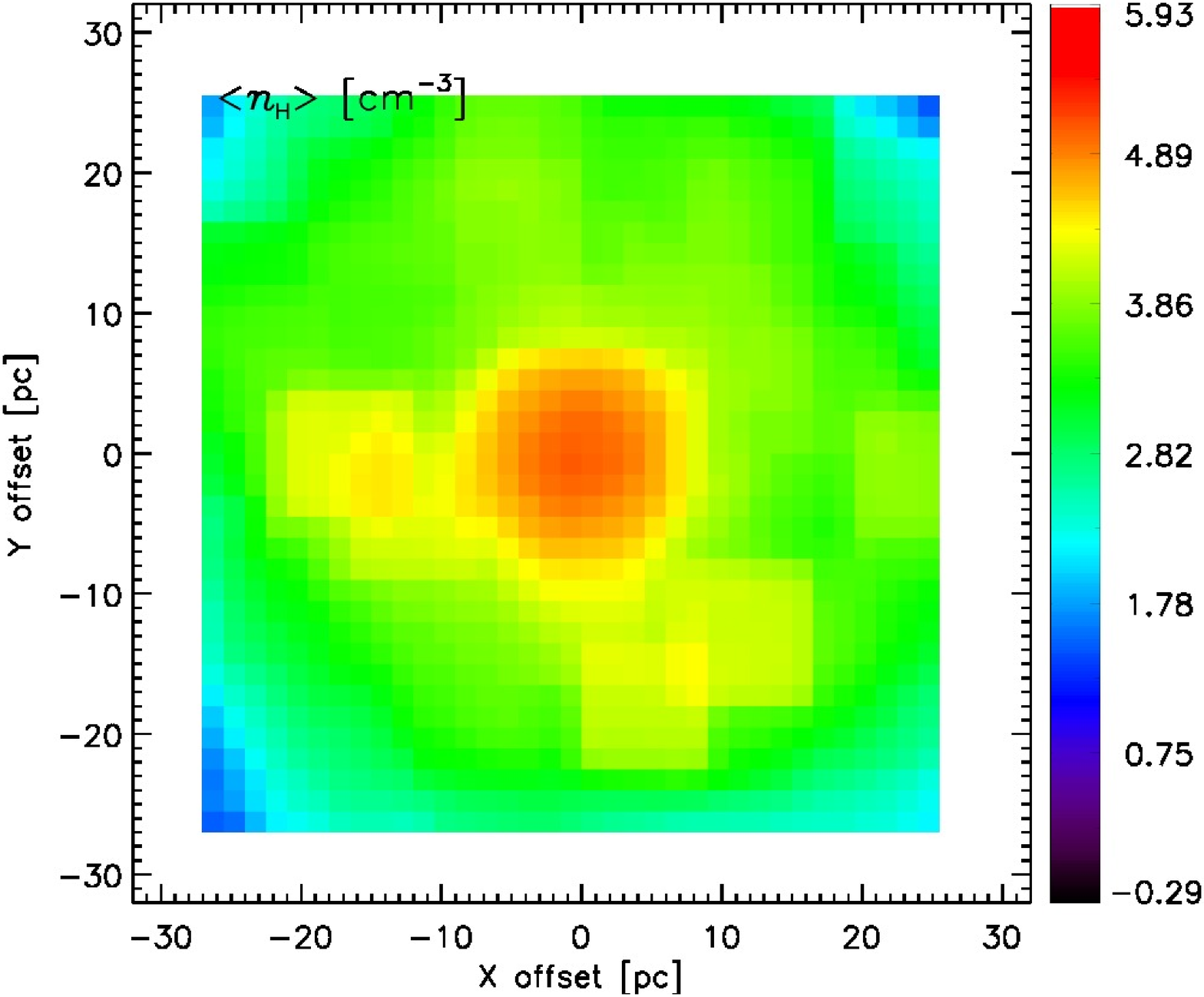}\hspace{-0.1cm}%
\hfill\includegraphics[angle=0,width=0.30\textwidth]{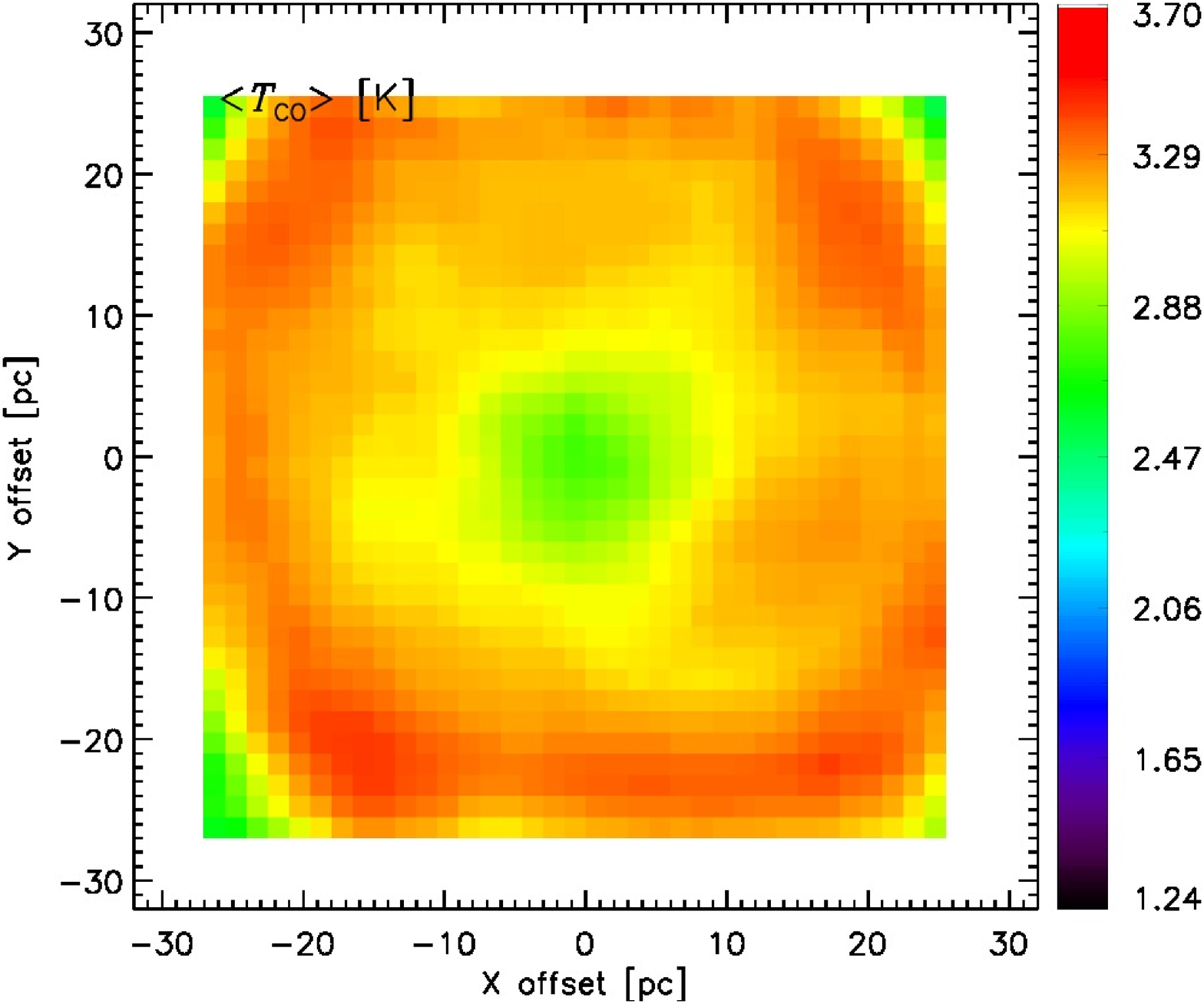}\hspace{-0.05cm}%
\hfill\includegraphics[angle=90,width=0.35\textwidth]{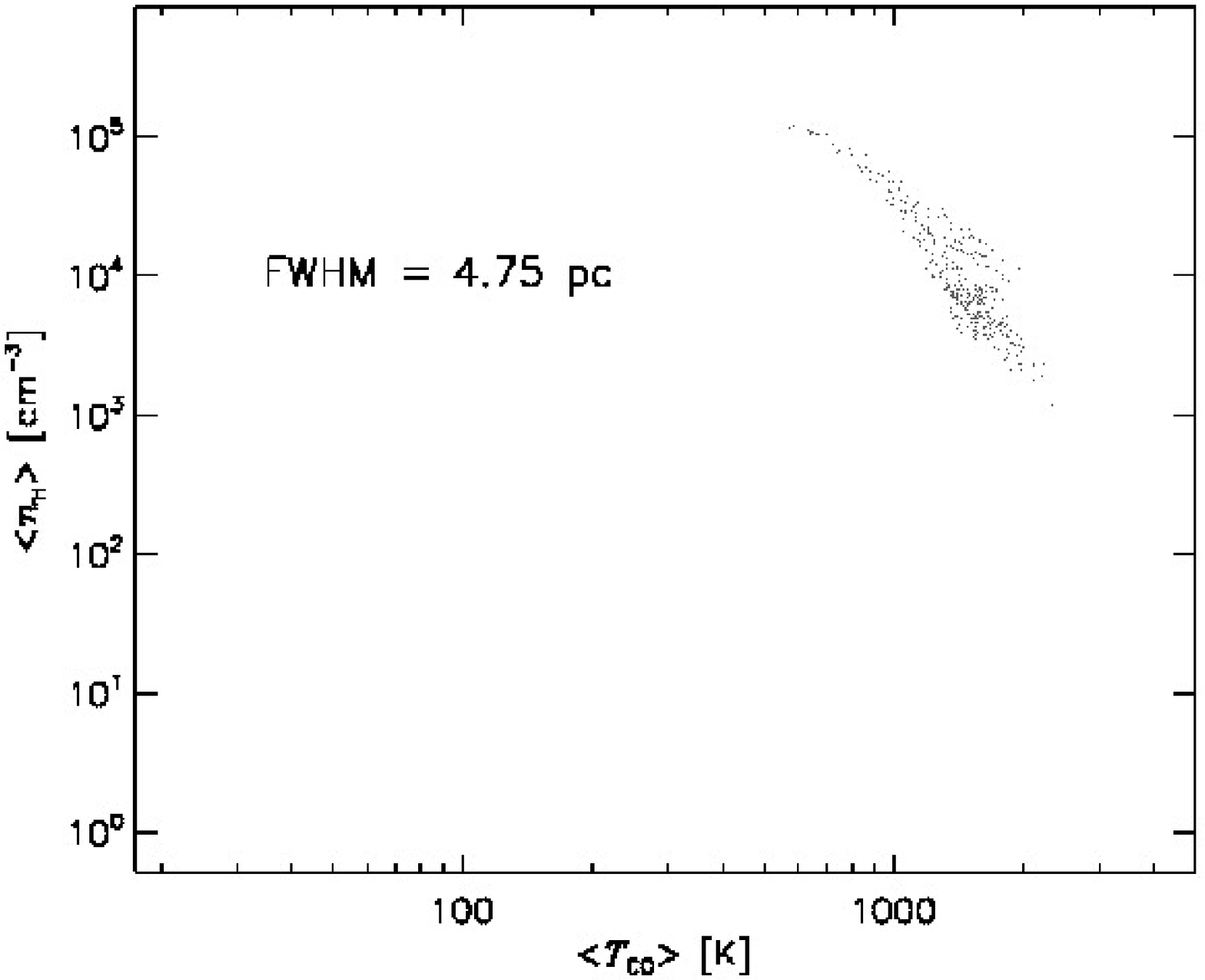}\hspace*{\fill}\\

\vspace{-0.5cm}

\hspace*{\fill}\includegraphics[angle=0,width=0.30\textwidth]{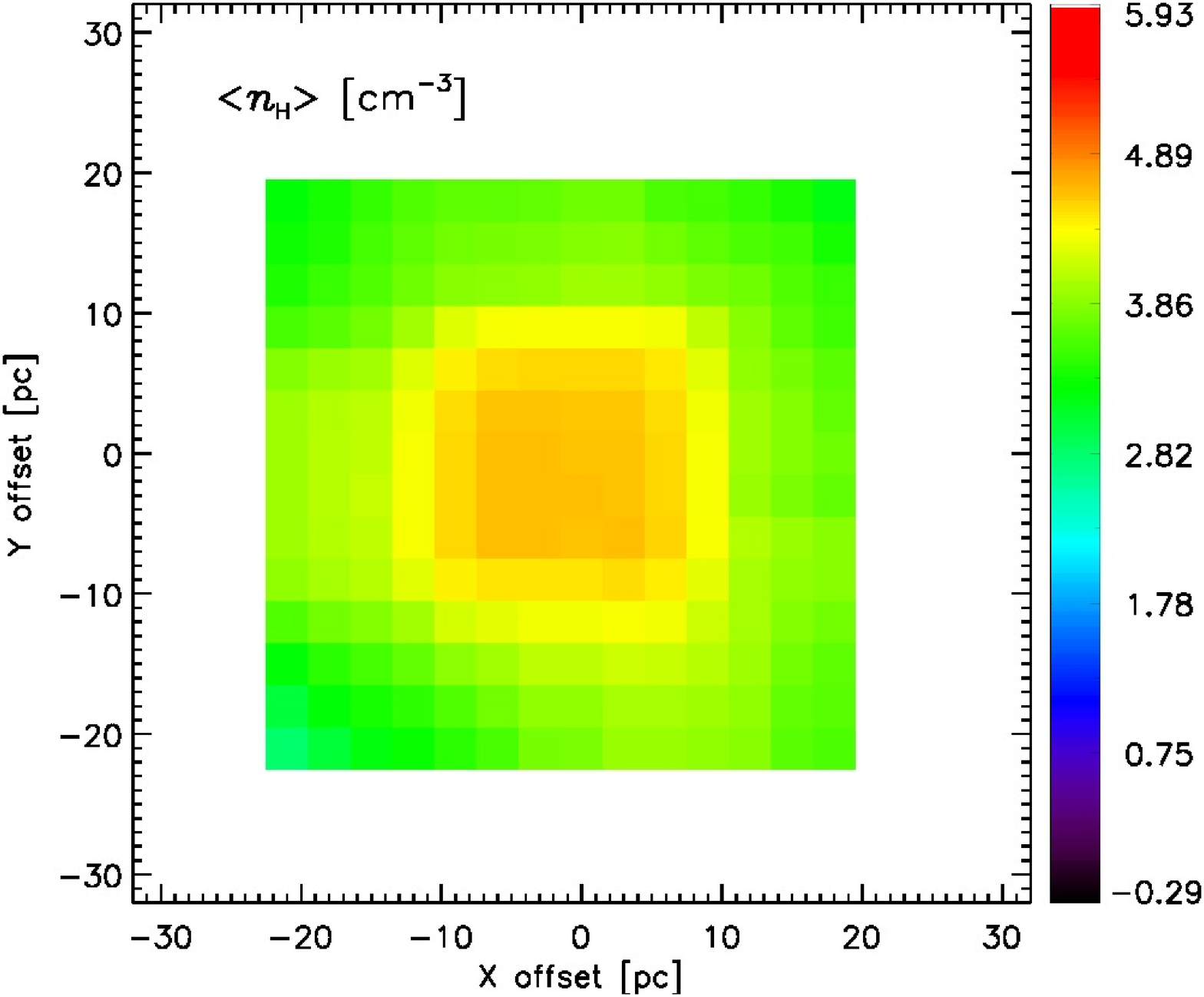}\hspace{-0.1cm}%
\hfill\includegraphics[angle=0,width=0.30\textwidth]{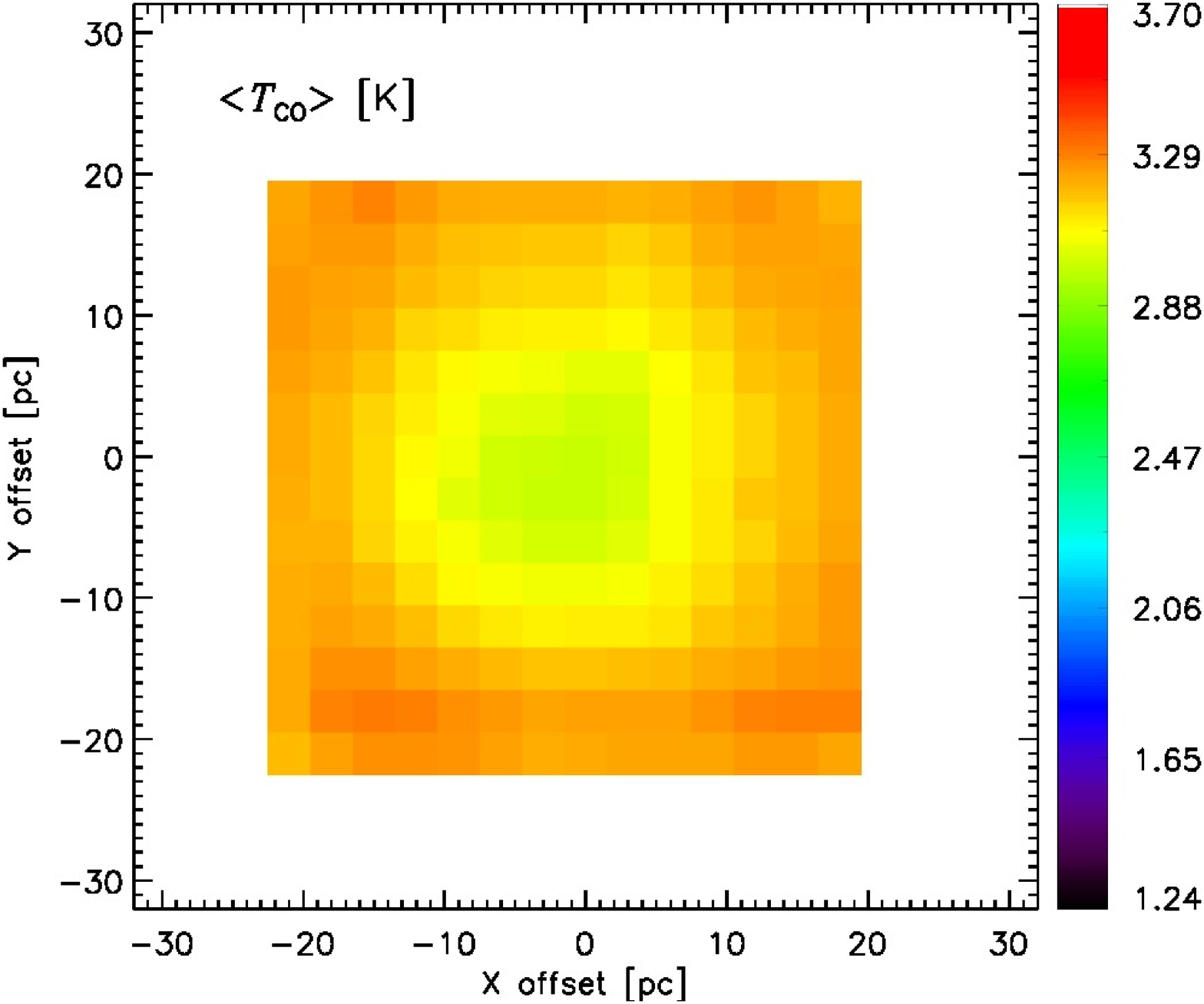}\hspace{-0.05cm}%
\hfill\includegraphics[angle=90,width=0.35\textwidth]{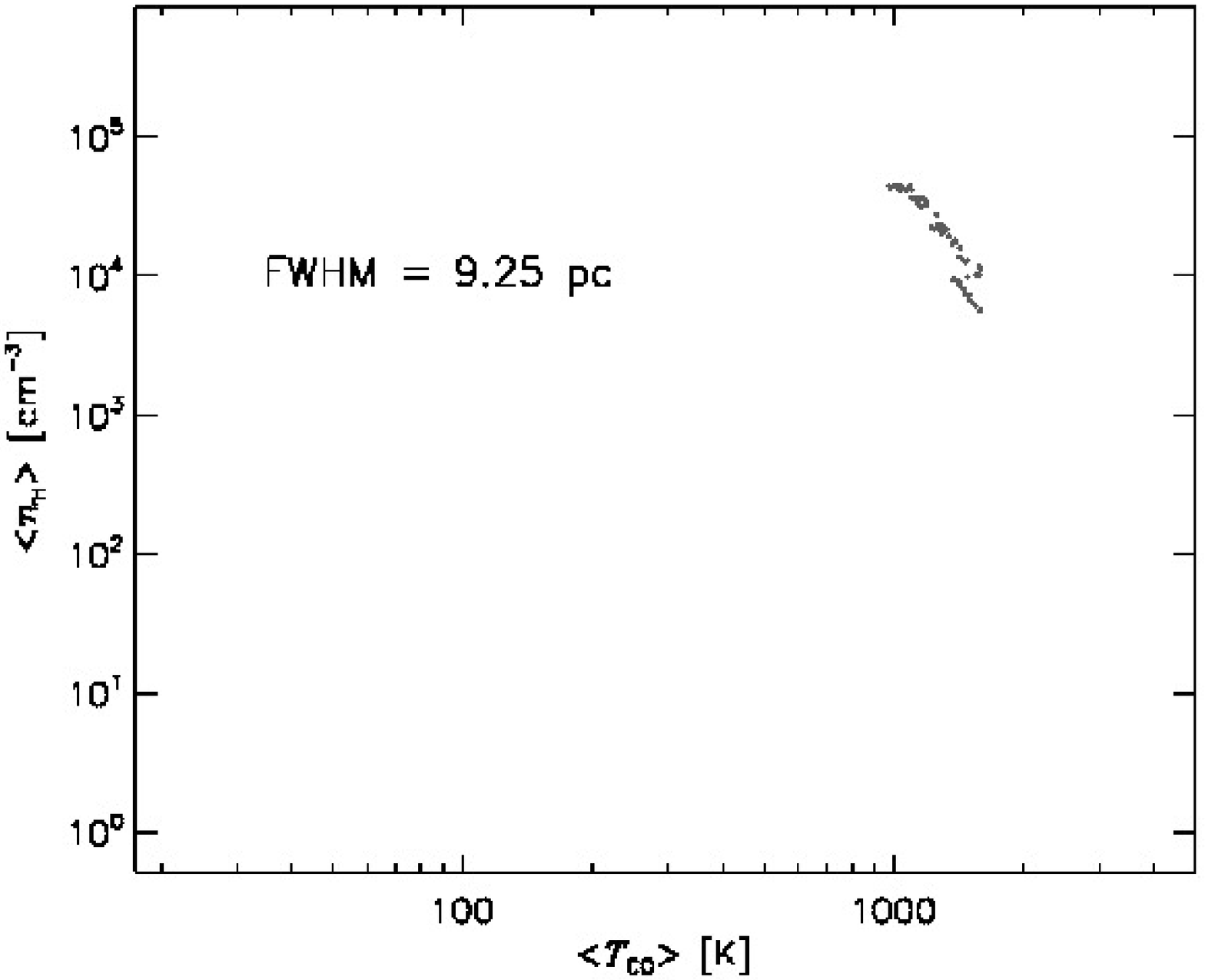}\hspace*{\fill}\\

\caption{{\footnotesize Two dimensional maps (at the original resolution of 0.25 pc) of the average density $<n_{\rm H}>$ (\textit{left panels}), the average abundance-weighted temperature $<T_{\rm CO}>$ of CO (\textit{middle panels}), and scatter plot of $<n_{\rm H}>$ versus $<T_{\rm CO}>$. From \textit{top} to \textit{bottom}: same as in the \textit{top panels} but convolving the maps with beam sizes (FWHM) equivalent to 1.75 pc, 4.75 pc, and 9.25 pc, for a distance of 3.82 Mpc to the source. For the lowest resolution maps the scatter plot is shown with larger symbols for better visualization. All the scatter plots show the expected inverse relation between $<n_{\rm H}>$ and $<T_{\rm CO}>$: lower temperatures correspond to higher densities, and they converge to a linear relation as the beam size increases.}}
\label{fig:Tk-nH}
\end{figure}

%-------------------------------------------------------------------------

Since we know exactly what is the gas mass in our models, we can explore the behaviour of the CO-to-H$_2$ conversion factor in our model of an AGN torus from the computed luminosities of several CO rotational lines.
{First we check the relation between the \textit{average} density $<n_{\rm H}>$ and the \textit{average} abundance-weighted temperature of CO $<T_{\rm CO}>$ (computed through the line of sight or column) for each pixel of the map at the original resolution. We apply a similar criteria as in Sec.2.2 when computing the average density and temperature. That is, we use only the grid points with total densities $n_{\rm H}$ larger than $100~\rm cm^{-3}$ and temperatures $T_{\rm CO}$ lower than 5000 K to compute $<n_{\rm H}>$ and $<T_{\rm CO}>$. This criteria, however, produces several pixels with $<n_{\rm H}>=0$, specially in the outher region of the maps. Then we generate lower resolution raster maps by convolving the original resolution (0.25 pc) maps of  $<n_{\rm H}>$ and $<T_{\rm CO}>$ with different beam sizes (FWHM) corresponding to 1.75 pc, 4.75 pc, and 9.25 pc (with step or pixel sizes of $\sim$FHWM/3), as it was done in Sec.~\ref{sec:co-maps}. Those pixels with $<n_{\rm H}>=0$ are masked out in both density and temperature maps during the convolution process. From all the pixels of the raster maps we create scatter plots of $<n_{\rm H}>$ versus $<T_{\rm CO}>$ at the different resolutions.

The density and temperature maps, as well as the corresponding scatter plots, are shown in Fig.~\ref{fig:Tk-nH}. The $<n_{\rm H}>$ and $<T_{\rm CO}>$ maps (first and second column in Fig.~\ref{fig:Tk-nH}, respectively) show denser and colder gas in the center of the AGN torus, indicating an \textit{inverse} relation between the average density and temperature (i.e., higher densities correspond to lower temperatures). The third column in Fig.~\ref{fig:Tk-nH} shows the scatter plots obtained considering all the pixels in the maps, where the expected \textit{inverse} relation between $<n_{\rm H}>$ and $<T_{\rm CO}>$ is reproduced at all the resolutions. The relation between density and temperature becomes tighter and almost linear in the maps convolved with larger beams.

\begin{figure}[htp]
%\centering

\hspace*{\fill}\includegraphics[angle=90,width=0.35\textwidth]{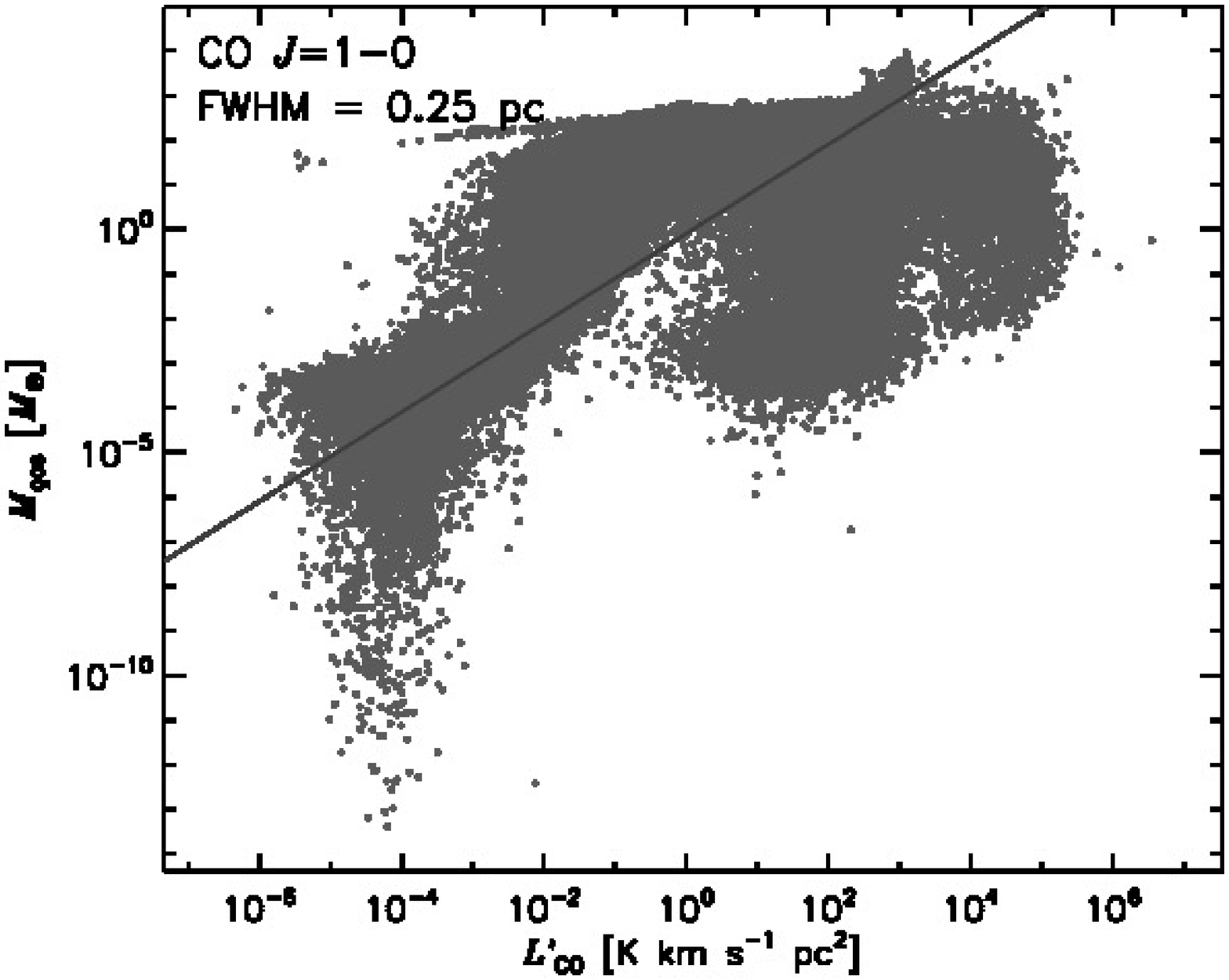}\hspace{-0.7cm}%
\hfill\includegraphics[angle=90,width=0.35\textwidth]{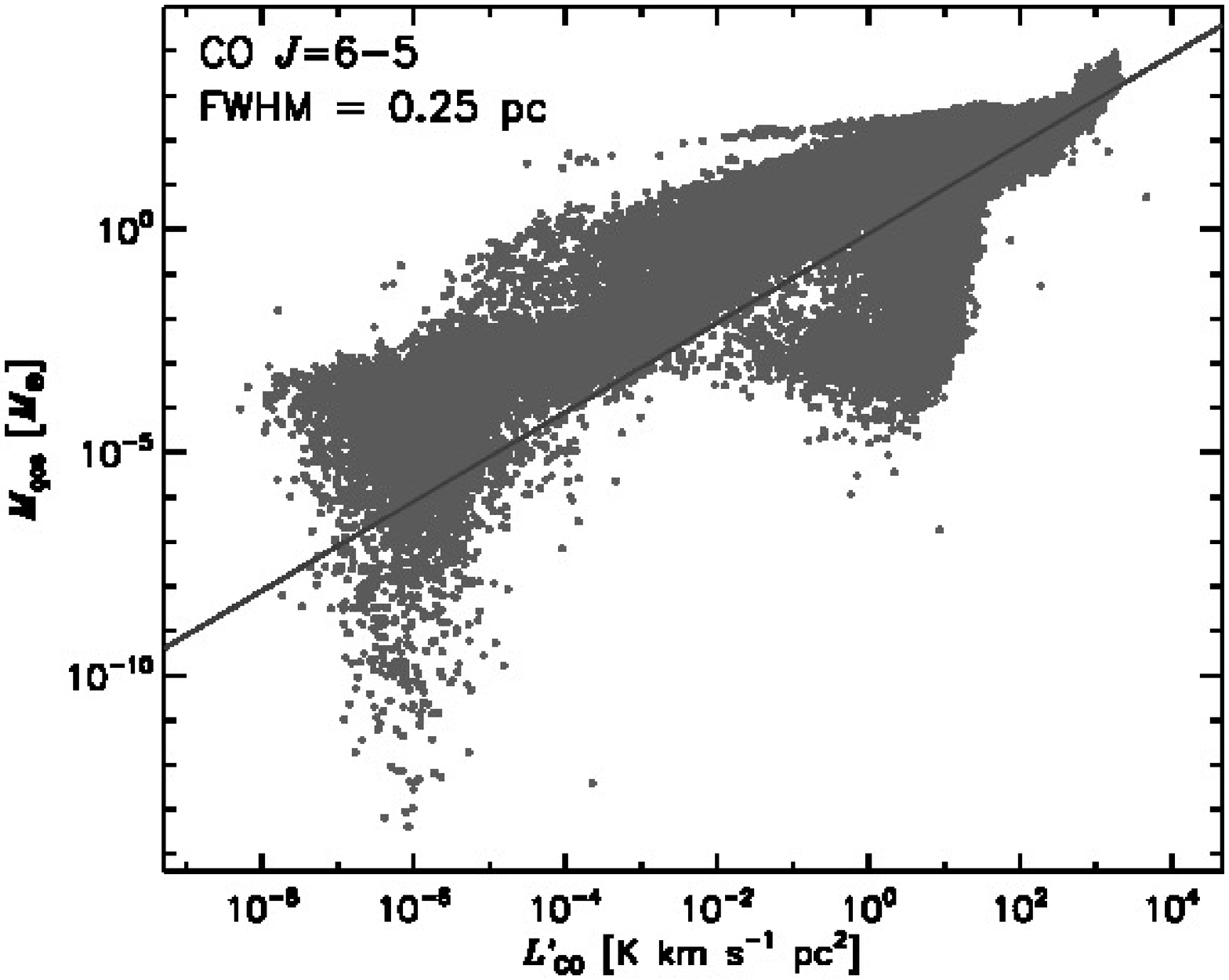}\hspace{-0.7cm}%
\hfill\includegraphics[angle=90,width=0.35\textwidth]{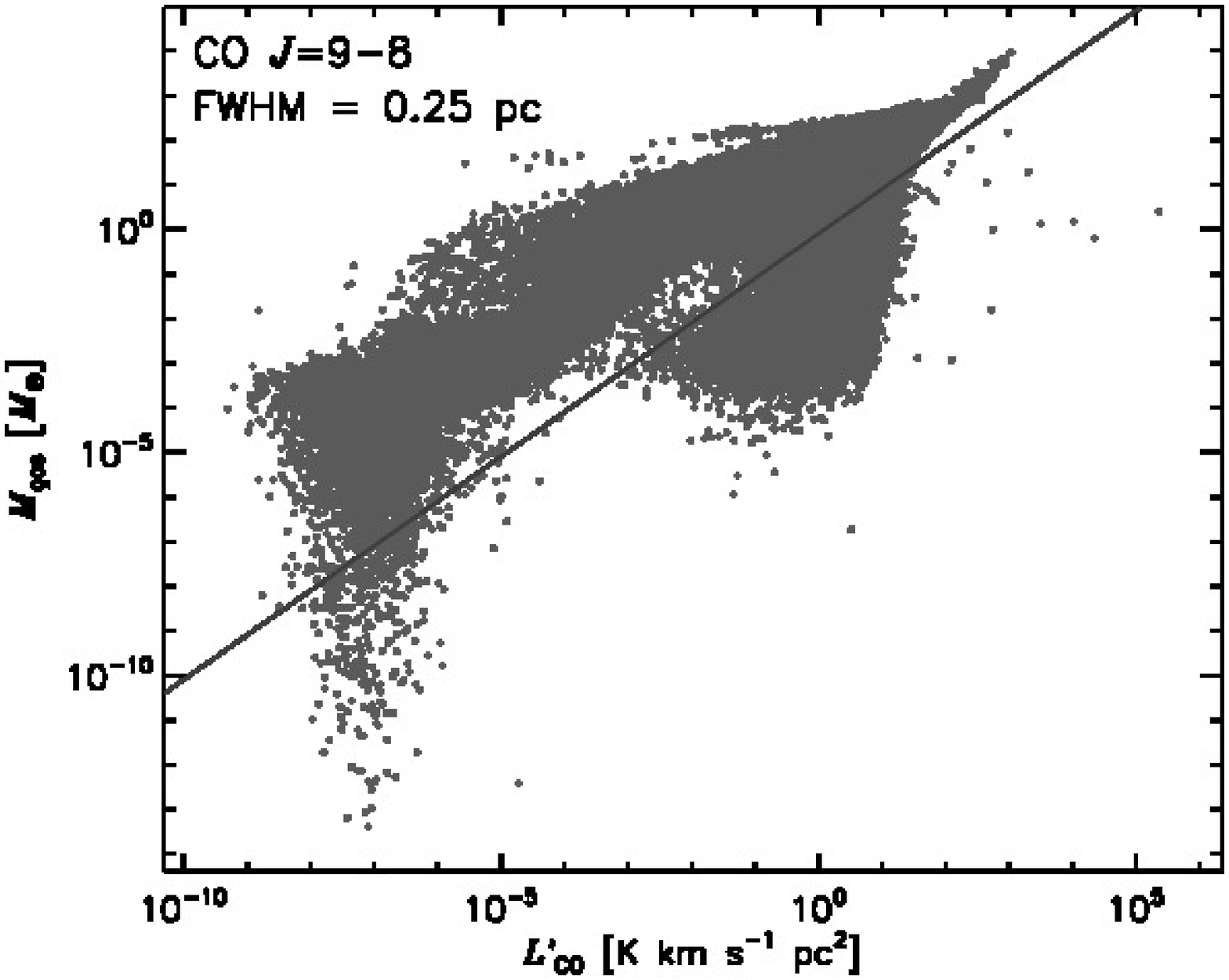}\hspace*{\fill}\\

\vspace{-0.5cm}

\hspace*{\fill}\includegraphics[angle=90,width=0.35\textwidth]{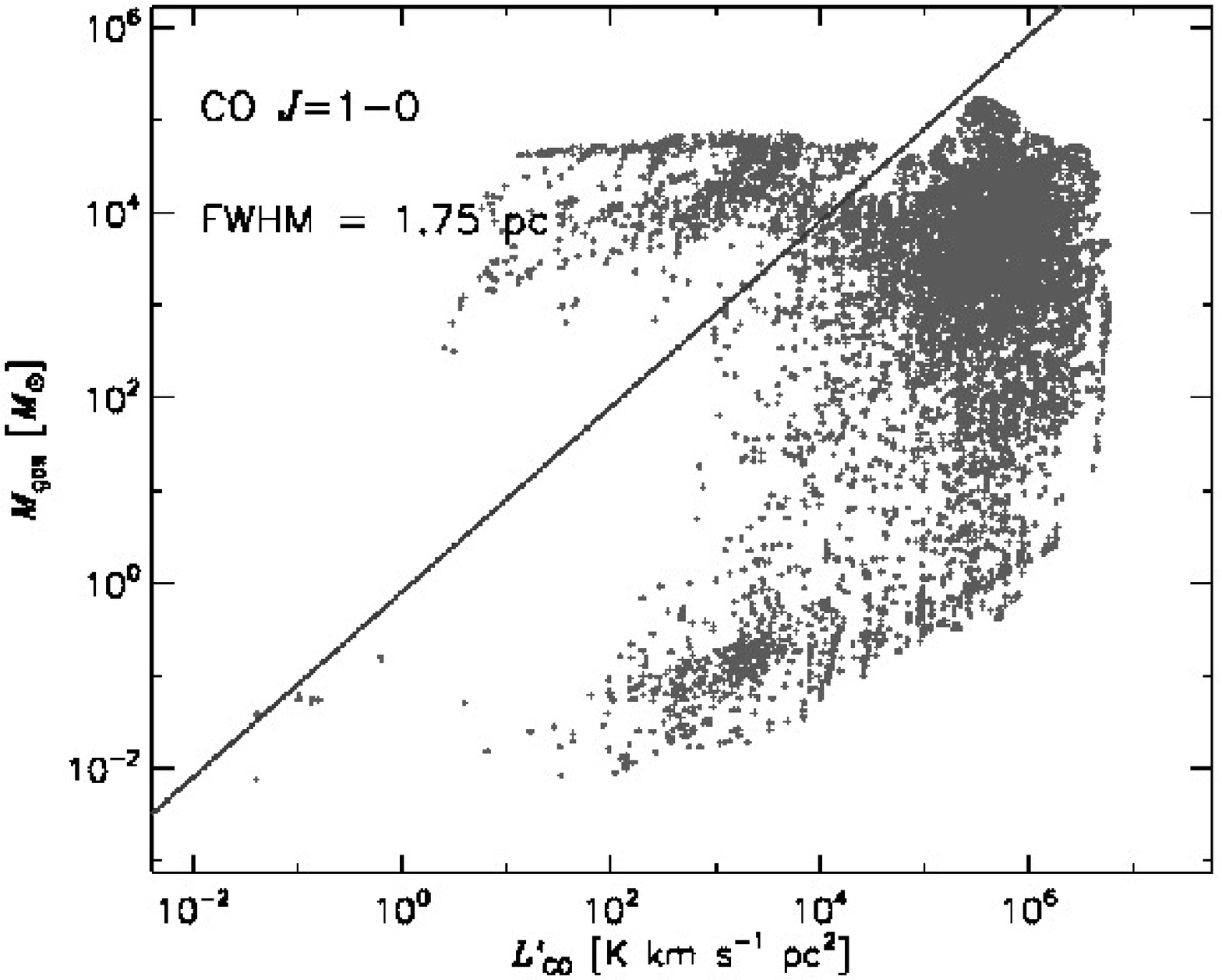}\hspace{-0.7cm}%
\hfill\includegraphics[angle=90,width=0.35\textwidth]{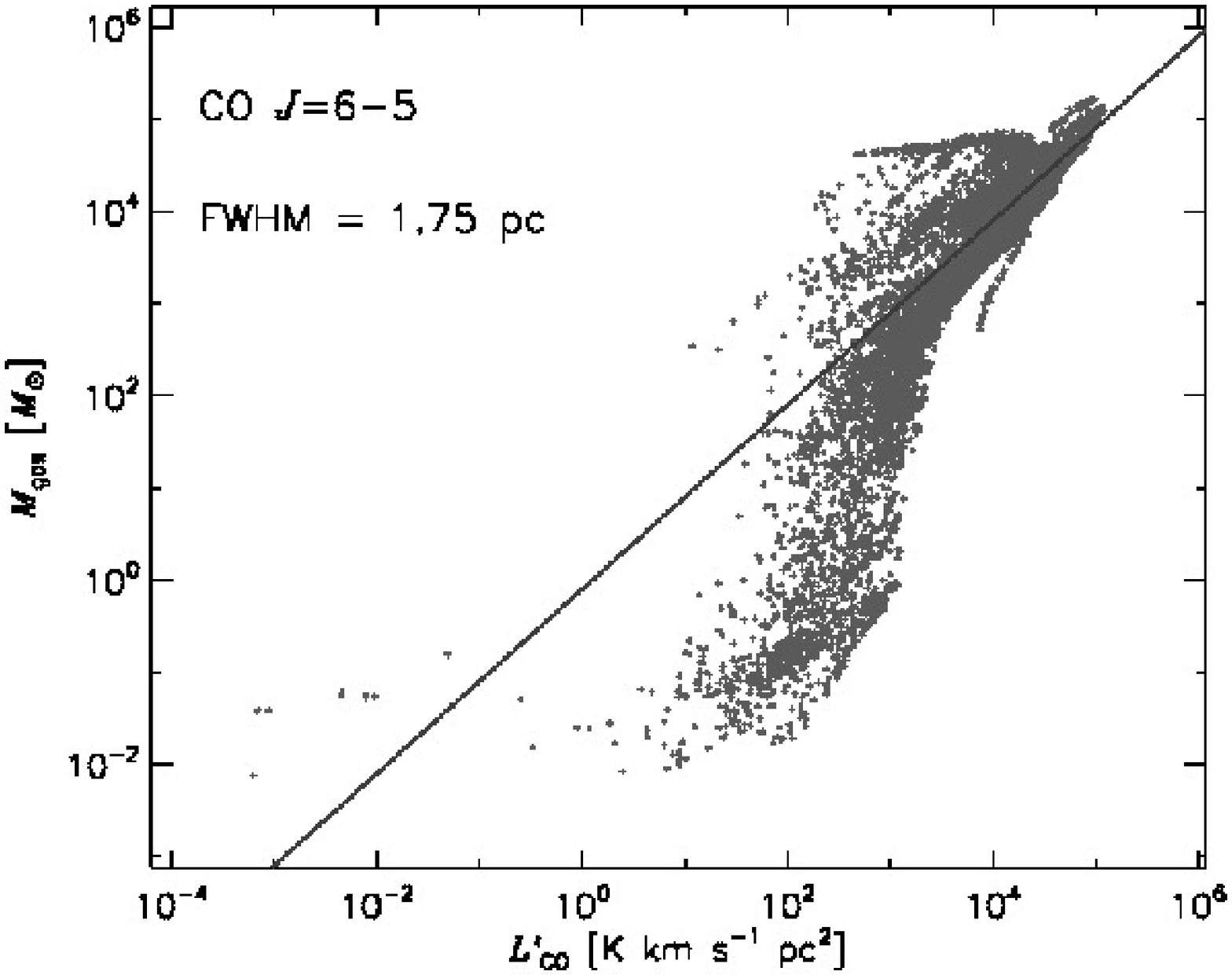}\hspace{-0.7cm}%
\hfill\includegraphics[angle=90,width=0.35\textwidth]{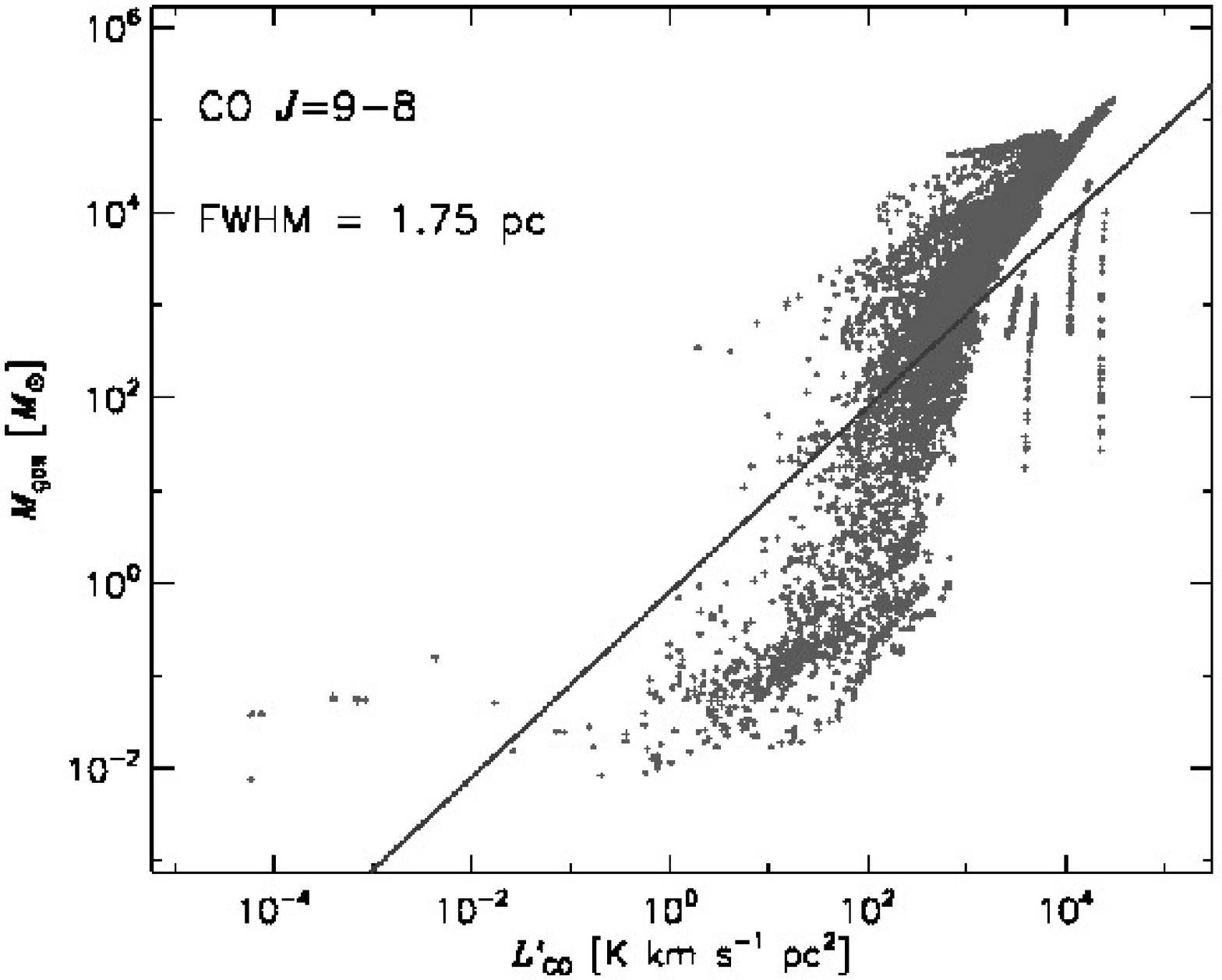}\hspace*{\fill}\\

\vspace{-0.5cm}

\hspace*{\fill}\includegraphics[angle=90,width=0.35\textwidth]{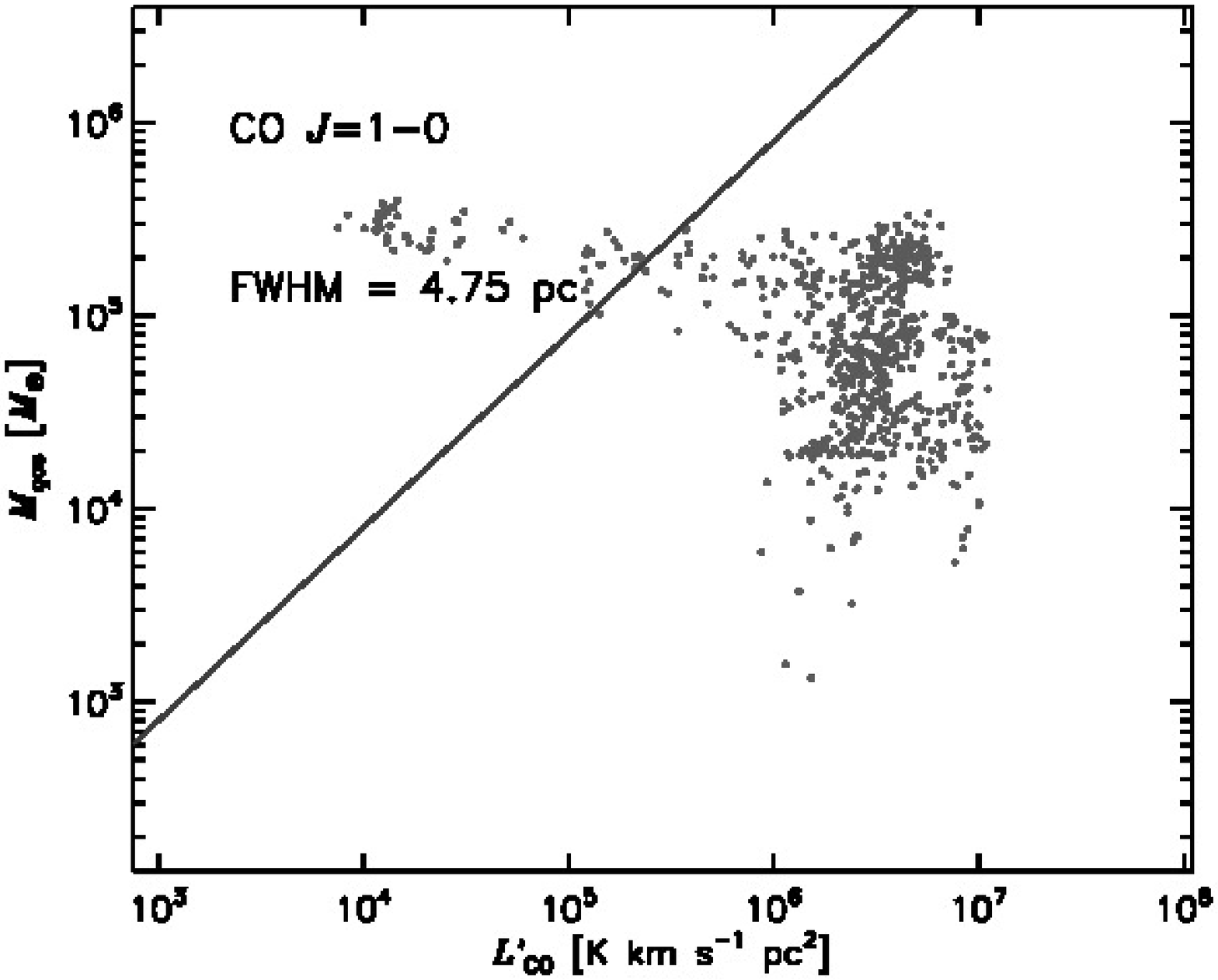}\hspace{-0.7cm}%
\hfill\includegraphics[angle=90,width=0.35\textwidth]{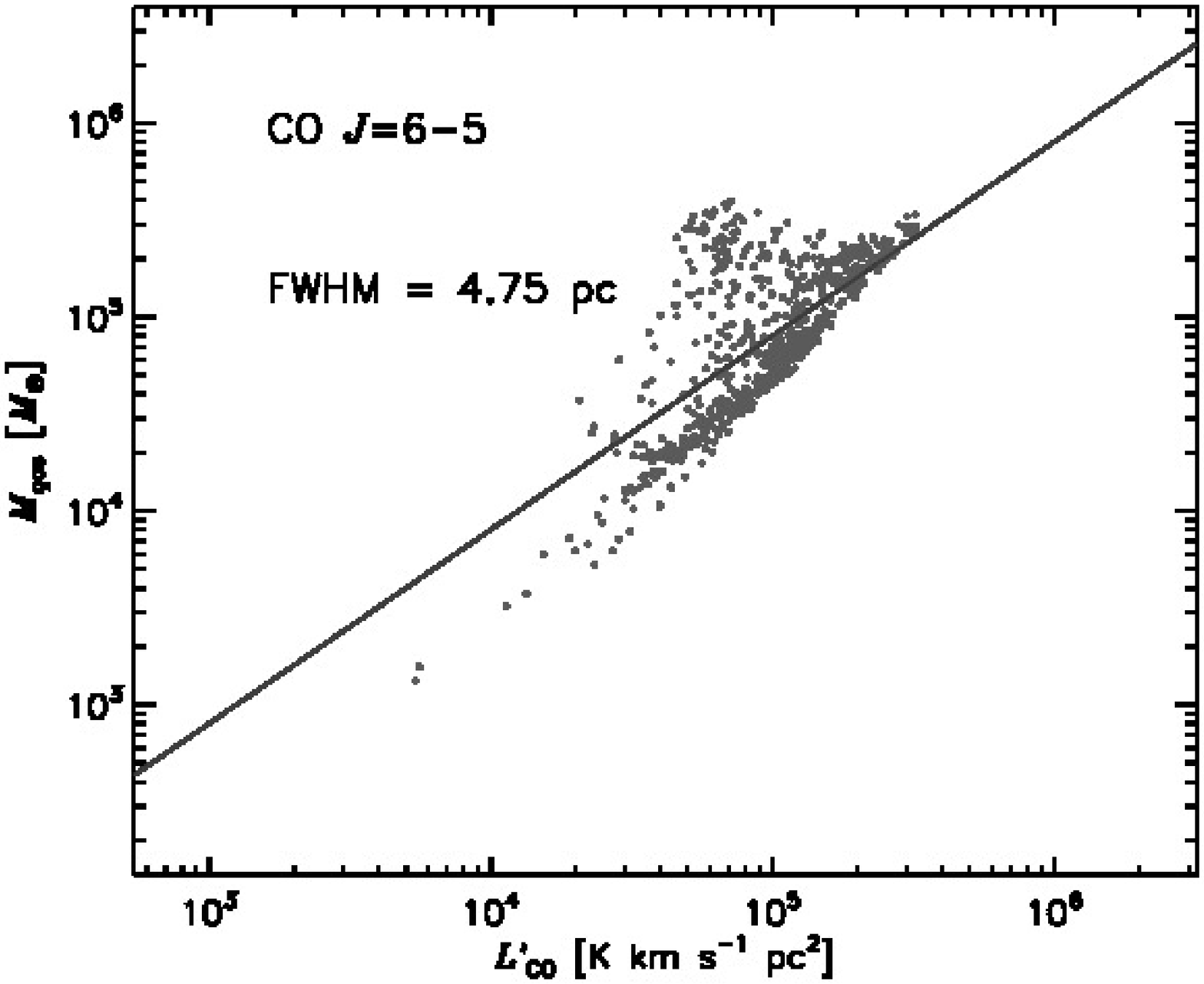}\hspace{-0.7cm}%
\hfill\includegraphics[angle=90,width=0.35\textwidth]{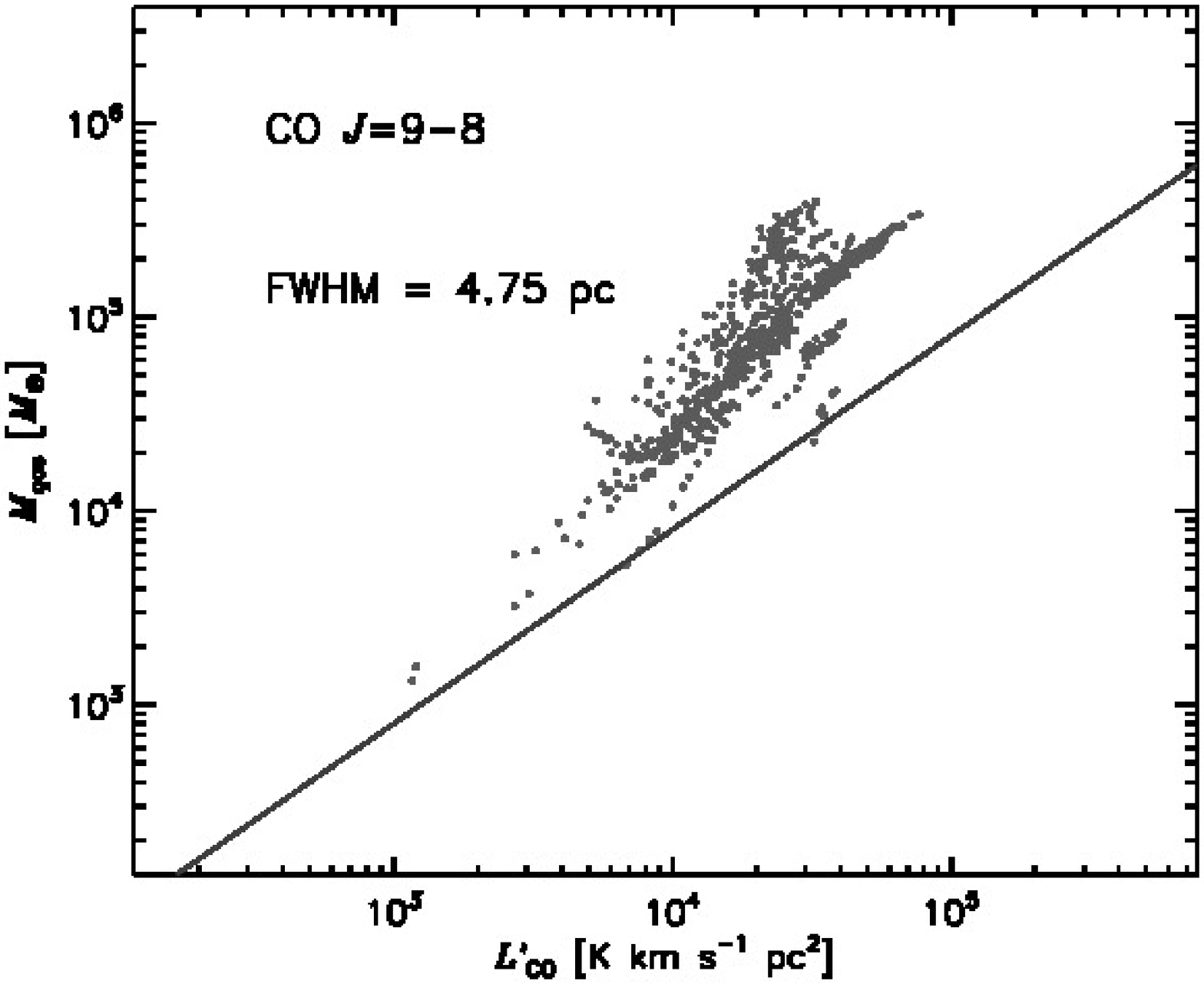}\hspace*{\fill}\\

\vspace{-0.5cm}

\hspace*{\fill}\includegraphics[angle=90,width=0.35\textwidth]{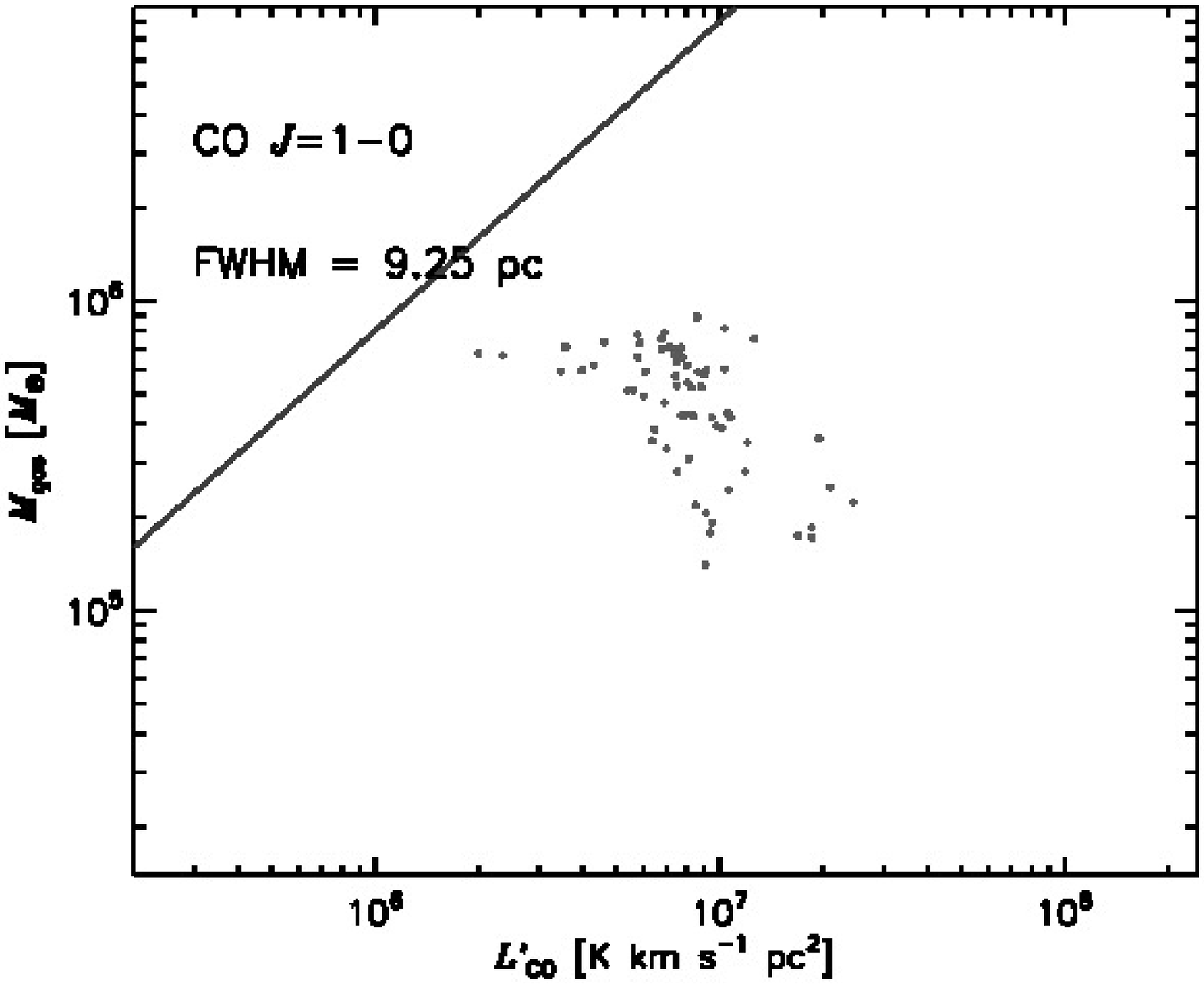}\hspace{-0.7cm}%
\hfill\includegraphics[angle=90,width=0.35\textwidth]{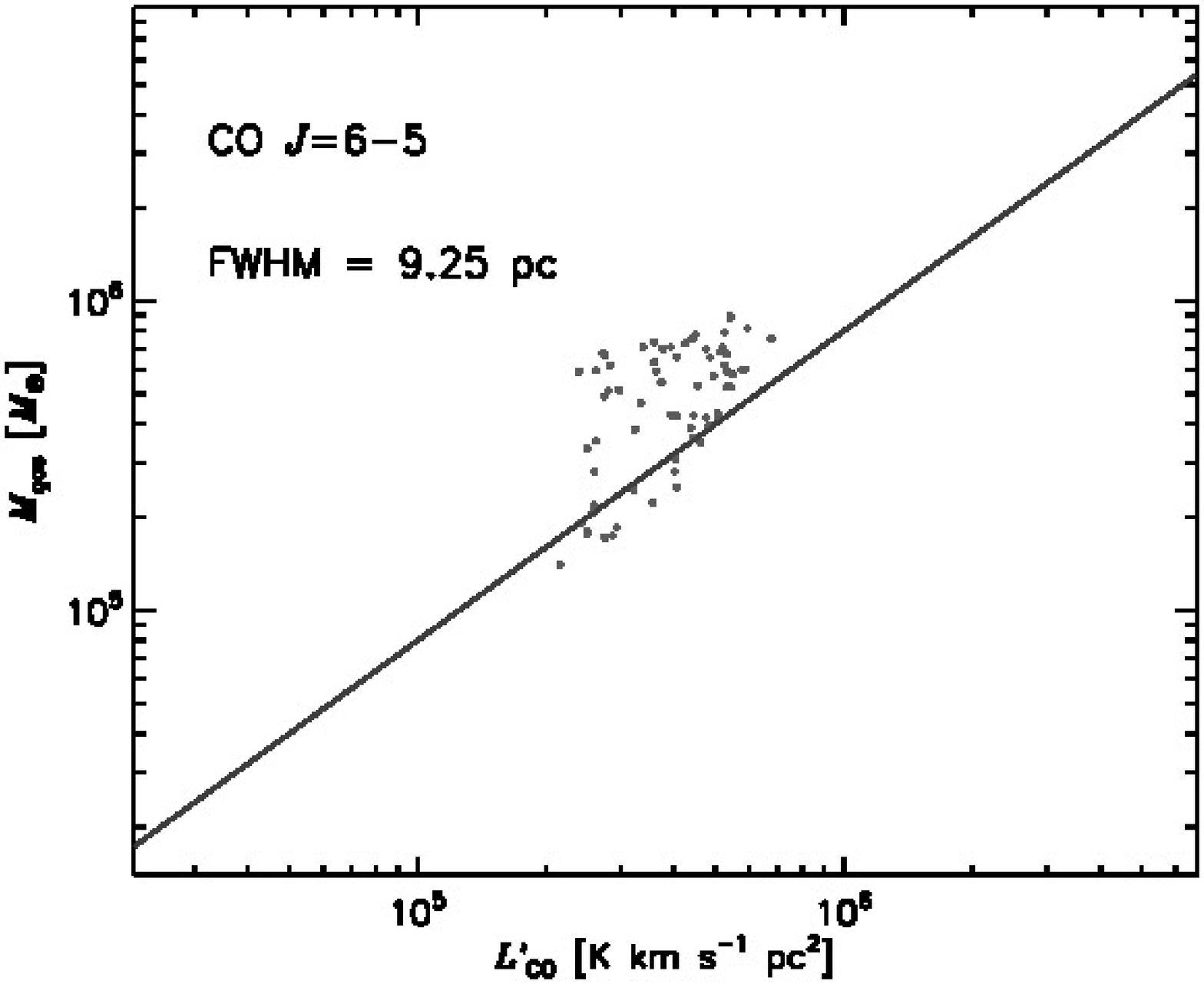}\hspace{-0.7cm}%
\hfill\includegraphics[angle=90,width=0.35\textwidth]{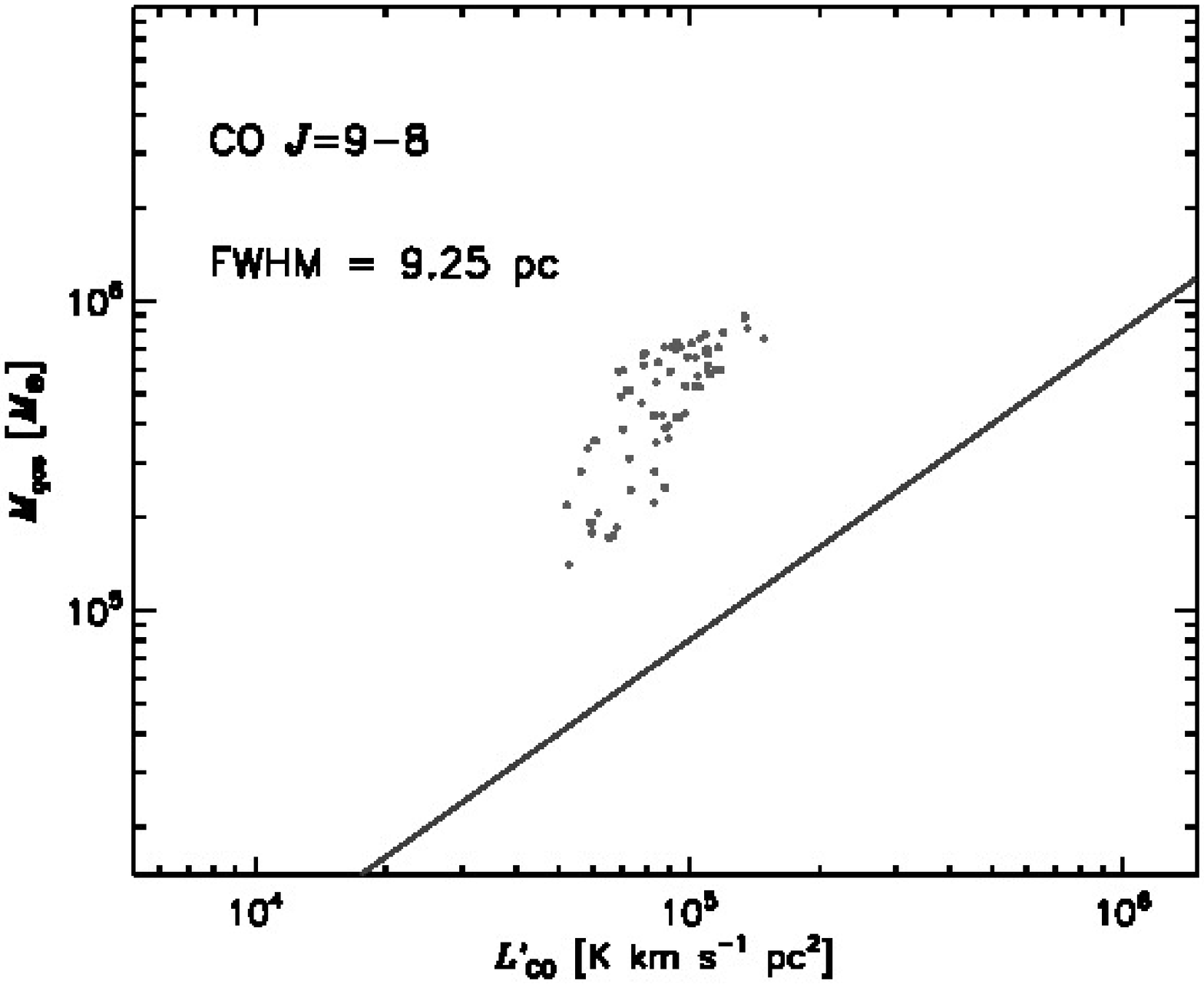}\hspace*{\fill}

\caption{{\footnotesize Scatter plots between the ensemble of gas masses ($M_{\rm gas}$ [\Msun]) and the CO luminosity ($L'_{\rm CO}$ [\Kkms\ $\rm pc^2$]) of the (from \textit{left} to \textit{right}) $J=1\rightarrow0$, $J=6\rightarrow5$, and $J=9\rightarrow8$ transitions derived from the low resolution ($0.25~\rm pc$) hydrodynamical model, and convolved (beam averaged) with beam sizes (FWHM) equivalent to 0.25 pc (the original scale of the model), 1.75 pc, 4.75 pc, and 9.25 pc (from \textit{top} to \textit{bottom}, respectively), for a distance of 3.82 Mpc to the source. The straight line corresponds to the gas mass estimated assuming a luminosity-to-gas mass conversion factor $\alpha$ of ${M_{\rm gas}}/{L'_{\rm CO}}=\alpha=0.8$ (e.g., Solomon et al. 1997; Downes \& Solomon 1998). The gas mass used here includes a 36\% correction to account for helium. {Note the different scales of the axis for different resolutions and $J$-lines.}
}}
\label{fig:Lco-MH2}
\end{figure}

%-------------------------------------------------------------------------

Similarly, for each pixel of the CO maps we compute the total gas mass $M_{\rm gas}$ by adding the individual masses of each grid point along the line of sight, and the CO luminosity $L'_{\rm CO}$ derived from the CO intensity maps obtained in Sec.~\ref{sec:co-maps}. In this case we do not apply the criteria $n_{\rm H}>100~\rm cm^{-3}$ and $T_{\rm CO}<5000~\rm K$ as before, in order to obtain the actual total $M_{\rm gas}$ of the hydrodynamical model along the line of sight. However, we use the previously obtained average density map in order to mask the pixels with $<n_{\rm H}>=0$ in the gas mass and CO luminosity maps as well. This is because those pixels correspond to CO emission that is more than five orders of magnitude lower than the peak CO emission produced from our model. Given that is unlikely to have such large dynamical range in a real detector, the lower emission would be even below the noise level that could be found in maps obtained from real observations. Thus, not masking these pixels would cause an artificial bias toward lower luminosities during the convolution process. An alternative approach to masking pixels is to add random background noise to our intensity maps. However, the noise level to be added is arbitrary, so the results can be made at least comparable, and we are actually more interested in to show what is really coming out from the hydrodynamical and XDR models.}

Figure~\ref{fig:Lco-MH2} shows the scatter plots of the gas masses ($M_{\rm gas}$) and the CO luminosity ($L'_{\rm CO}$ [\Kkms\ $\rm pc^2$]) of the $J=1\rightarrow0$, $J=6\rightarrow5$ and $J=9\rightarrow8$ transitions as obtained from the low resolution ($0.25~\rm pc$) hydrodynamical model. {We consider} a 36\% mass correction to account for the mass of helium atoms. The straight line corresponds to the gas mass estimated assuming a luminosity-to-gas mass conversion factor $\alpha$ of ${M_{\rm gas}}/{L'_{\rm CO}}=\alpha=0.8$ \citep[e.g.][]{solomon97, downes98}. The result of the original resolution shows a large scatter of gas masses for a given CO luminosity. When considering the source at a nominal distance of 3.82 Mpc, however, the beam averaged masses and luminosities result in a tighter correlation, {but still with a considerable scatter. 
The relation between $L'_{\rm CO}$ and $M_{\rm gas}$ is clustered just at the higher luminosity and mass ranges as the beam size increases. That is, the pixels with lower mass and luminosity are missing (beam smeared) in the lower resolution maps.}
The higher the CO $J$-line and the larger the beam used, the closer is the relation found with our models to the {luminosity-to-gas mass conversion factor} proposed in the literature. {The $\alpha$} factor was estimated based on the lower transitions ($J=1\rightarrow0$, $J=2\rightarrow1$, $J=3\rightarrow2$), which are way off in our models because there is very little cold ($T_{\rm CO}<100~\rm K$) gas in the inner $60~\rm pc$. {Besides, the relatively large scatter observed at all resolutions indicates that many of the clumps where the CO emission emerges from are not in virial equilibrium.}
{On the other hand,} the CO $J=6\rightarrow5$ and $J=9\rightarrow8$ transitions show similar trends. This implies that the higher-$J$ line does not provide significant additional information compared to the $J=6\rightarrow5$ line.

We can confirm then a {tight} correlation between the {beam averaged luminosity of the} higher CO $J$-lines (from $J=5\rightarrow4$ to $J=9\rightarrow8$) and the gas mass of the AGN torus. However, the larger scatter seen when using higher resolution (smaller beam sizes) should be taken as a warning sign for future higher resolution observations like ALMA, which may no longer show a linear correlation between CO luminosity and gas mass, or may introduce a larger range of correlations for ensembles of clouds with different density and high temperature structures. So, using only mid-$J$ CO lines {(like $J=6\rightarrow5$)} for gas mass determinations in AGN tori {would} be the most reliable approach.

\subsection{The \cii\ 158$\mum$ fine structure line}\label{sec:CII-to-CO}

A considerable amount of gas close to the SMBH is predicted to be at very high temperatures ($T_k>1000$ K) in the 3-D hydrodynamical model. At these temperatures hydrogen is found mostly in atomic form, and other atoms like carbon are mostly ionized. Therefore, in this section we present the results of the radiative transfer calculations done for \cii.
We use the collision rate coefficients of \cii\ from the LAMDA database. In contrast with the CO molecule, we considered not only the molecular hydrogen as collision partner, but also the atomic hydrogen and the electrons, using the collision data reported by \citet{flower77}, \citet{launay77}, and \citet{wilson02}. These are the most relevant collision partners for \cii\ in an AGN environment, since their densities are expected to be higher than $n(\rm H_2)$ in the very hot gas ($T_k>1000$ K) regions. As described in {Secs.~\ref{sec:abundances} and \ref{sec:beta3D}}, the fractional abundance of the electrons is derived from the XDR model. Whilst the density $n(\rm H)$ is computed from the hydrodynamical model as $n({\rm H})=n_{{\rm H}}-2n(\rm H_2)$ \citep{wada09}.

\begin{figure}[tp!]
%\centering

\hspace*{\fill}\includegraphics[angle=0,width=0.33\textwidth]{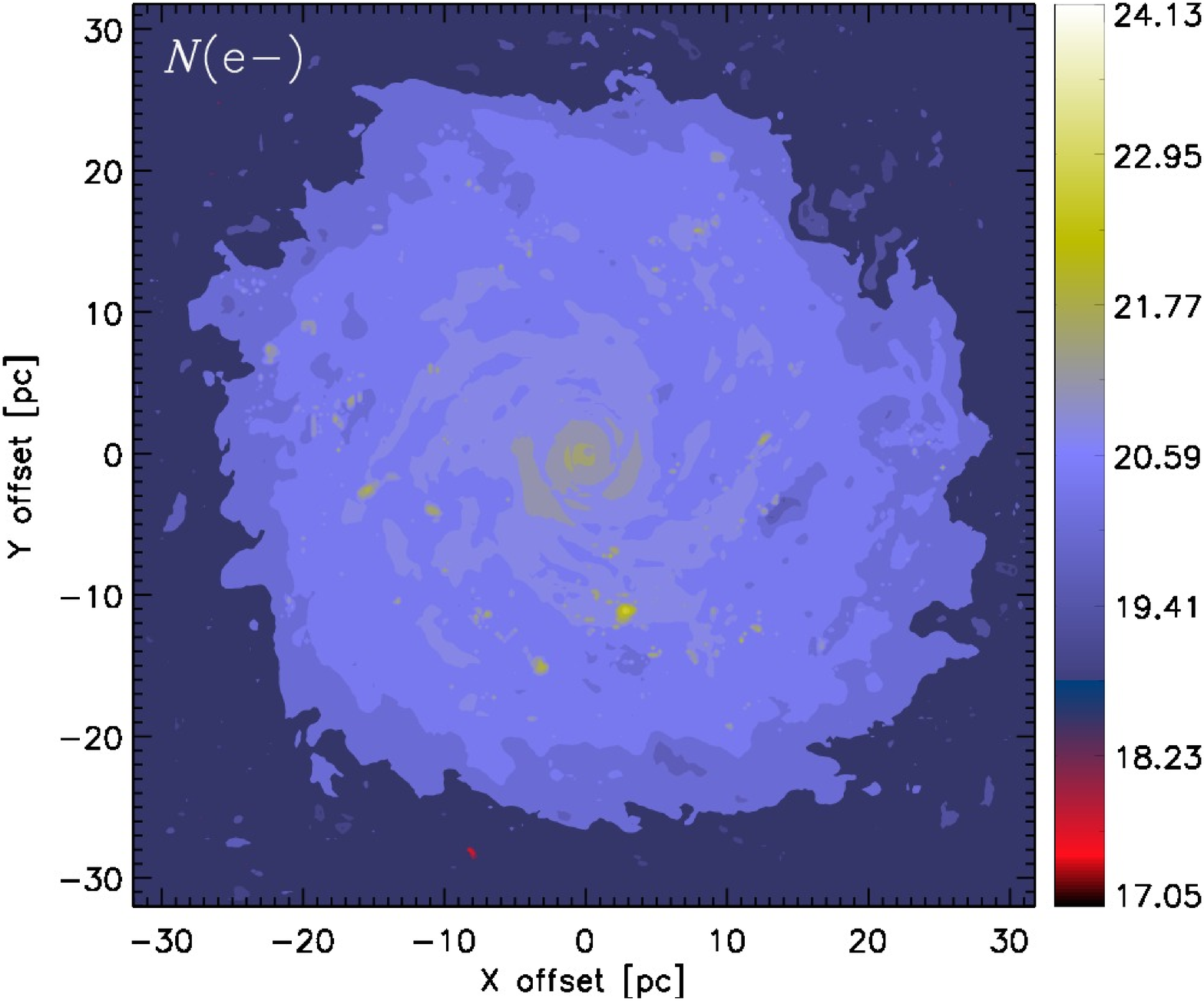}%
\hfill\includegraphics[angle=0,width=0.33\textwidth]{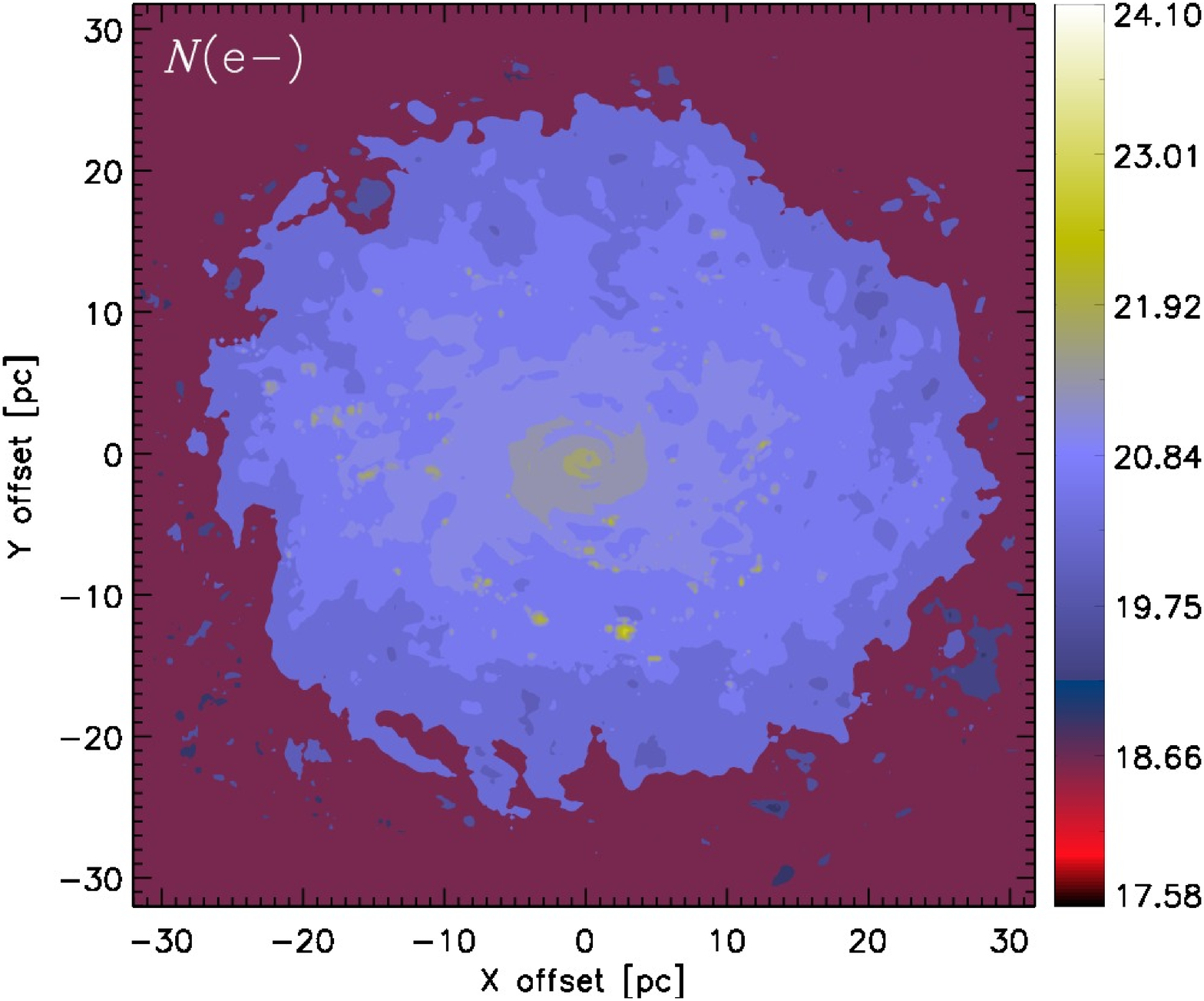}\hspace*{\fill}\\

\vspace{-0.6cm}

\hspace*{\fill}\includegraphics[angle=0,width=0.33\textwidth]{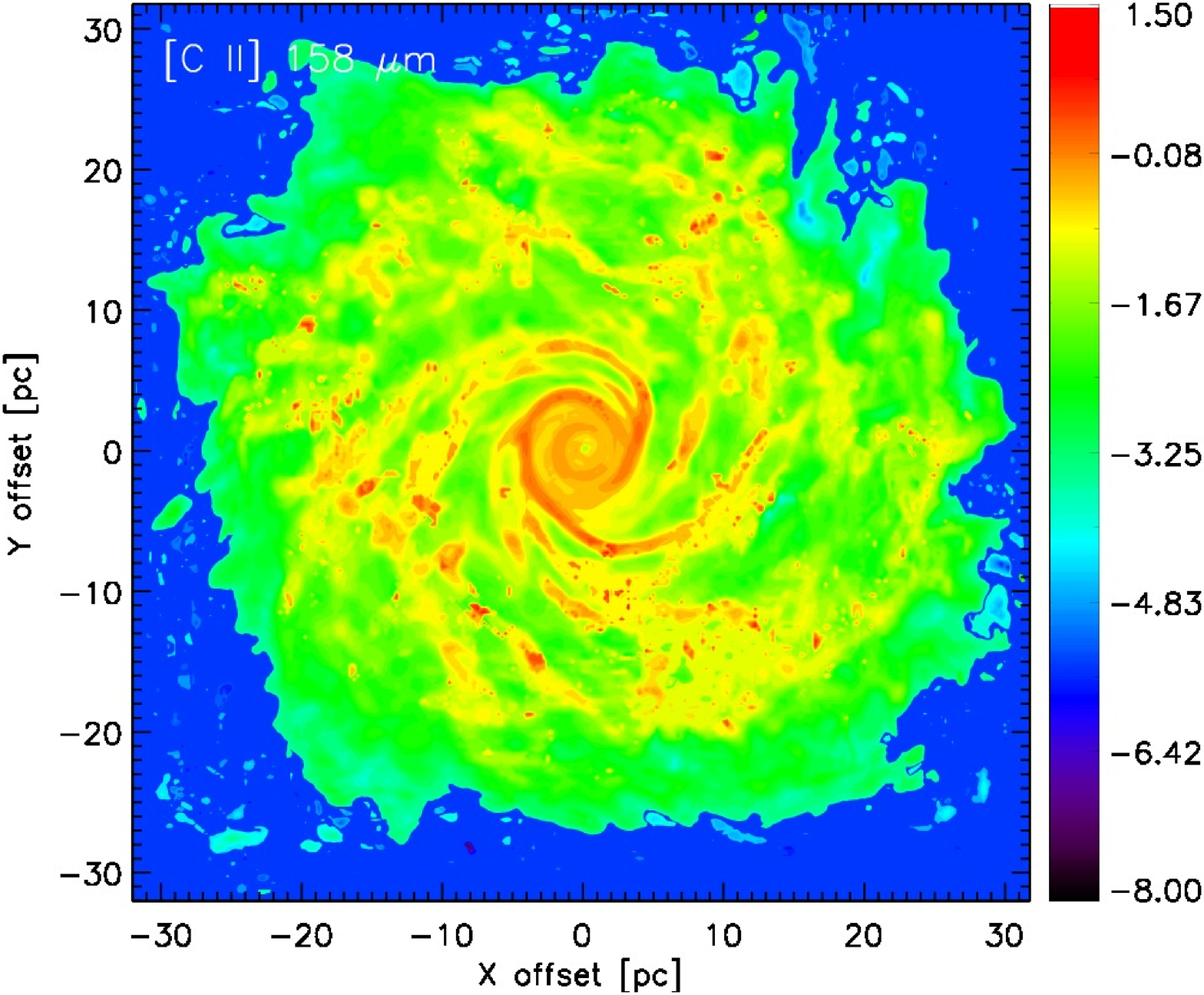}%
\hfill\includegraphics[angle=0,width=0.33\textwidth]{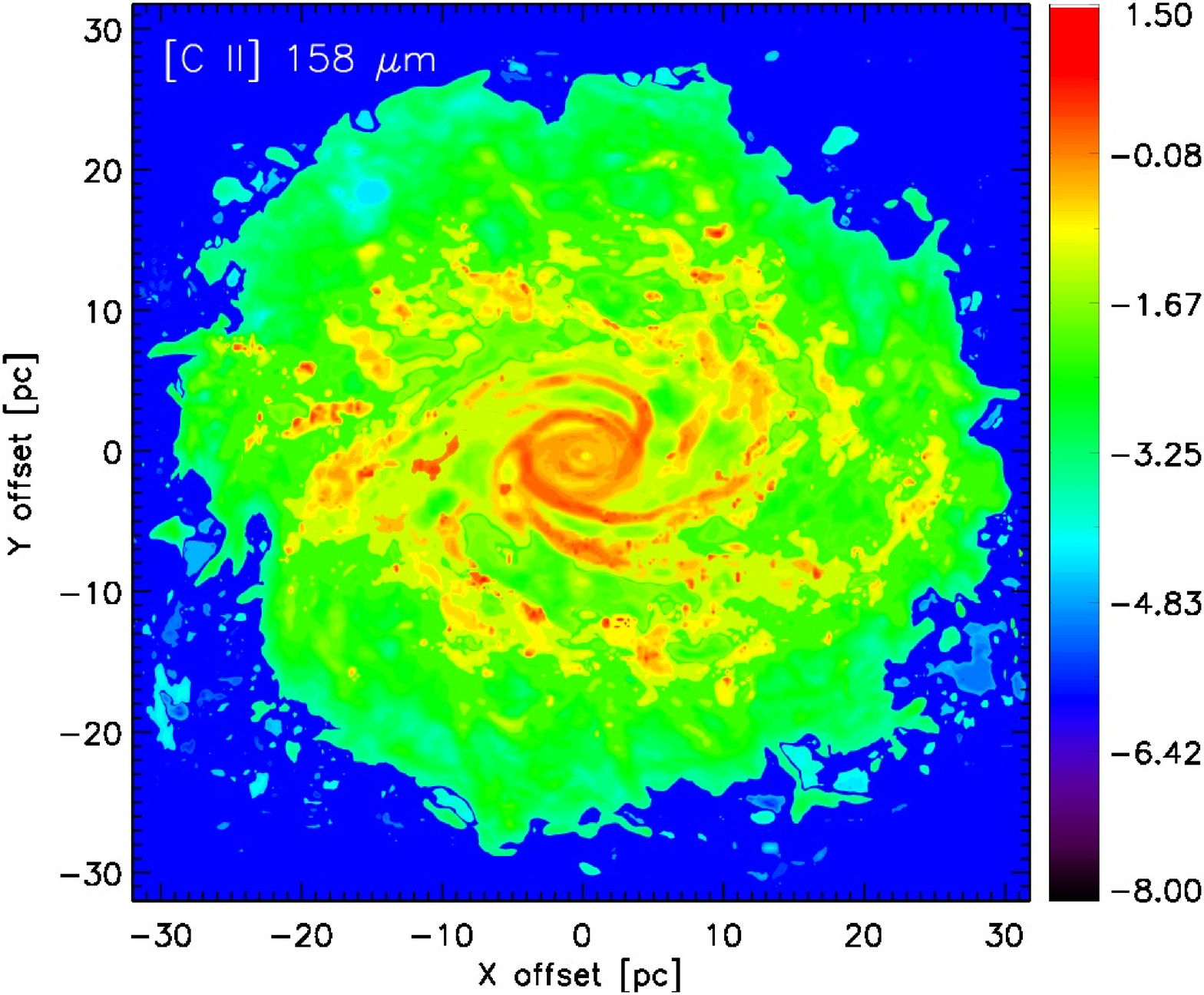}\hspace*{\fill}\\

\vspace{-0.6cm}

\hspace*{\fill}\includegraphics[angle=0,width=0.33\textwidth]{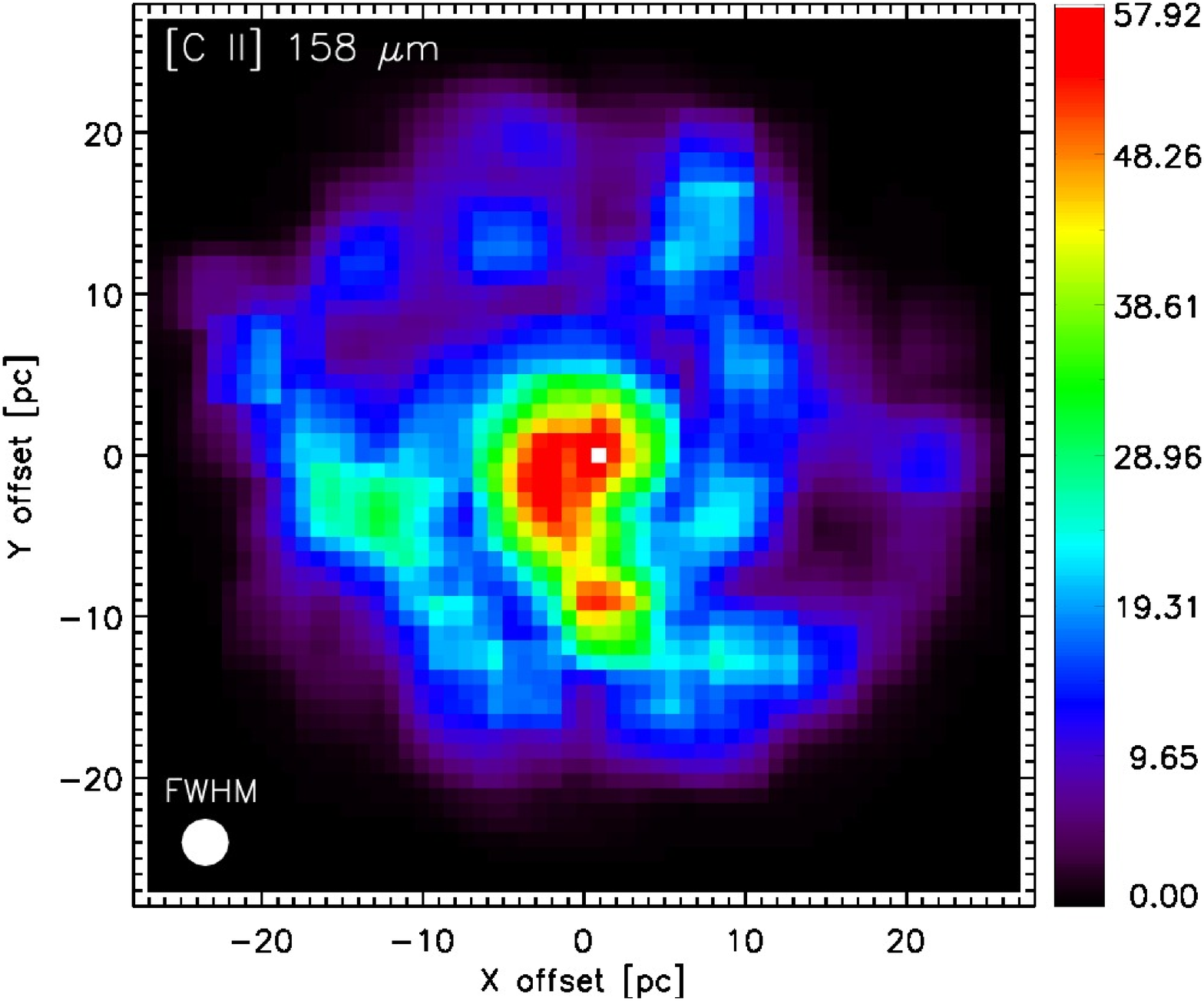}%
\hfill\includegraphics[angle=0,width=0.33\textwidth]{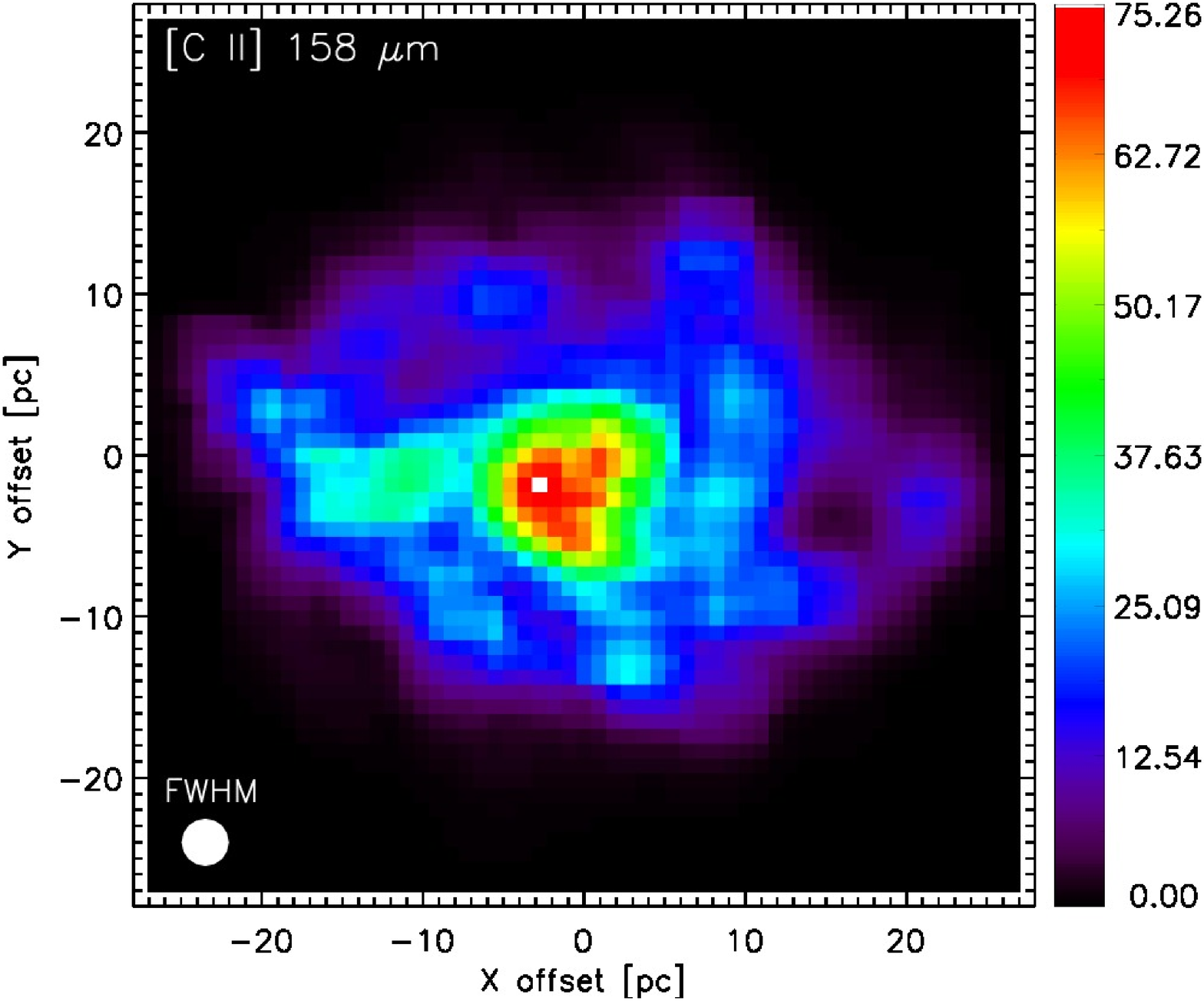}\hspace*{\fill}\\

%\vspace{-0.6cm}

%\hspace*{\fill}\includegraphics[angle=0,width=0.33\textwidth]{C+1_fwhm_0.15_log.eps}%
%\hfill\includegraphics[angle=0,width=0.33\textwidth]{C+1_fwhm_0.15_45x_log.eps}\hspace*{\fill}\\

\caption{{\footnotesize \textit{Left panels} - Face-on view (from top to bottom) of the total column density $N({\rm e^-})$ ($\2cm$) of electrons in logarithmic scale, the face-on view of the surface brightness (in $log_{10}$ scale and units of $\ergscmsr$) of the \cii\ $158~\mum$ emission, and the same map as above but convolved with a single dish beam of FWHM=$0.15''$ ($\sim2.8~\rm pc$ at an adopted distance $D=3.82~\rm Mpc$ to the source), which gives a flux in units of $10^{-15} \ergscm$.
\textit{Right panels} - Same as in the left panels, but with an {inclination (viewing) angle of $45^o$ about} the X-axis.}}
\label{fig:CII}
\end{figure}

Figure~\ref{fig:CII} shows the face-on view (\textit{left panels}) of the total column density $N({\rm e^-})$ ($\2cm$) of electrons, the surface brightness (in $log_{10}$ scale and units of $\ergscmsr$) of the \cii\ $158~\mum$ emission at the original spatial resolution ($0.25~\rm pc$), and at $\sim2.8~\rm pc$ resolution (for a distance $D=3.82~\rm Mpc$ to the source) after convoling the original map with a single dish beam of FWHM=$0.15''$. This results in a flux with units of $10^{-15} \ergscm$ after multiplying the surface brightness by the solid angle as described in Sec.~\ref{sec:co-maps}. Comparing with the CO emission shown in Figs.~\ref{fig:column-density} and \ref{fig:co-flux-fwhm}, it is clear that the \cii\ $158~\mum$ emission traces mostly the central region {(inner NLR)} of the AGN torus, and it is a better tracer of the hot regions than the mid-$J$ CO lines.

The \textit{right panels} of Fig.~\ref{fig:CII} shows the same as in the left panels, but with an inclination angle of $45^o$ about the X-axis. The viewing angle produces a slightly larger column of \cii\ observed through the line of sight, which in turn results in a $\sim30\%$ brighter \cii\ $158~\mum$ peak emission. The \cii\ $158~\mum$ emission (rest frequency $\sim1900~\rm GHz$) from galaxies at redshift $z\gtrsim1$ will be observable with ALMA. {Unfortunately, we cannot obtain high resolution maps of \cii\ with the current specifications of ALMA. Even if we consider both the largest baseline of 16 km and the highest frequency of 950 GHz (band 10) that ALMA is expected to have in the future, we could get a resolution of FWHM\footnote{http://science.nrao.edu/alma/specifications.shtml}$\sim$60/16(km)/950(GHz)$\sim$0.004''. This in turn gives a spatial scale of ~32 pc for a source at z=1 (with an equivalent angular distance of 1658.6 Mpc, in a LambdaCDM cosmology with H$_0$=71 km/s/Mpc). That is, the whole region we show in Fig.~\ref{fig:CII} would correspond to just one or two pixels in the ALMA maps of sources at $z=1$.}

\subsection{Temperature and density driven by X-rays}\label{sec:temps-dens}

In eq.\ref{eq:X-ray-flux} we assumed that the X-ray flux is spherically symmetric with respect to the central SMBH. If we assume instead that the X-rays are emitted in a preferential direction perpendicular to the mid-XY-plane of the accretion disk (and of the AGN torus as a whole), we can consider the X-ray flux emerging from a Lambertian object. That is, the radiation flux impinging on each grid point of the cube is proportional to the cosine of the viewing angle $\theta$, with respect to the vertical Z-axis. This means that the X-ray flux would be negligible in the disk (mid-plane) of the AGN torus. 

In order to compare the impact that the two different X-ray flux distributions have on the temperature and molecular hydrogen density of the gas, we compare the temperature $T_{\rm HYD}$ obtained from the hydrodynamical model with the H$_2$ abundance-weighted average temperature $T_{\rm XDR}$ (eq.\ref{eq:abun-wa-temp}) derived from the XDR chemical model using both the \textit{spherical} and \textit{Lambertian} X-ray fluxes. We take a strip volume of $64\times1.25\times1.5~\rm pc^3$ along the X-axis, and around the center of the Y-axis ($\Delta \rm Y=0$) and Z-axis ($\Delta \rm Z=0$) of the 3-D cube. We use this \textit{thin} volume so we can have similar X-ray fluxes (decreasing mostly with radial distance) impinging at each grid element of the $1.25\times1.5~\rm pc^2$ slices of the volume. We computed the average temperature and H$_2$ density of the $1.25\times1.5~\rm pc^2$ slices at each $\Delta \rm X$ grid element. At the resolution of $0.25~\rm pc/element$ we have 30 grid elements per slice, which is a good compromise between a representative number of grid elements, and a fairly constant impinging X-ray flux at each slice.

\begin{figure}[!htp]

\hspace*{\fill}\includegraphics[angle=90,width=0.5\textwidth]{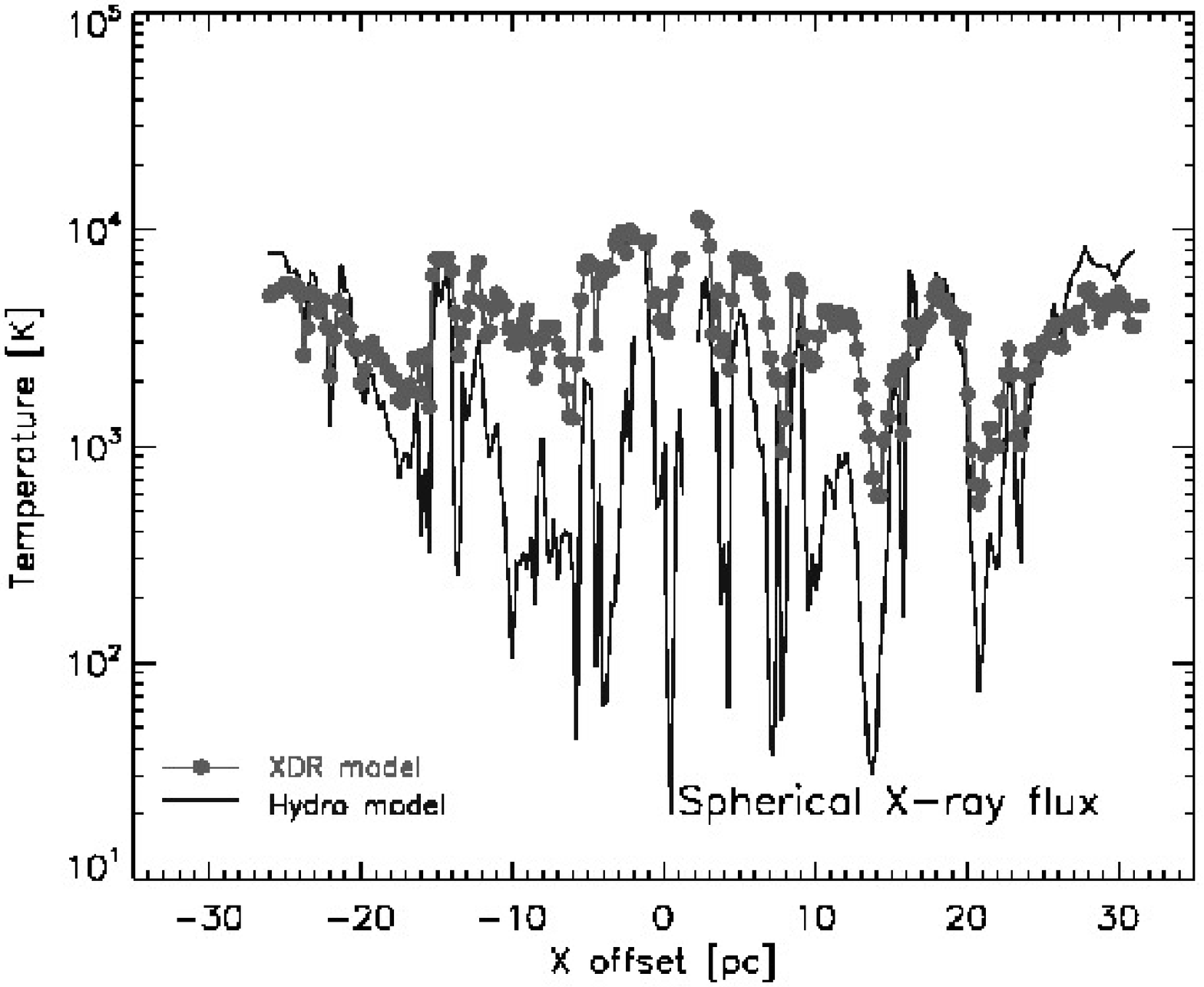}%
\hfill\includegraphics[angle=90,width=0.5\textwidth]{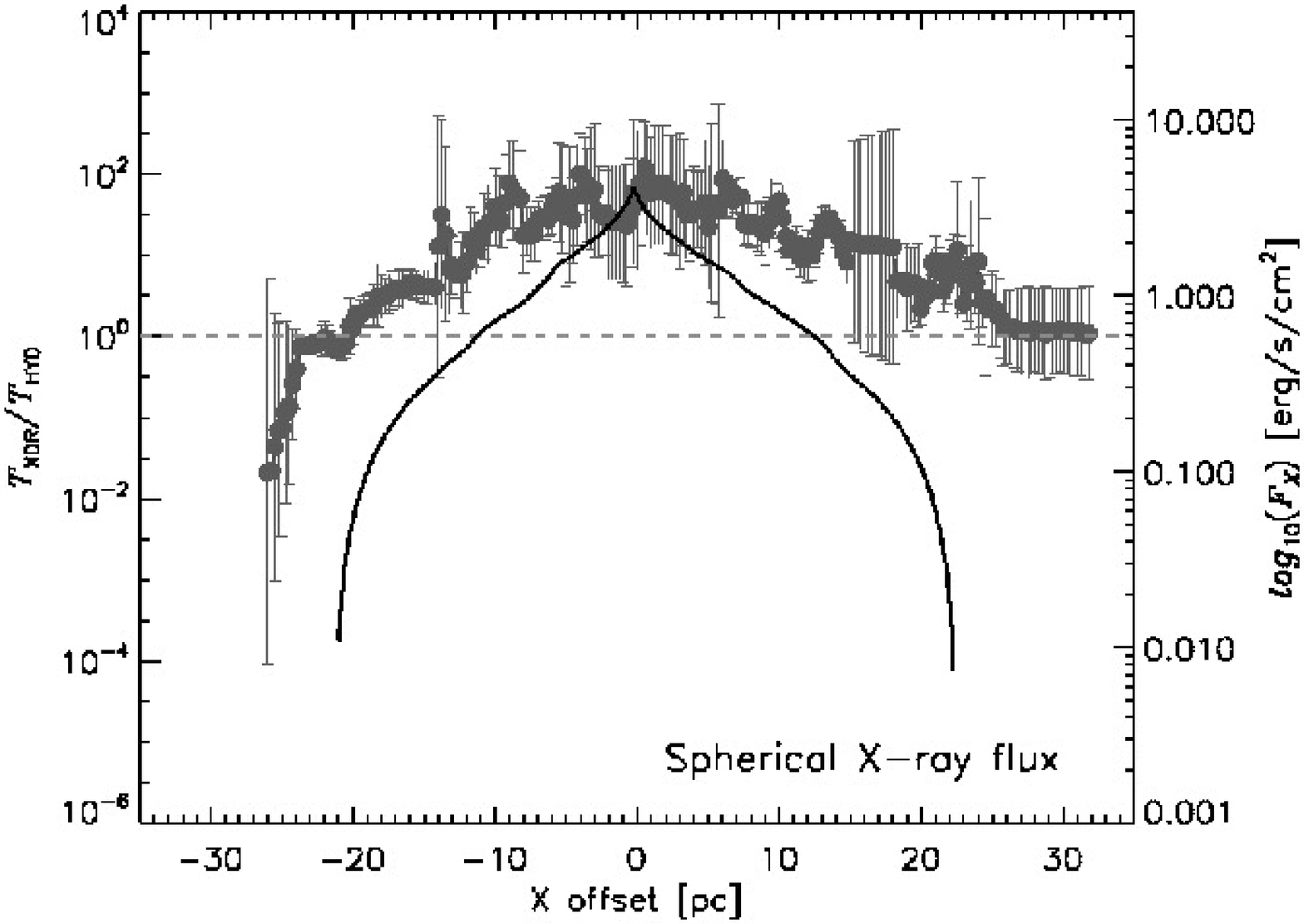}\hspace*{\fill}\\

\vspace{-0.5cm}

\hspace*{\fill}\includegraphics[angle=90,width=0.5\textwidth]{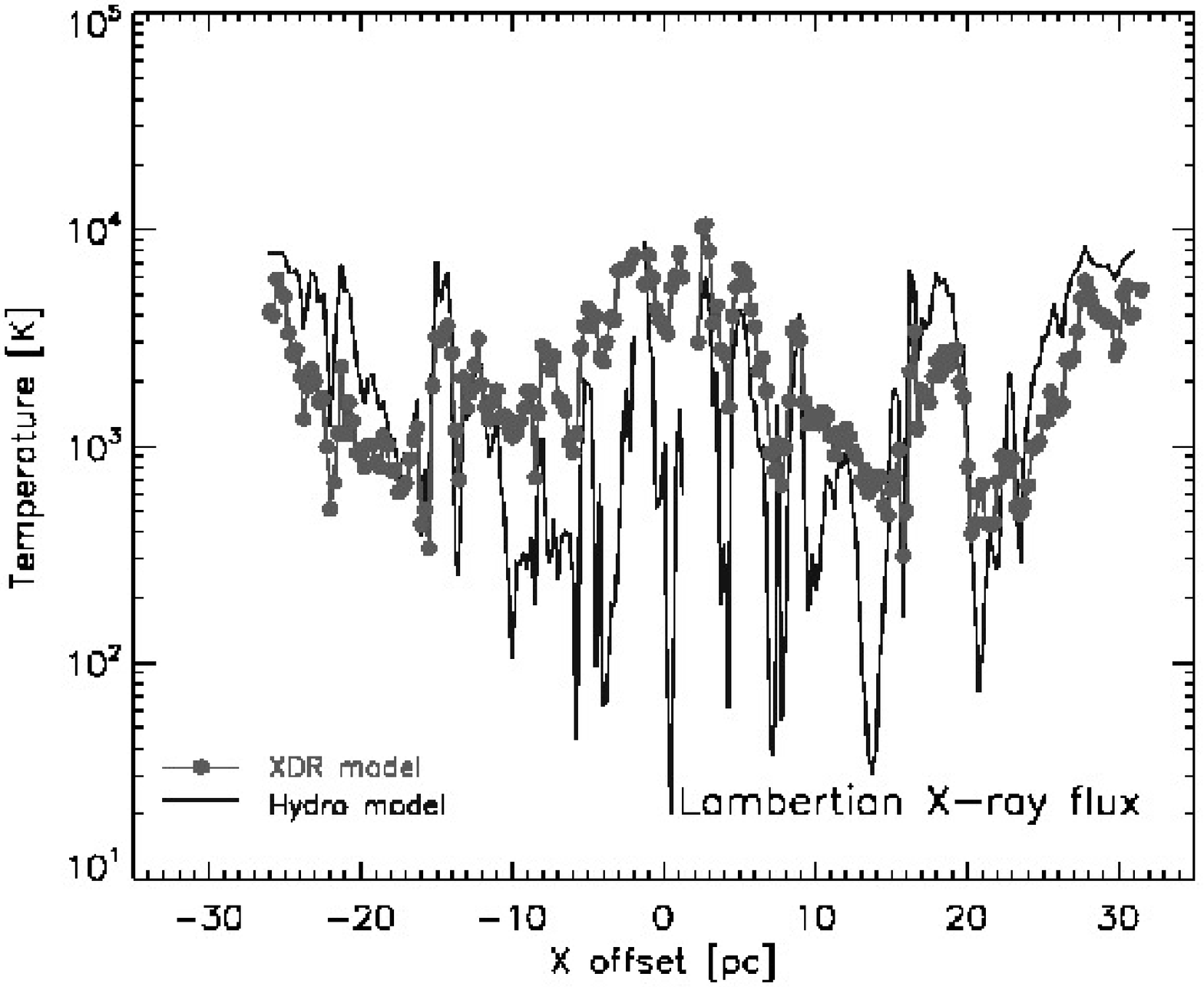}%
\hfill\includegraphics[angle=90,width=0.5\textwidth]{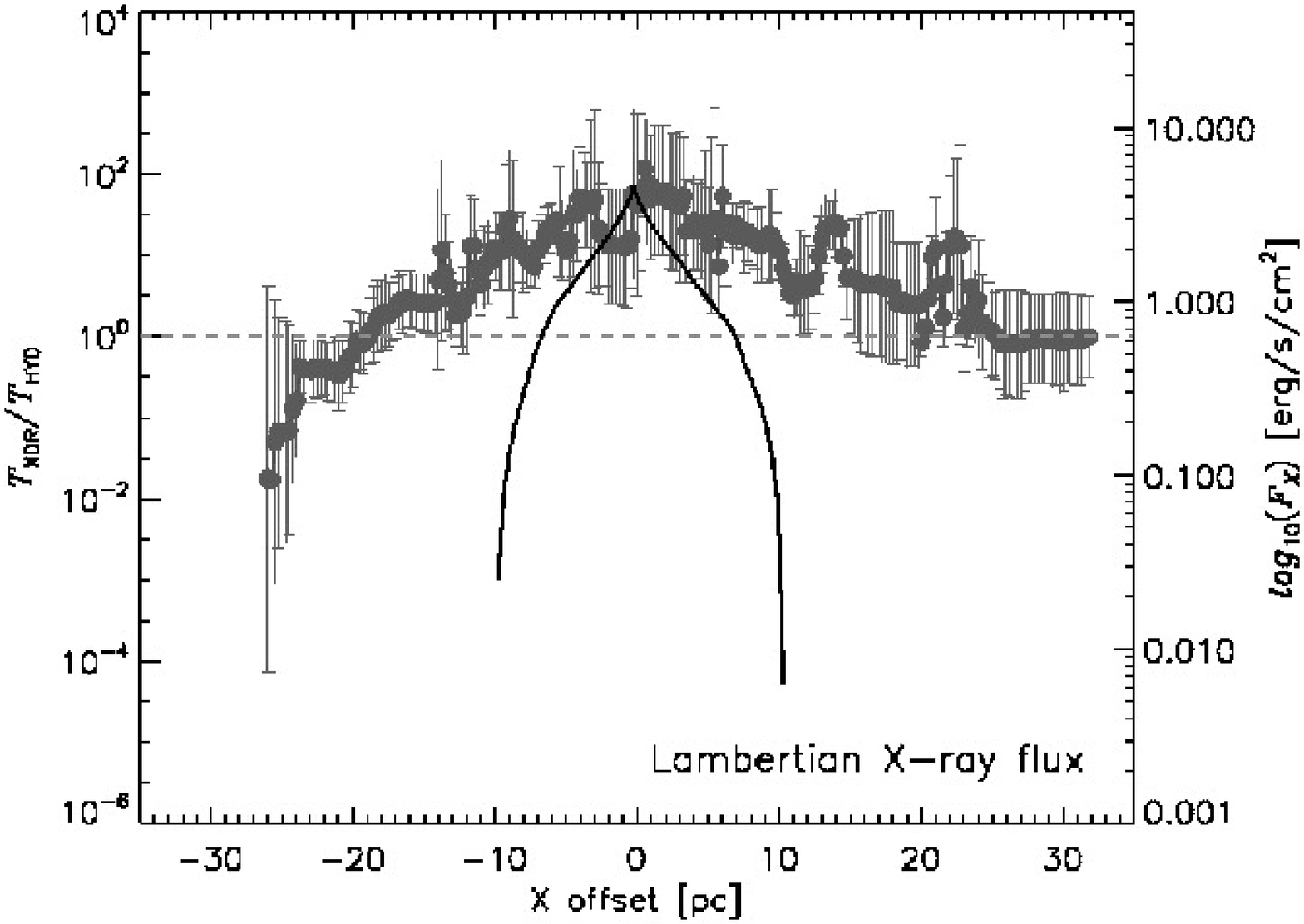}\hspace*{\fill}

\vspace{-0.3cm}

\caption{{\footnotesize \textit{Top left} - Average temperature $T_{\rm HYD}$ (in units of $\rm K$) of the gas along a $64\times1.25\times1.5~\rm pc^3$ strip volume, as estimated in the 3-D hydrodynamical model (\textit{black line}) and the corresponding H$_2$ abundance-weighted average temperature $T_{\rm XDR}$ obtained from the XDR chemical model (\textit{gray line}) using the \textit{spherical} X-ray flux (eq.\ref{eq:X-ray-flux}). The filled circles show the actual data points obtained with the XDR model in grid cells with $T_{\rm HYD}<10^4~\rm K$. \textit{Top right} - Average $T_{\rm XDR}/T_{\rm HYD}$ ratio (\textit{gray line + filled circles}) and the average \textit{spherical} X-ray flux (\textit{solid black line}) $F_X$ ($\ergscm$) in $log_{10}$ scale. The standard deviation at each $\Delta \rm X$ offset is shown by the error bars. The relative average temperature is directly related to the impinging flux at each grid point. The average temperature $T_{\rm XDR}$ is predominantly higher than $T_{\rm HYD}$ in the inner $20~\rm pc$ around the center of the AGN torus. The \textit{dashed line} indicates where $T_{\rm XDR}/T_{\rm HYD}=1$. The \textit{Bottom panels} show the same as above, but for the \textit{Lambertian} X-ray flux. The average temperature $T_{\rm XDR}$ is still higher than $T_{\rm HYD}$, but the difference is smaller than for the \textit{spherical} X-ray flux, particularly beyond $\pm5~\rm pc$ from the central SMBH.}}
\label{fig:xdr-hyd-temps}
\end{figure}

\begin{figure}[!htp]

\hspace*{\fill}\includegraphics[angle=90,width=0.5\textwidth]{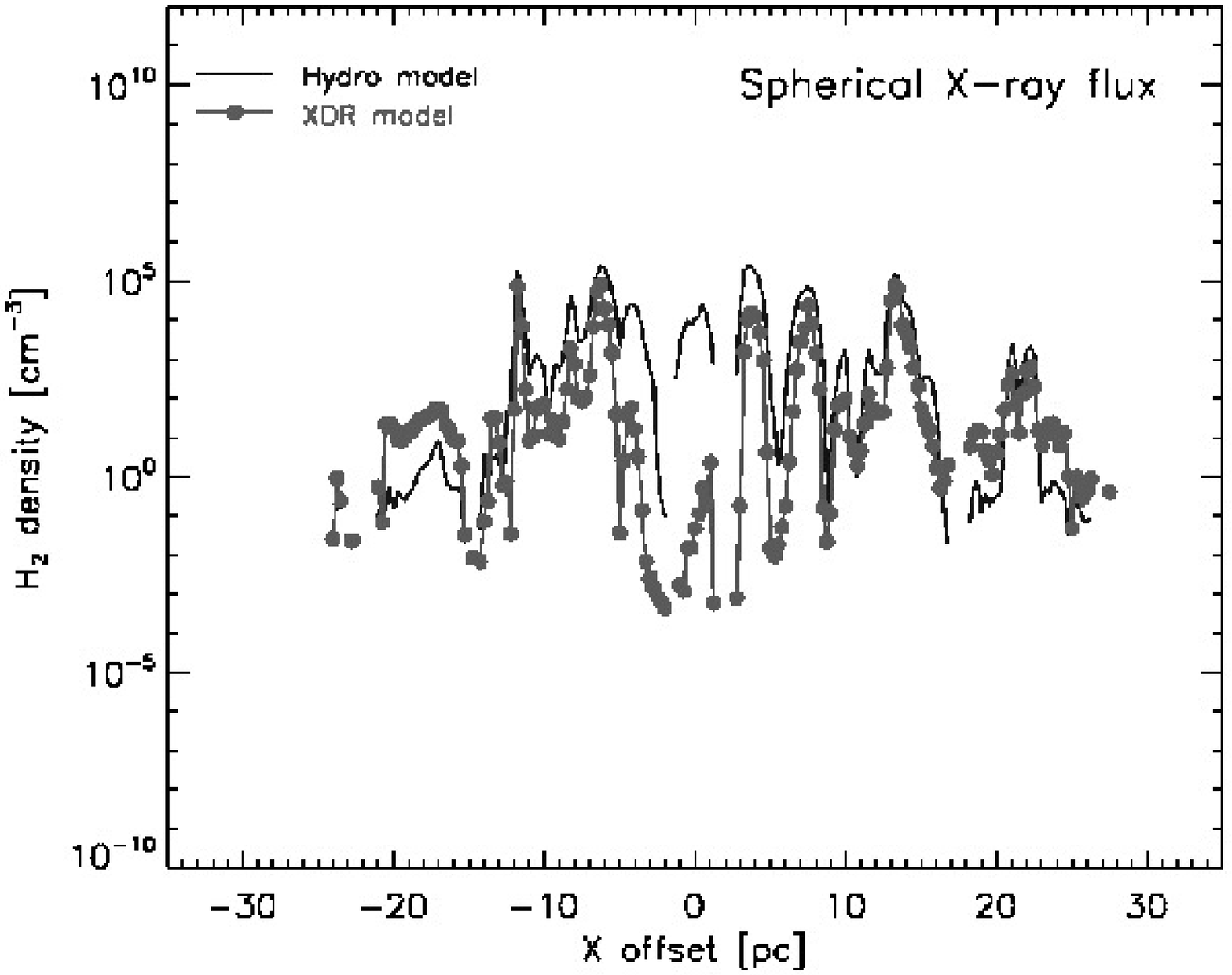}%
\hfill\includegraphics[angle=90,width=0.5\textwidth]{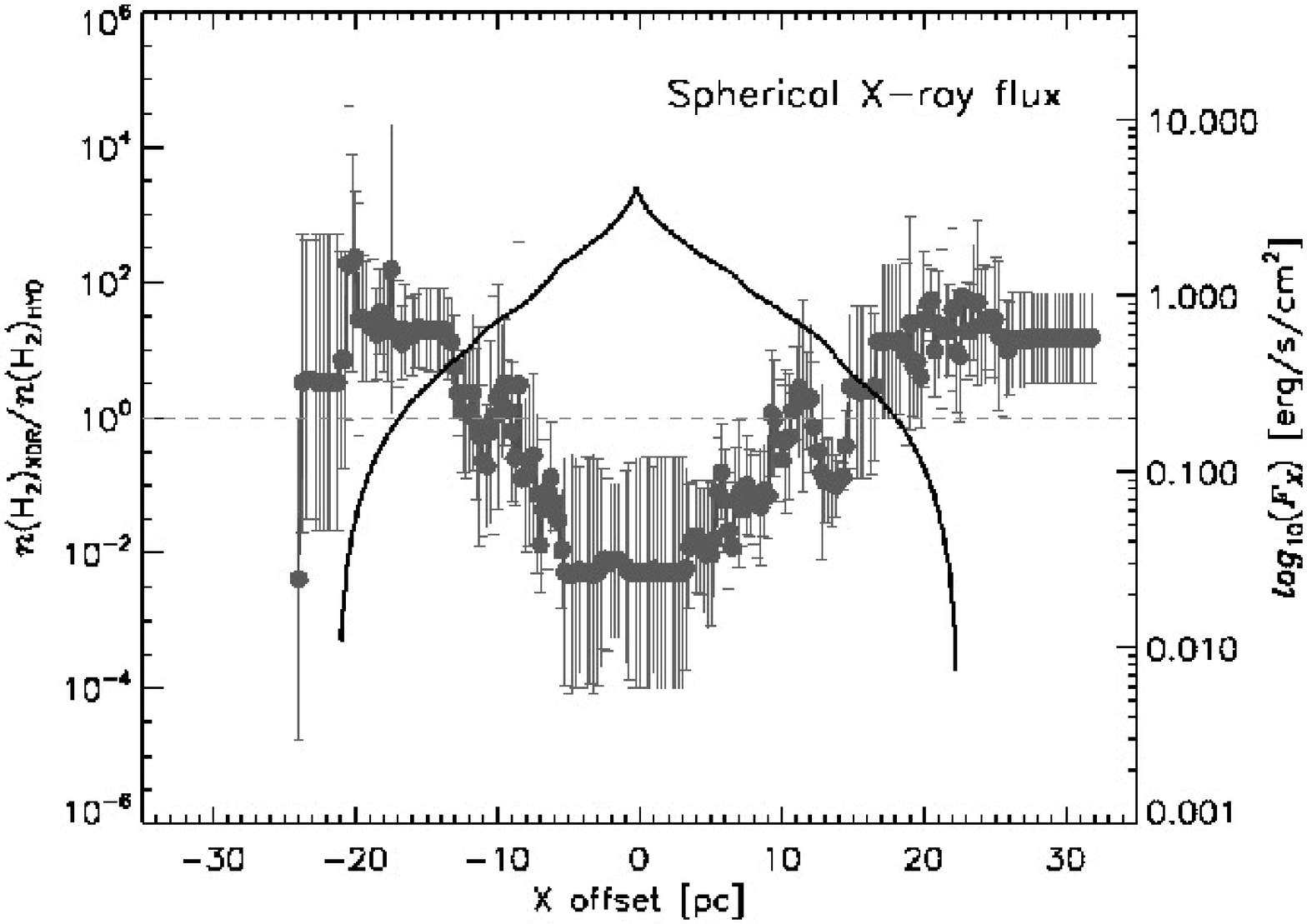}\hspace*{\fill}\\

\vspace{-0.5cm}

\hspace*{\fill}\includegraphics[angle=90,width=0.5\textwidth]{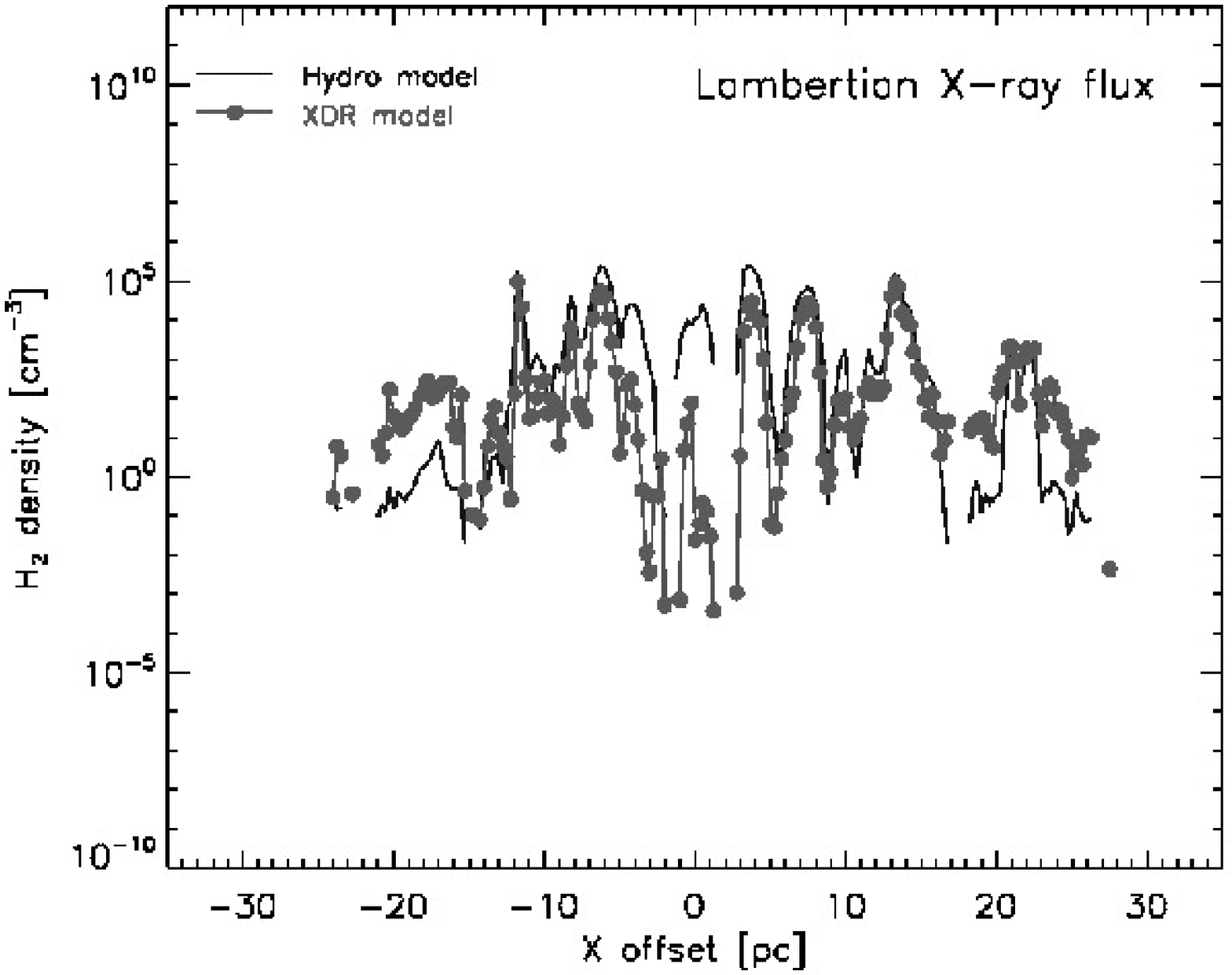}%
\hfill\includegraphics[angle=90,width=0.5\textwidth]{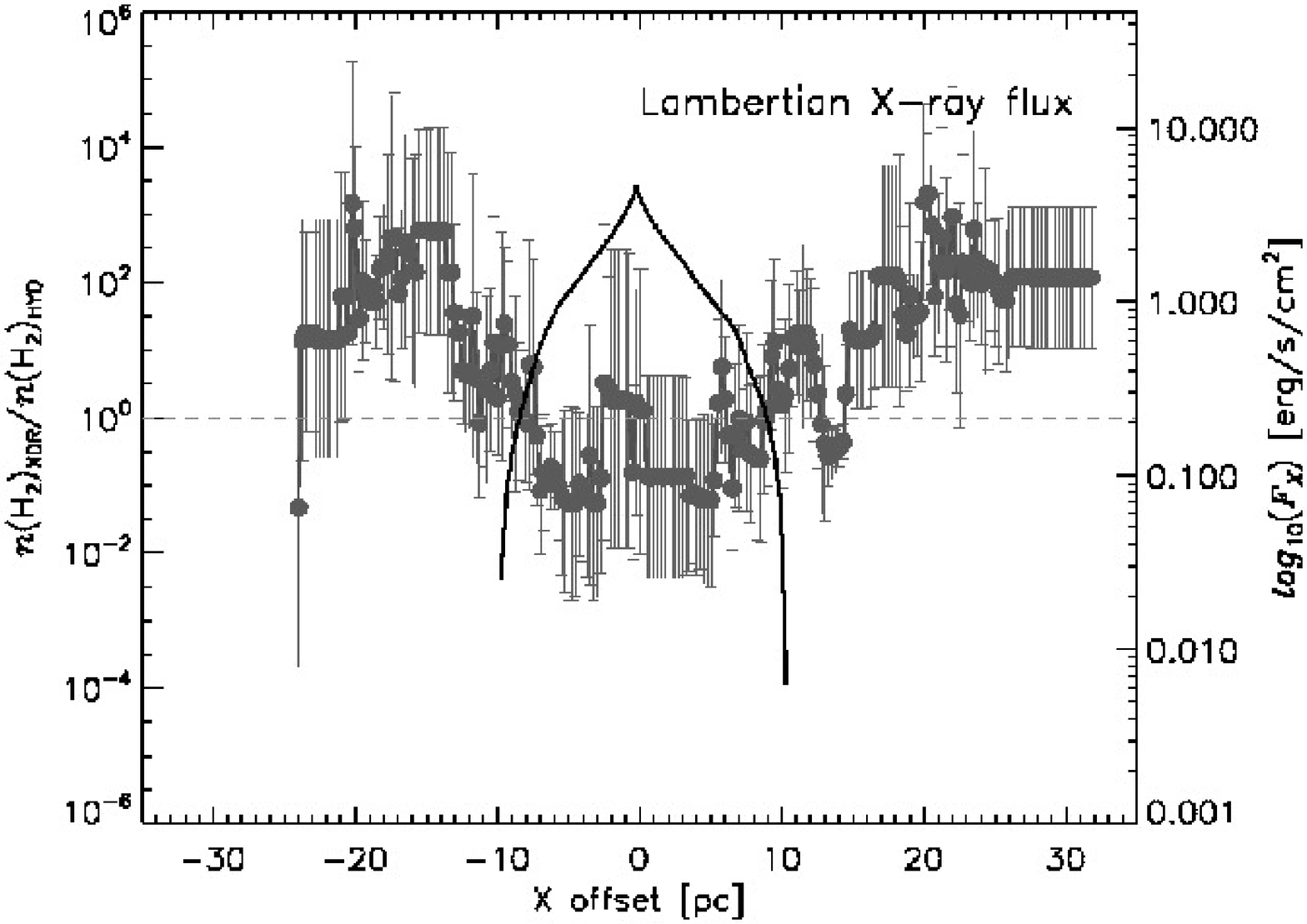}\hspace*{\fill}\\

\vspace{-0.3cm}

\caption{{\footnotesize \textit{Top left} - Average density of molecular hydrogen $n(\rm H_2)_{\rm HYD}$ (in units of $\3cm$) of the gas along a $64\times1.25\times1.5~\rm pc^3$ strip volume, as estimated in the 3-D hydrodynamical model (\textit{solid line}) and the corresponding average $n(\rm H_2)_{\rm XDR}$ density obtained from the XDR chemical model (\textit{gray line + filled circles}) using the \textit{spherical} X-ray flux (eq.\ref{eq:X-ray-flux}). The filled circles show the actual data points obtained with the XDR model in grid cells with $T_{\rm HYD}<10^4~\rm K$. \textit{Top right} - Average $n(\rm H_2)_{\rm XDR}$/$n(\rm H_2)_{\rm HYD}$ ratio (\textit{gray line + filled circles}) and the average \textit{spherical} X-ray flux (\textit{solid black line}) $F_X$ ($\ergscm$) in $log_{10}$ scale.  The standard deviation of the relative density at each $\Delta \rm X$ offset is shown by the error bars. The \textit{dashed line} shows where $n(\rm H_2)_{\rm XDR}$/$n(\rm H_2)_{\rm HYD}=1$. 
The \textit{Bottom panels} show the same as above, but for the \textit{Lambertian} X-ray flux. The turnover in the relation of the hydrogen density is still at about $\Delta \rm X=10~\rm pc$ from the central SMBH, but $n(\rm H_2)_{\rm XDR}$ can be up to $10^3$ times higher than for the \textit{spherical} X-ray flux beyond $\pm10~\rm pc$.}}
\label{fig:xdr-hyd-h2}
\end{figure}

The \textit{top panel} of Fig.~\ref{fig:xdr-hyd-temps} shows the average temperature $T_{\rm HYD}$ ($\rm K$) estimated in the 3-D hydrodynamical model (\textit{solid line}) at each $\Delta \rm X$ grid element. The corresponding average (of the H$_2$ abundance-weighted average) temperature $T_{\rm XDR}$ ($\rm K$) obtained from the XDR chemical model is shown with a \textit{gray line}. Only the grid elements with $T_{\rm HYD}<10^4~\rm K$ were used in the XDR model, and they are shown with filled circles. The average $T_{\rm XDR}/T_{\rm HYD}$ ratio (\textit{gray line + filled circles}) and the average X-ray flux (\textit{solid black line}) $F_X$ ($\ergscm$) are shown in the \textit{bottom panel} of Fig.~\ref{fig:xdr-hyd-temps}. The average relative temperature is directly related to the impinging flux at each $\Delta \rm X$ grid element, and decreases as the X-ray flux decreases. The temperature derived from the XDR model is higher ($T_{\rm XDR}/T_{\rm HYD}>1$) than the one estimated in the hydrodynamical model. Hence, the presence of X-rays has an undeniable effect on the thermodynamics of the AGN torus, up to at least $60~\rm pc$.

%Both temperatures tend to be equal ($T_{\rm XDR}/T_{\rm HYD}\approx1$, \textit{dashed line}) at radial positions where the X-ray flux decreases to $F_X\sim1.1~\ergscm$. This corresponds to the inner $\pm20~\rm pc$ (X-axis offset) of the torus, which is consistent with an effective number (attenuation) density $n_{eff}$ larger than $10^5~\3cm$, for an equivalent radiation field of $G_0\sim0.7\times10^3$ in Habing units \citep[their Fig.2]{schleicher10}.

In Fig.~\ref{fig:xdr-hyd-h2} the average density of molecular hydrogen $n(\rm H_2)_{\rm HYD}$ along a $64\times1.25\times1.5~\rm pc^3$ strip volume is shown. The density computed for the 3-D hydrodynamical model is shown by the \textit{solid line}, and the corresponding average $n(\rm H_2)_{\rm XDR}$ density obtained from the XDR chemical model, using the \textit{spherical} (\textit{top panels}) and the \textit{Lambertian} (\textit{bottom panels}) X-ray fluxes, is shown with the \textit{gray line + filled circles}. Only the data points for grid cells with $T_{\rm HYD}<10^4~\rm K$ and $n(\rm H_2)_{\rm HYD}>10^{-2}~\3cm$ are shown in the figure. 

The relative H$_2$ density seems to be inversely related to the impinging flux.
That is, the average $n(\rm H_2)_{\rm XDR}$ density derived from the XDR model is lower (by factors up to $\sim10^{4}$) than the average H$_2$ density of the hydrodynamical model. This is observed in the inner $\pm10~\rm pc$ region around the center of the AGN torus. This is mostly the consequence of relatively \textit{thin} slabs ($N_{\rm H}<10^{23}~\2cm$) being irradiated by a rather strong ($F_X>1.6~\ergscm$) X-ray flux in the proximity of the torus center, as described in Sec.~\ref{sec:abundances} and shown in Fig.\ref{fig:XDR-model}.
However, beyond $\Delta \rm X\sim10~\rm pc$ the XDR H$_2$ abundance exceeds the one in the pure hydrodynamical model. This change is explained by the background star formation considered in the hydrodynamical model at those distances. The FUV is a more efficient destroyer of the H$_2$ gas than X-rays (Meijerink \& Spaans 2005).

With a \textit{spherical} X-ray flux $n(\rm H_2)_{\rm XDR}$ can be up to $10^4$ times higher than $n(\rm H_2)_{\rm HYD}$, while it can be about $10^5$ higher if a \textit{Lambertian} X-ray flux is considered instead. In the inner region ($\Delta \rm X\sim-2~\rm pc$.) a density $n(\rm H_2)_{\rm XDR}$ a few times higher than $n(\rm H_2)_{\rm HYD}$ is observed. This is the consequence of a weaker X-ray flux, with respect to the \textit{spherical} radiation flux.
This is consistent with the fluctuations of the column density distribution of H$_2$ explored by \citetalias{wada09} for different viewing angles and, as expected, the largest $N(\rm H_2)$ columns are found at a viewing angle $\sim0$ degrees (i.e., edge-on). These facts imply that molecules will tend to disappear in the central ($\lesssim10~\rm pc$) region. But, depending on the viewing angle, and for total hydrogen columns $\gtrsim10^{24}~\2cm$ like in the case of, for instance, the LIRG NGC~4945, molecules can survive and emission lines like, e.g., high-$J$ CO, \cii, \neii\ and \nev, can be bright.
In all, we conclude that the H$_2$ abundance in the AGN torus is strongly affected by the black hole ($\le10~\rm pc$) and star formation ($\ge10~\rm pc$) \citep[see also][]{schleicher10}.

We note, though, that the nature of the models used here to derive temperature and density of the gas (a static X-ray driven chemical model and an hydrodynamical X-ray free model) are different, and a comparison between their corresponding derived temperatures and densities is merely intended to motivate the need for a joint XDR-hydrodynamical model for the thermodynamics of AGN tori. A first attempt to this effect has been made by \citet{hocuk10} for individual $\sim$1 $\rm pc$ molecular clouds close to a SMBH.

\section{Final remarks}\label{sec:remarks}

We compared the total hydrogen column density, $N(\rm CO)$ and CO $J=1\rightarrow0$ to $J=9\rightarrow8$ line intensities, and found that the mid-$J$ CO lines are excellent probes of density and dynamics, but the low-$J$ CO lines are not good tracers of $n_{\rm H}$ in the central ($\lesssim60~\rm pc$) region of the AGN torus. The analysis of the $X_{\rm CO}/\alpha$ conversion factors indicated that only the higher-$J$ CO lines will show a linear correlation with the gas mass in AGN tori at lower spatial resolutions ($\sim9~\rm pc$). But at higher resolution ($<5~\rm pc$) different proportionality factors (or no correlation at all) pertained between the CO lines and the total gas mass in AGN tori. We also determined that the \cii\ $158~\mum$ emission will trace mostly the central region of AGN tori, detectable {(but not resolved)} by ALMA in $z\gtrsim1$ galaxy nuclei.

We found that the presence of X-rays has an undeniable effect on the thermodynamics of the AGN torus, up to at least $60~\rm pc$.
An important implication of this is that circum-nuclear star formation could be suppressed in the central $\sim$5 $\rm pc$. This can shed light on the starburst-AGN connection. Self-consistent UV/X-ray radiation-chemical-hydrodynamical simulations \citep[e.g.][]{hocuk10} will allow us to explore this theoretically, and their predictions can be confirmed (and used for data interpretation) by ALMA in the near future.

With a rest frequency of 691.5 GHz the CO $J=6\rightarrow5$ line can be observed with the ALMA band 9 receivers, which will be the highest frequency band available when the early science begins with $\ge16$ antennas. Considering a minimum baseline of $250~\rm m$ for the compact configuration, we would have an angular resolution\footnote{http://science.nrao.edu/alma/earlyscience.shtml} of FWHM$\approx$0.35$''$ that will allow us to resolve structures of $\sim$7 $\rm pc$ at a nearby distance of $4~\rm Mpc$ (about the distance to NGC~4945) and of $\sim$25 $\rm pc$ at a distance of $15~\rm Mpc$ (roughly the distance to NGC~1068).
In the near future, however, the higher sensitivity (with $\ge$50 antennas) and the availability of longer baselines of up to $\sim$15 $\rm km$ will provide angular resolutions of FWHM$\approx$0.006$''$ with ALMA band 9, which will allow the study of structures between $\sim$0.1 $\rm pc$ and $\sim$0.4 $\rm pc$, respectively, at the distances mentioned above. 
Therefore, the spatial scales ($\ge$0.25 $\rm pc$) that we probe with our simulations match the angular resolutions provided by ALMA.

%Using the flux maps of CO will allow us to compute the actual H$_2$ mass to CO $J=1\rightarrow0$ luminosity ratio of our models, and to compare with the ratio given by the higher-$J$ CO lines. We also would like to explore the morphology and distribution of the warm ionized medium traced by the \cii~158~\mum\ fine-structure line, estimated assuming a two-level atom system \citep[e.g.,][]{hollenbach91}. Its correlation with the CO $J=1\rightarrow0$ can also be addressed \citep[e.g.,][]{crawford85, stacey91}.

%% Included in this acknowledgments section are examples of the
%% AASTeX hypertext markup commands. Use \url without the optional [HREF]
%% argument when you want to print the url directly in the text. Otherwise,
%% use either \url or \anchor, with the HREF as the first argument and the
%% text to be printed in the second.

\acknowledgments

We thank the referee for his/her constructive and insightful remarks that helped to improve this work.
We are grateful to Aycin Aykutalp and Seyit Hocuk for their help and advise in using the Gemini supercomputers at the Kapteyn Institute. We are also thankful to Dieter Poelman for initial discussions and help with the original $\beta3D$ radiative transfer code. Dr. J.P. P\'erez-Beaupuits (MPIfR) is sponsored by the Alexander von Humboldt Foundation. The hydrodynamical model was computed on NEC SX-9 at Center for Computational Astrophysics, CfCA, of National Astronomical Observatory of Japan. The radiative transfer and line tracing were computed on Cray SV1e at the Centre for High Performance Computing and Visualisation, HPC/V, of the University of Groningen, The Netherlands. Molecular Databases that have been helpful include the NASA/JPL, LAMDA and NIST.


%% in parentheses, after the keyword, but they will not be verified.

%{\it Facilities:} \facility{Nickel}, \facility{HST (STIS)}, \facility{CXO (ASIS)}.



\appendix

\section{Rotational excitation of CO by He}\label{sec:appendix}

We used the rate coefficients for pure rotational de-excitation of CO by collisions with He atoms reported in \citet{cecchi02}. The original rate coefficients are given for the first 15 rotational levels and for ten different temperatures from 5 to 500 K.
In order to extend the available rates to higher rotational levels and temperatures, we followed the methodology for linear molecules described by \citet{schoier05}, which was used to produce the LAMDA database.

We first extrapolated the downward collisional rate coefficients ($\Delta J = J_u \rightarrow J_l, J_u>J_l$ ) in temperature (up to 2000 K) using the modified version of the analytic approximation given by \citet{dejong75} and presented by \citet{bieging98}:

\begin{equation}
\gamma_{ul}=A(\Delta J)~y~exp\left[ -B(\Delta J)~y^{1/4}\right] \times exp\left[ -C(\Delta J)~y^{1/2}\right],
\end{equation}

\noindent
where $y = \Delta E_{ul}/kT$ and the three parameters $A$, $B$, and $C$ are determined by least-squares fits to the original set of \citet{cecchi02} rate coefficients for each $\Delta J$. Then we extrapolated the collisional rate coefficients to include higher rotational levels (up to $J=40$) by fitting the rate coefficients (in natural logarithmic scale) connecting the ground rotational state to a second order polynomial

\begin{equation}
ln(\gamma_{J0})=a+bJ+cJ^2,
\end{equation}

\noindent
with $a$, $b$, and $c$ parameters determined from the fit, for each temperature. The Infinite Order Sudden (IOS) approximation \citep[e.g.,][]{goldflam77} was used to calculate the whole matrix of state-to-state rate coefficients from the coefficients connecting the ground state $\gamma_{L0}$

\begin{equation}\label{eq:IOS}
 \gamma_{J J'}=(2J'+1)\sum_{L=|J-J'|}^{L=J+J'}(2L+1)\left(\begin{array}{ccc} J & J' & L\\0 & 0 & 0 \end{array} \right)^2 \gamma_{L0},
%~~L\le J_{max}
\end{equation}

\noindent
\begin{equation}
{\rm where~the~term}~~~\left(\begin{array}{ccc} J & J' & L\\0 & 0 & 0 \end{array} \right),
\end{equation}
\noindent
is the Wigner 3-$j$ symbol that designates the Clebsch-Gordan coefficients \citep[e.g.,][]{tuzun98}, and references therein). The IOS approximation provides an accurate description of the collisional rates if the rotational energy differences are small compared to the kinetic energy of the colliding molecules. In cases where this condition is not satisfied, it is possible to approximately correct for the deviations by multiplying the summation in eq.(\ref{eq:IOS}) with the adiabaticity correction factor given by \citet{depristo79} and \citet{mckee82}

\begin{equation}
 A(L,J)=\frac{6+(\alpha L)^2}{6+(\alpha J)^2},~~~{\rm with}~\alpha=0.13~B_0~l~\left( \frac{\mu}{T} \right)^{1/2},
\end{equation}

\noindent
where $B_0$ is the rotational constant of the colliding molecule in cm$^{-3}$ ($B_0=1.9225$ cm$^{-3}$ for CO), $l=3$ \AA~is a typical scattering length, $\mu$ is the reduced mass of the colliding system in amu ($\mu\approx3.5$ amu for CO-He), and $T$ is the kinetic temperature.

\begin{figure}[tp!]
%\centering

\hspace*{\fill}\includegraphics[angle=0,width=0.33\textwidth]{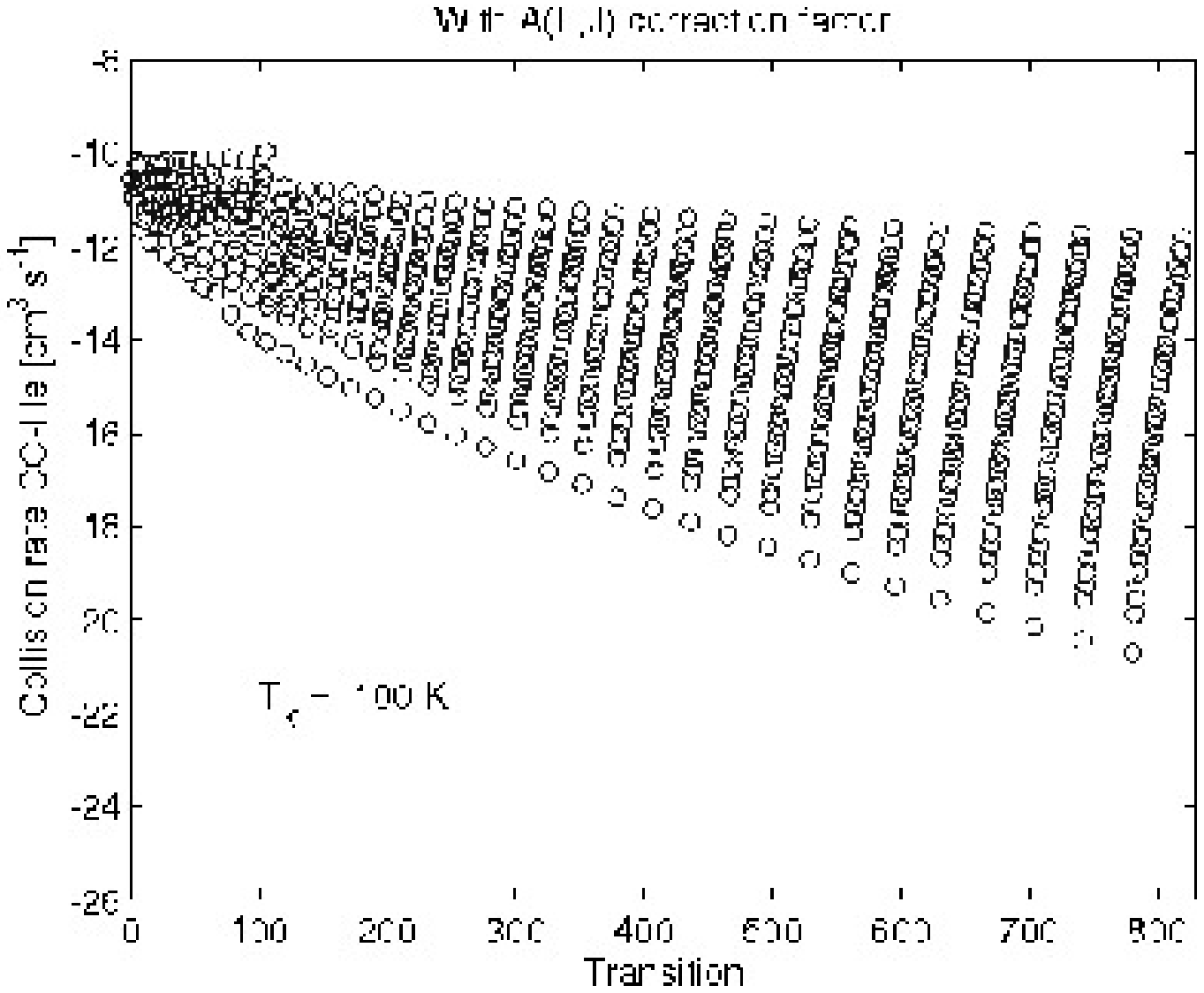}%
\hfill\includegraphics[angle=0,width=0.33\textwidth]{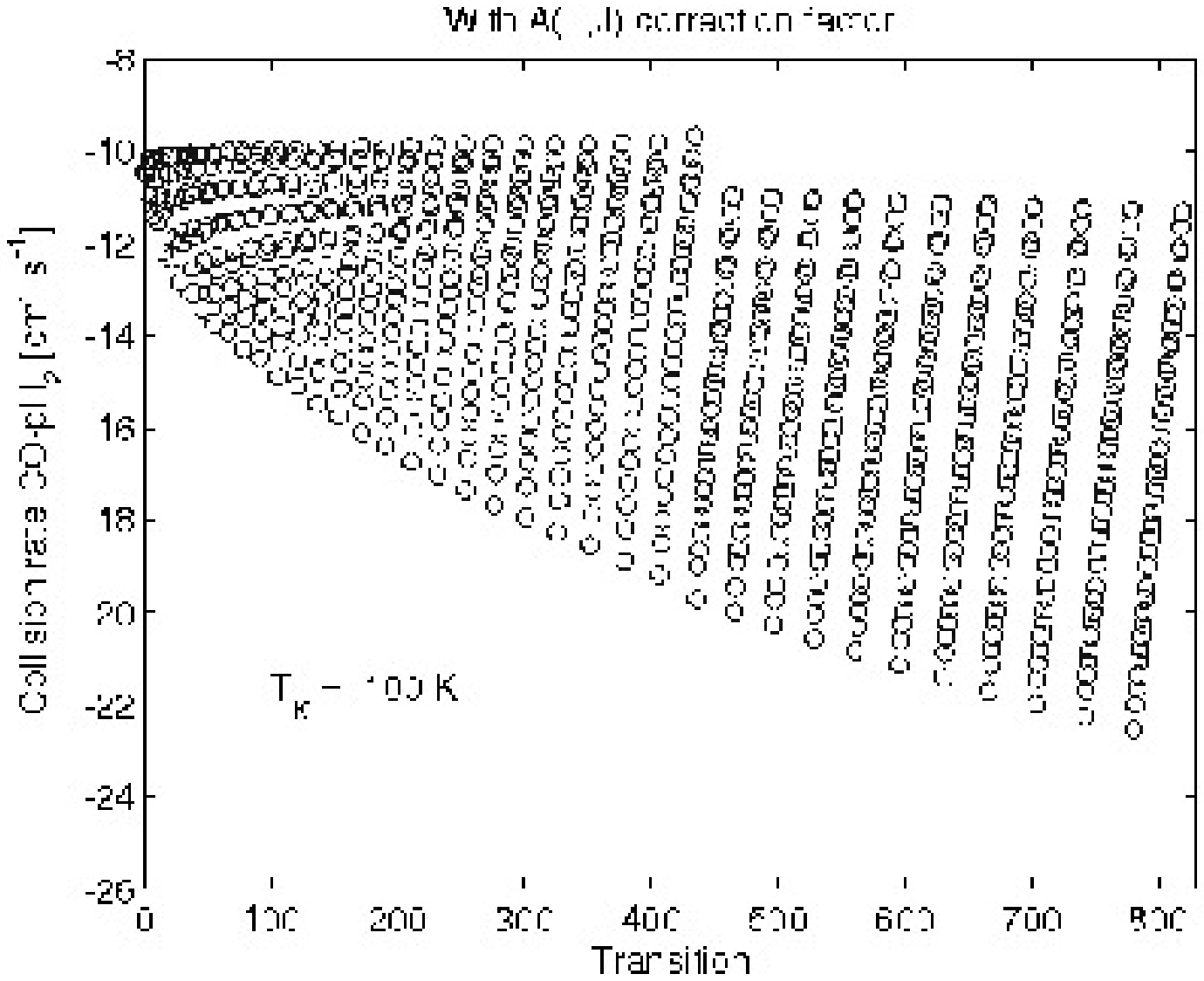}%
\hfill\includegraphics[angle=0,width=0.33\textwidth]{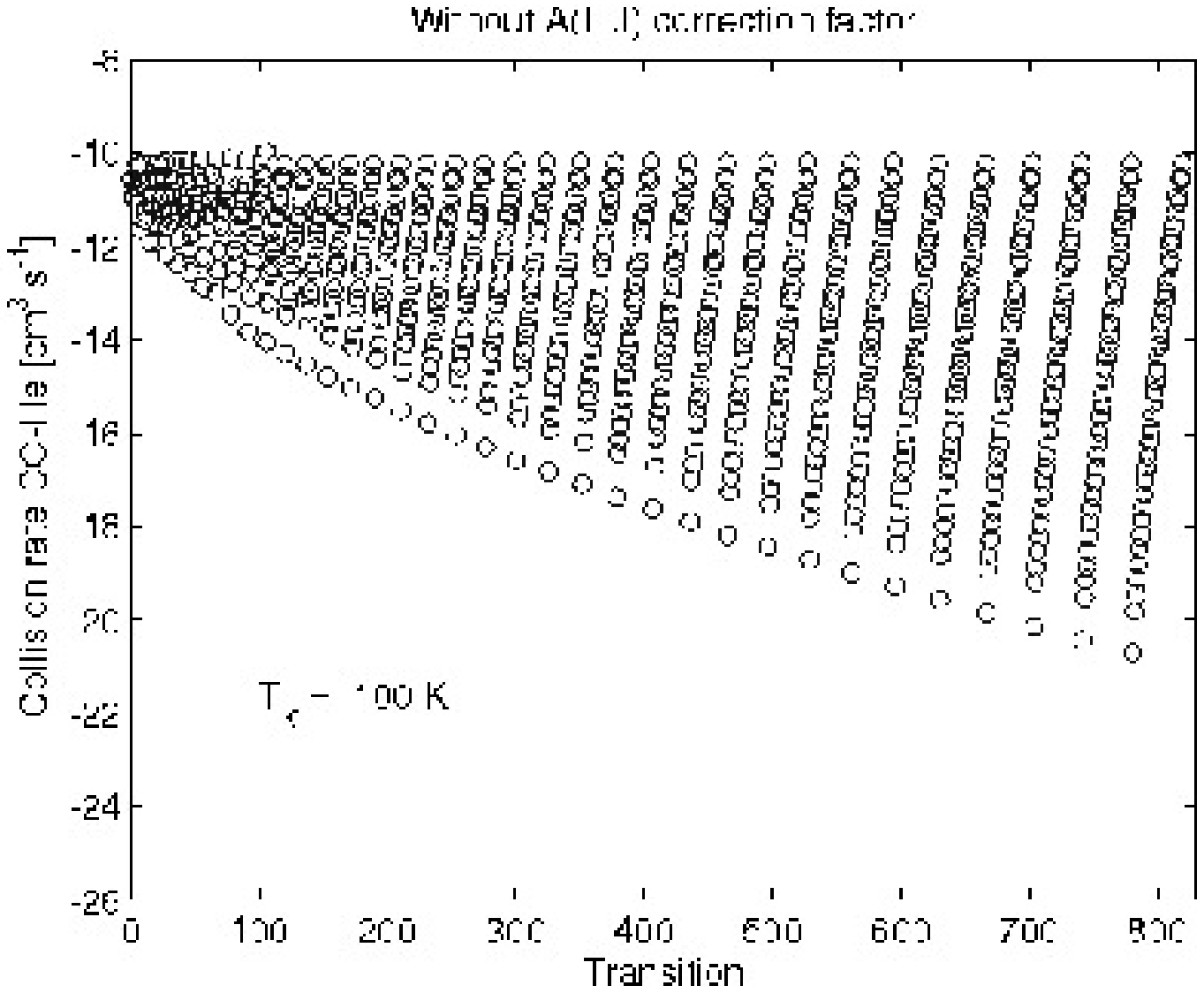}\hspace*{\fill}

%\vspace{-0.5cm}

\caption{{\footnotesize \textit{Top left panel} - Collisional rate coefficients (cm$^3$ s$^{-1}$) for CO with para-H$_2$ as collision partner. \textit{Top right panel} - Collisional rate coefficients for CO colliding with He with the wrong Wigner 3-j function. \textit{Bottom panel} - Collisional rate coefficients for CO-He collision partners with the correct Wigner 3-j function.}}
\label{fig:crates}
\end{figure}

However, the $A(L,J)$ correction factor should be used only if $E_L>E_J$ and $\left( E_L-E_J\right)\gg E_K$, where $E_{J,L}$ is the energy of the CO rotational levels $L,J$ and $E_K$ is the kinetic energy of the collision partners. The top left panel of Fig.\ref{fig:crates} shows the deviations introduced by the $A(L,J)$ factor when used arbitrarily to extrapolate the rate coefficients of CO colliding with He. The top right panel of Fig.\ref{fig:crates} shows similar discontinuities in the extrapolated rate coefficients between CO and para-H$_2$ presented in the current LAMDA molecular data. Similar deviations are observed for ortho-H$_2$. This means that the extrapolated LAMDA molecular data for CO need to be corrected. Although, the original CO-H$_2$ rate coefficients obtained from \citet{flower01} and \citet{wernli06} go up to $J=29$, and since we do not explore CO transitions above $J=20$ we can still use the LAMDA molecular data without corrections.

In the case of the CO-He colliding system the conditions for using the $A(L,J)$ adiabaticity factor are not satisfied for the temperatures and energy levels considered here, and the IOS approximation given by eq.(\ref{eq:IOS})
yields results with 10\%--15\% accuracy \citep{goldflam77}. The final extrapolated rate coefficients used in this work for the system CO-He are shown in the bottom panel of Fig.\ref{fig:crates}.

\section{CO maps at $45^o$ inclination}\label{sec:appendix-45x}

\begin{figure}[tp!]
%\centering

\hspace*{\fill}\includegraphics[angle=0,width=0.33\textwidth]{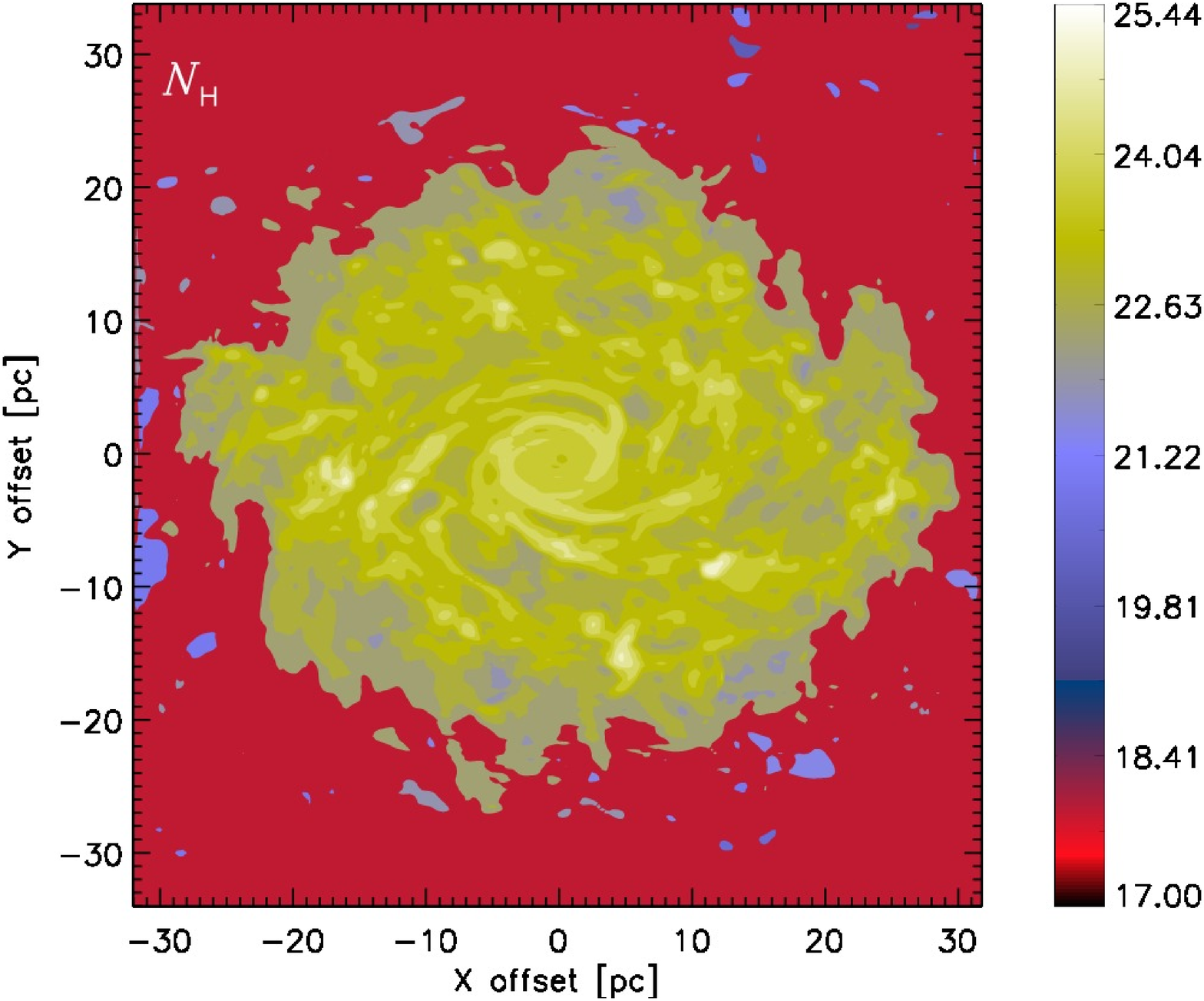}%
\hfill\includegraphics[angle=0,width=0.33\textwidth]{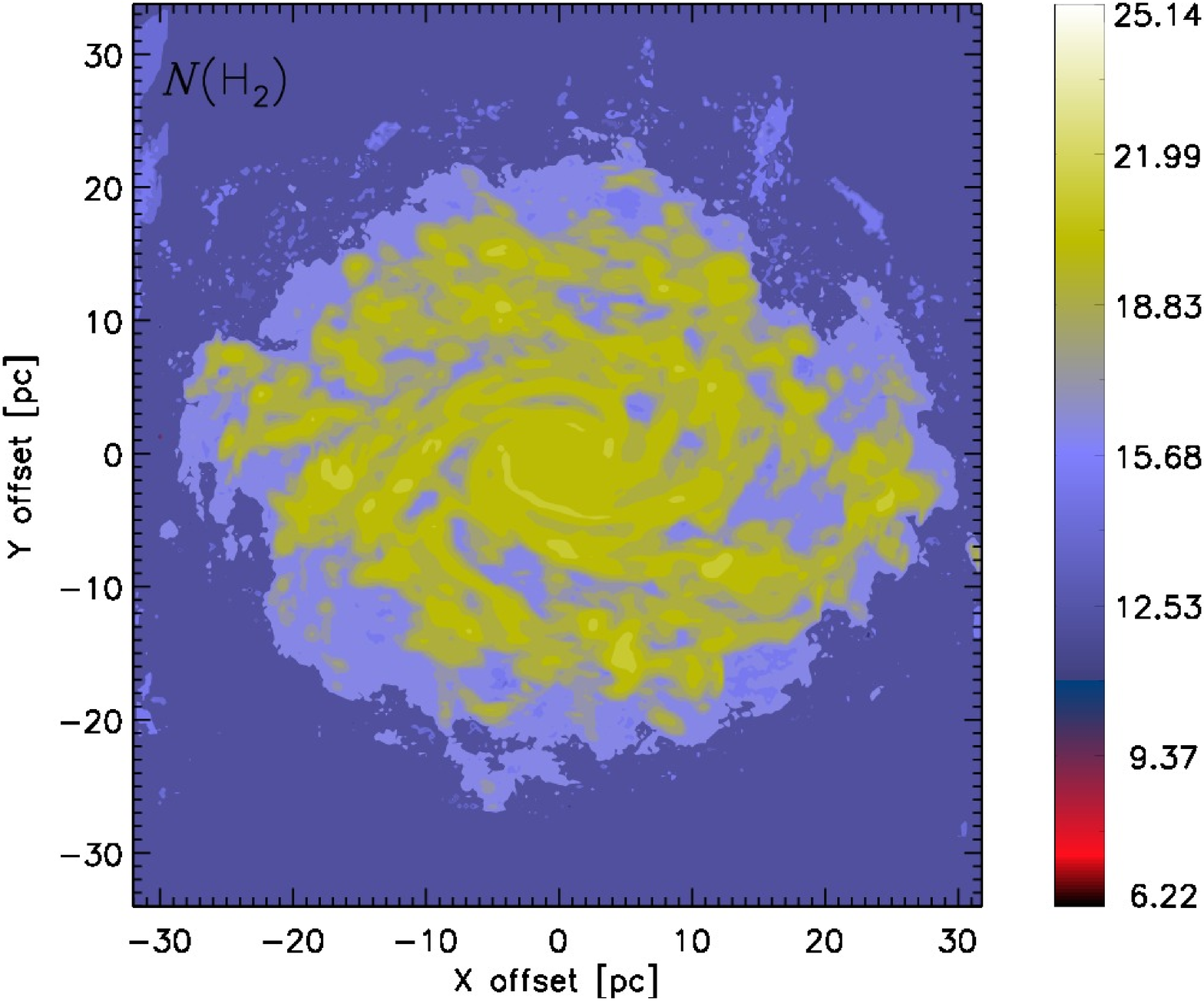}%
\hfill\includegraphics[angle=0,width=0.33\textwidth]{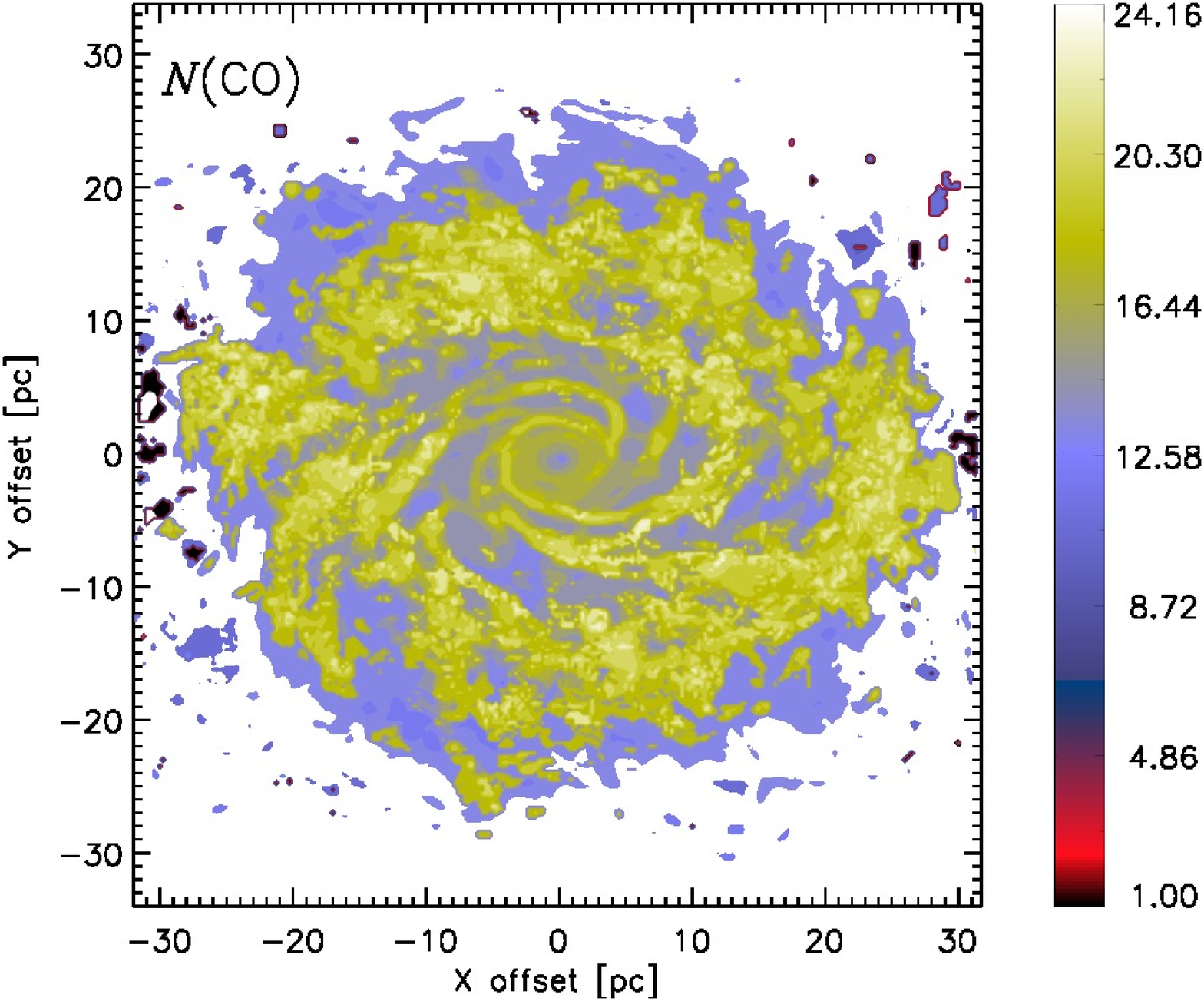}\hspace*{\fill}\\

\vspace{-0.5cm}

\hspace*{\fill}\includegraphics[angle=0,width=0.33\textwidth]{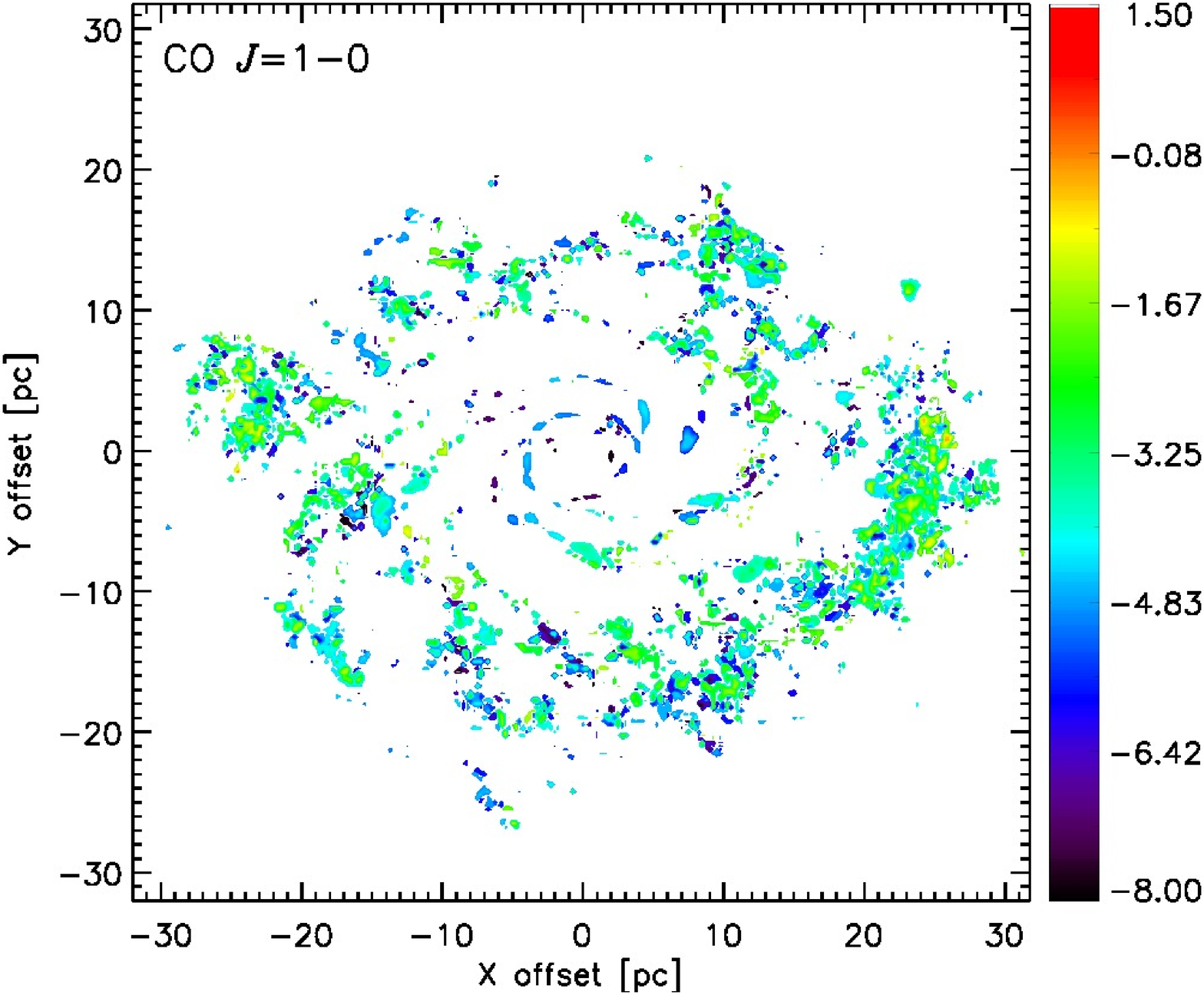}%
\hfill\includegraphics[angle=0,width=0.33\textwidth]{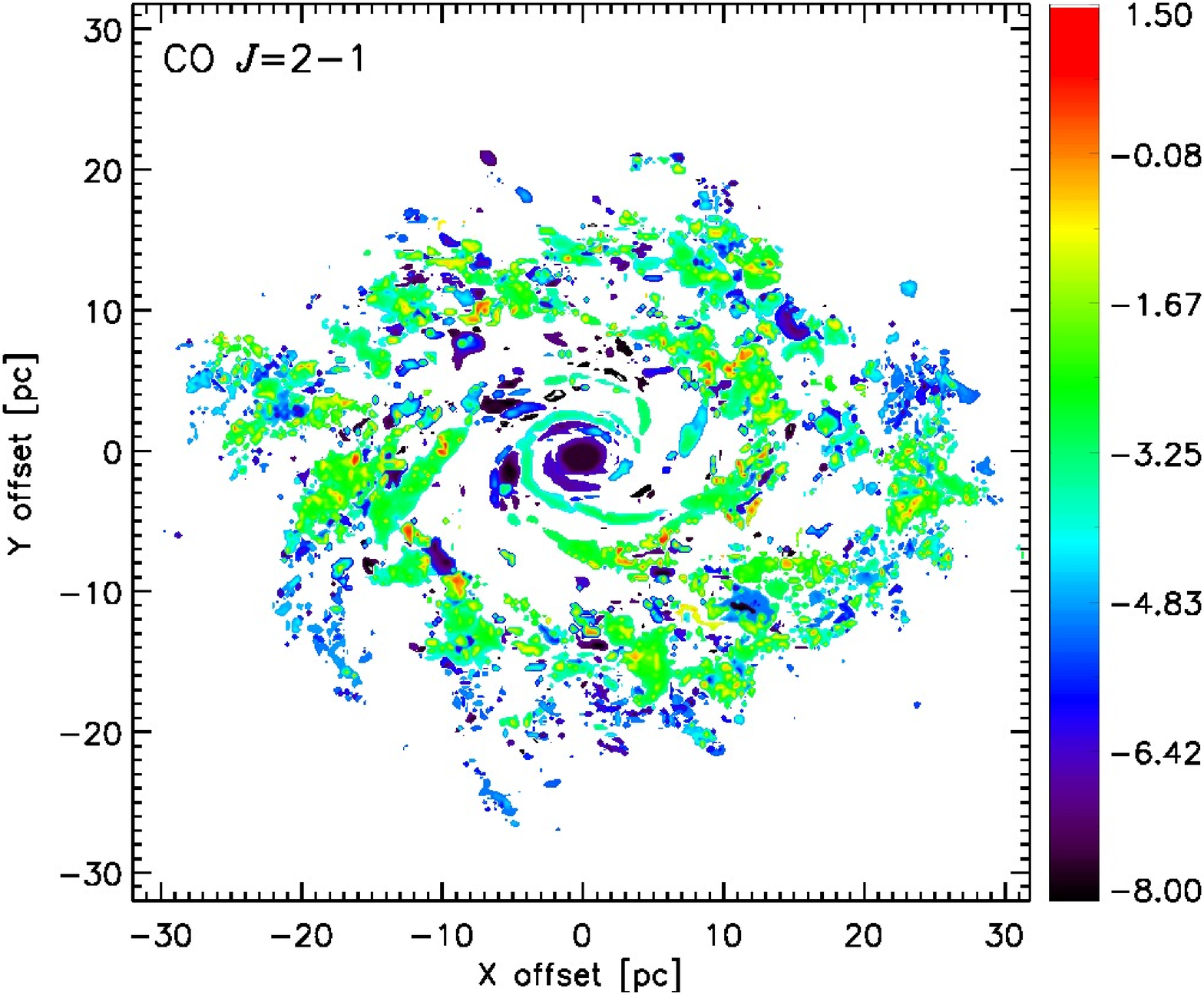}%
\hfill\includegraphics[angle=0,width=0.33\textwidth]{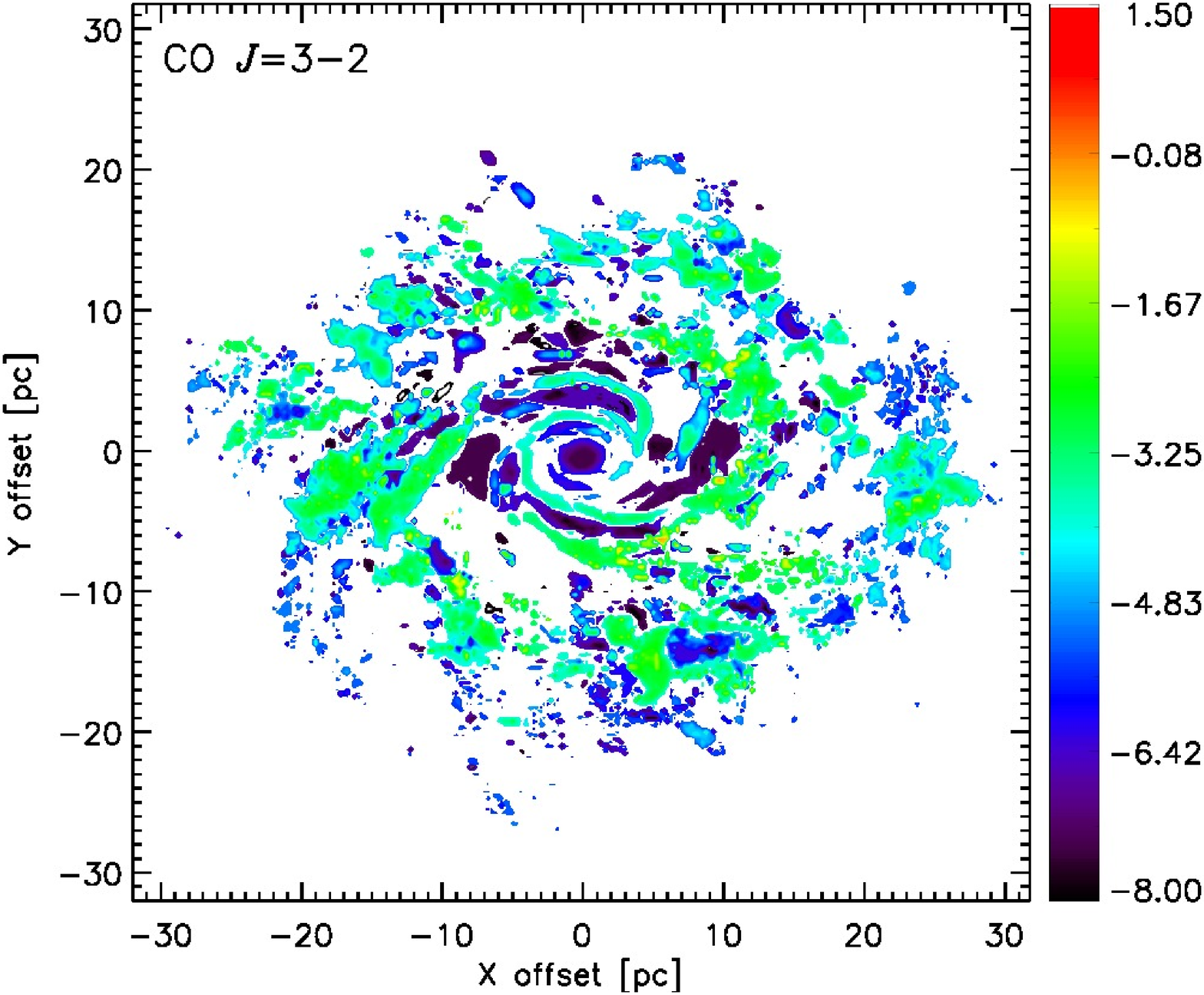}\hspace*{\fill}\\

\vspace{-0.5cm}

\hspace*{\fill}\includegraphics[angle=0,width=0.33\textwidth]{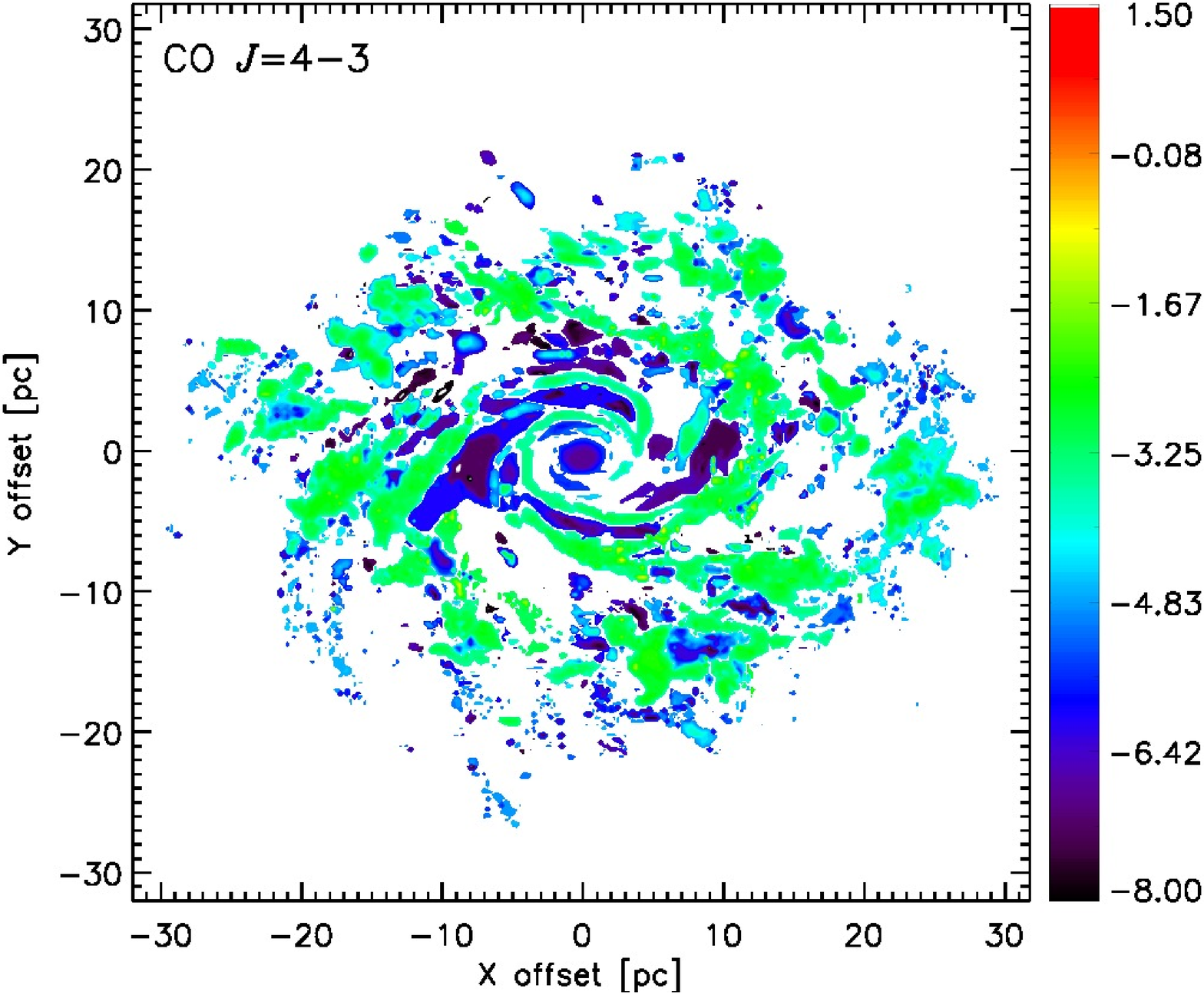}%
\hfill\includegraphics[angle=0,width=0.33\textwidth]{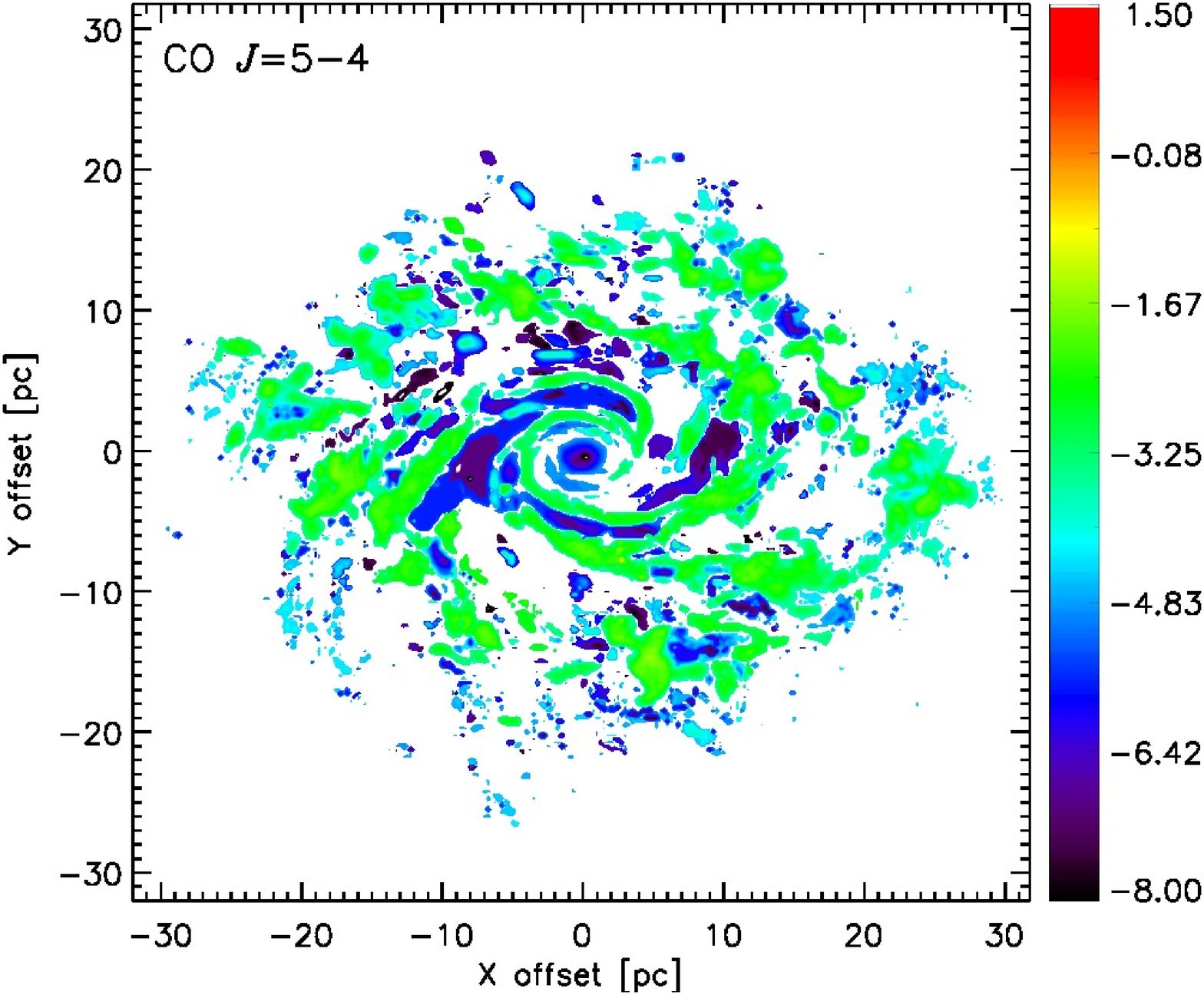}%
\hfill\includegraphics[angle=0,width=0.33\textwidth]{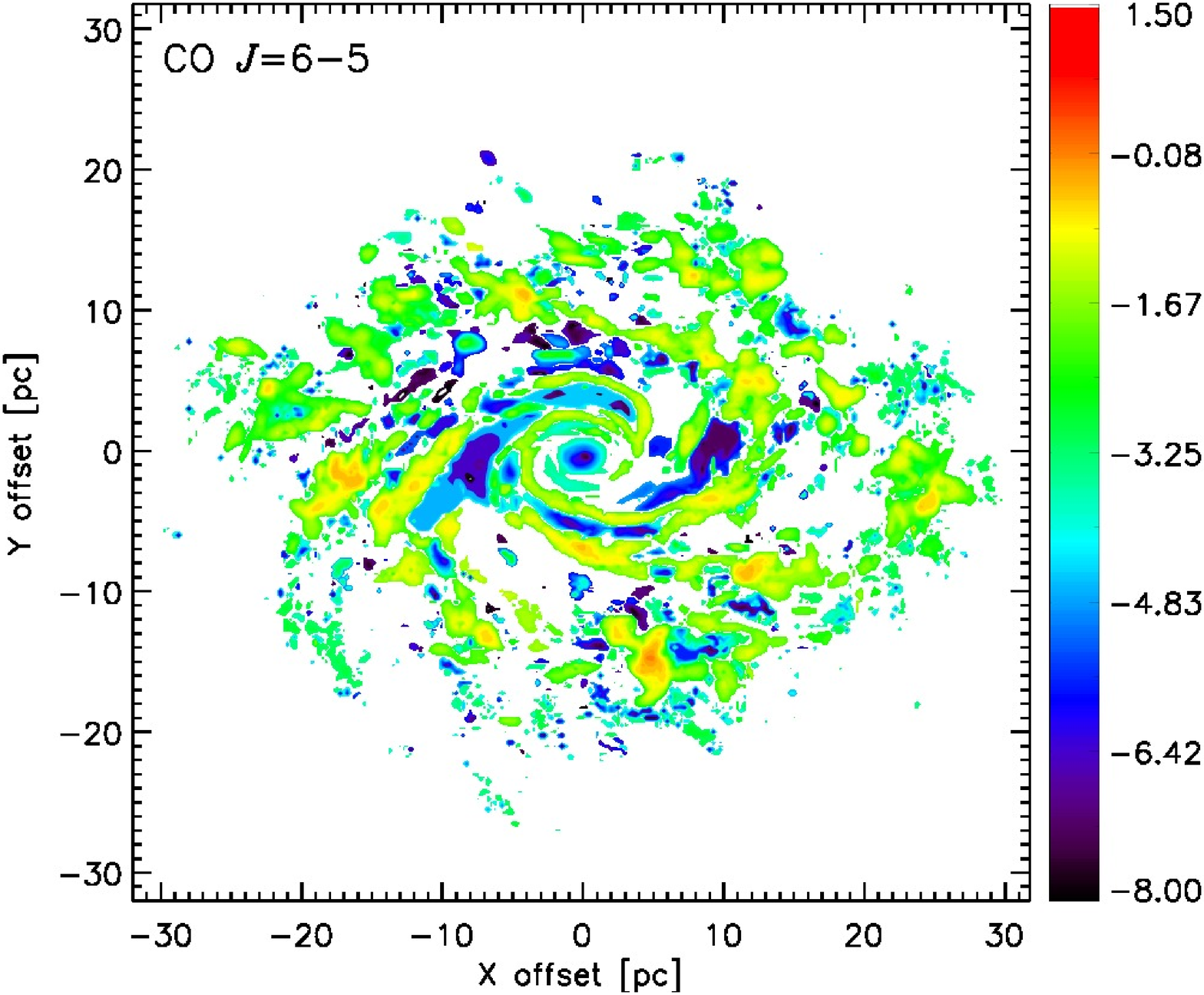}\hspace*{\fill}\\

\vspace{-0.5cm}

\hspace*{\fill}\includegraphics[angle=0,width=0.33\textwidth]{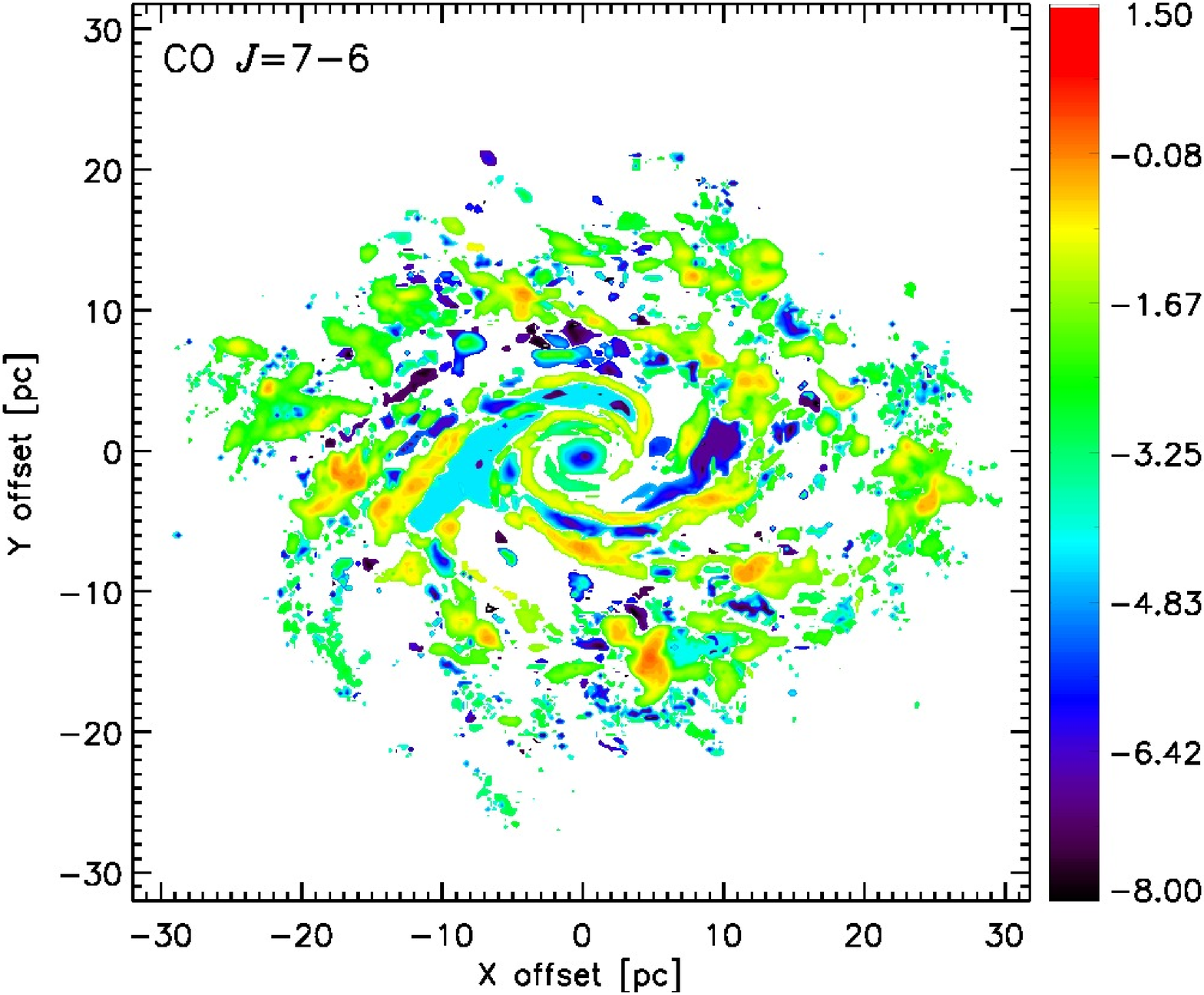}%
\hfill\includegraphics[angle=0,width=0.33\textwidth]{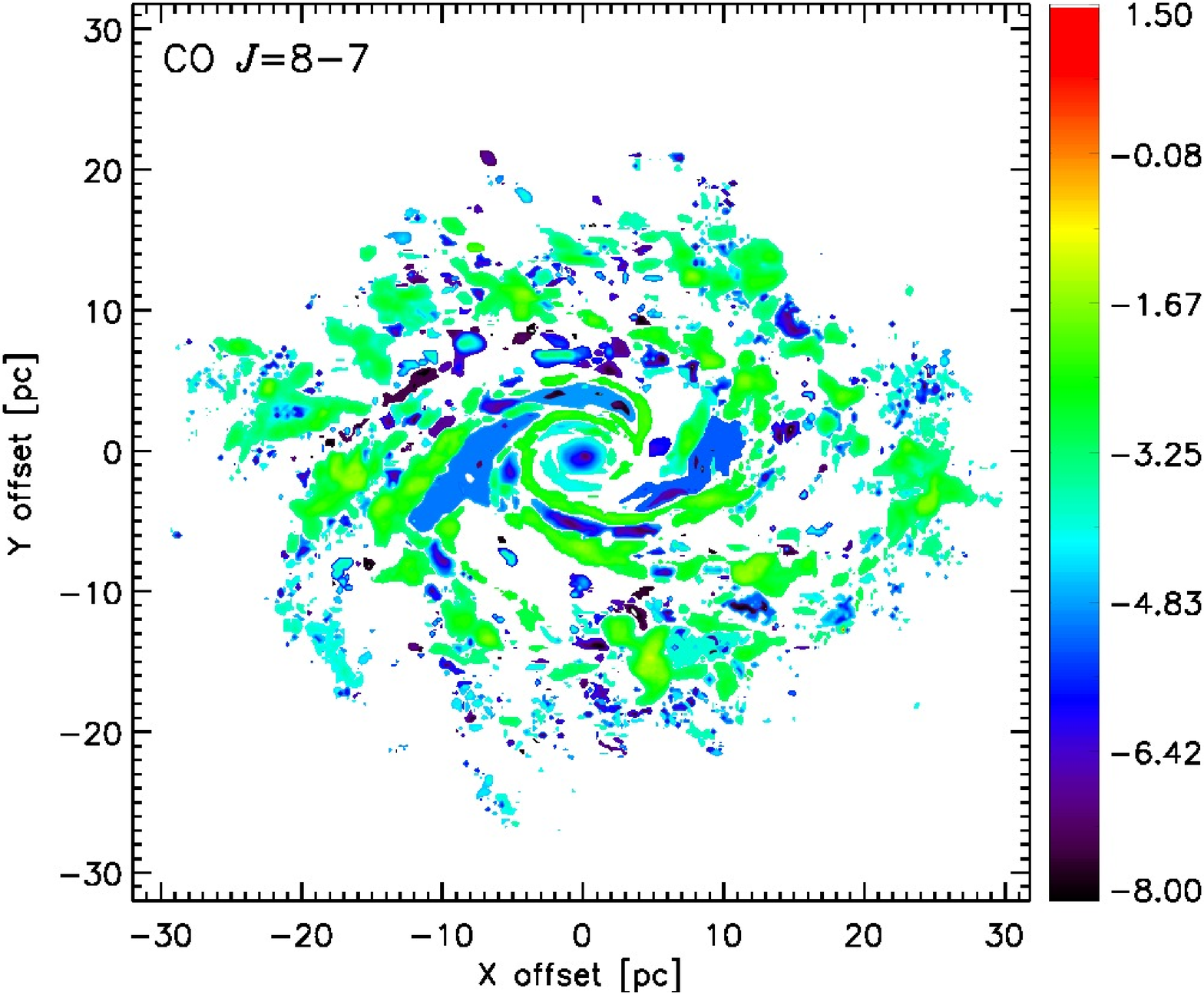}%
\hfill\includegraphics[angle=0,width=0.33\textwidth]{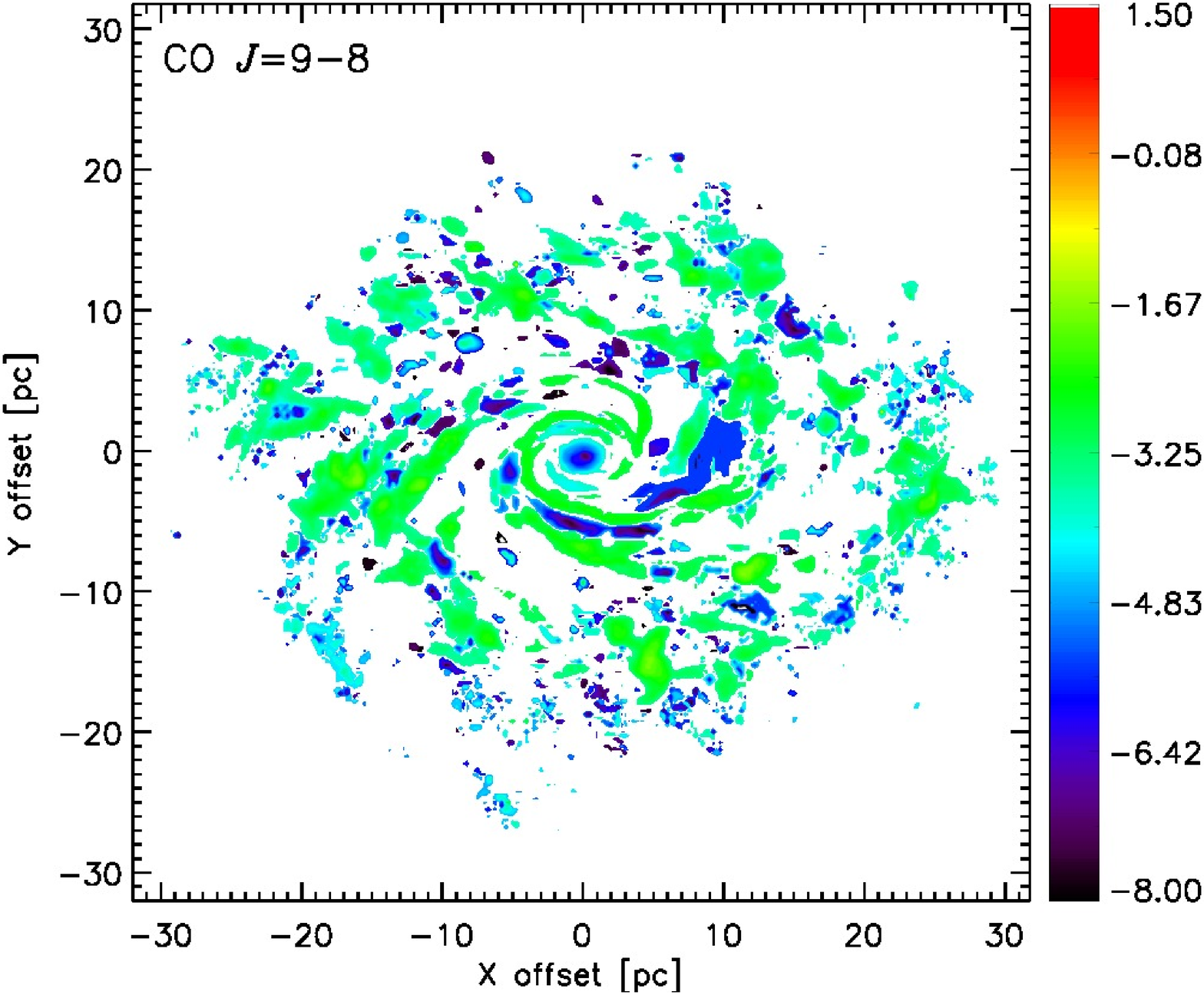}\hspace*{\fill}

\caption{{\footnotesize \textit{Top panels} - Maps with a $45^o$ inclination about the X-axis of the total column density (units of $\2cm$) $N_{\rm H}$ (\textit{left}), the column density of molecular hydrogen $N({\rm H}_2)$ (\textit{middle}) and the CO column $N({\rm CO})$ (\textit{right}) in logarithmic scale. 
\textit{Bottom panels} - Surface brightness maps of the $J=1\rightarrow0$ to $J=9\rightarrow8$ transitions of CO (in $log_{10}$ scale and units of $\ergscmsr$), as observed at the surface of the face-on data cube. The larger columns seen through the line of sight with the $45^o$ inclination produce higher emissions of the CO transitions with respect to the face-on maps of Fig.~\ref{fig:column-density}. However, the higher CO lines (from $J=4\rightarrow3$ up to $J=9\rightarrow8$) are still better tracers of the inner region of the torus.}}
\label{fig:maps-45x}
\end{figure}

\clearpage

\begin{figure}[htp!]
%\centering

\hspace*{\fill}\includegraphics[angle=0,width=0.33\textwidth]{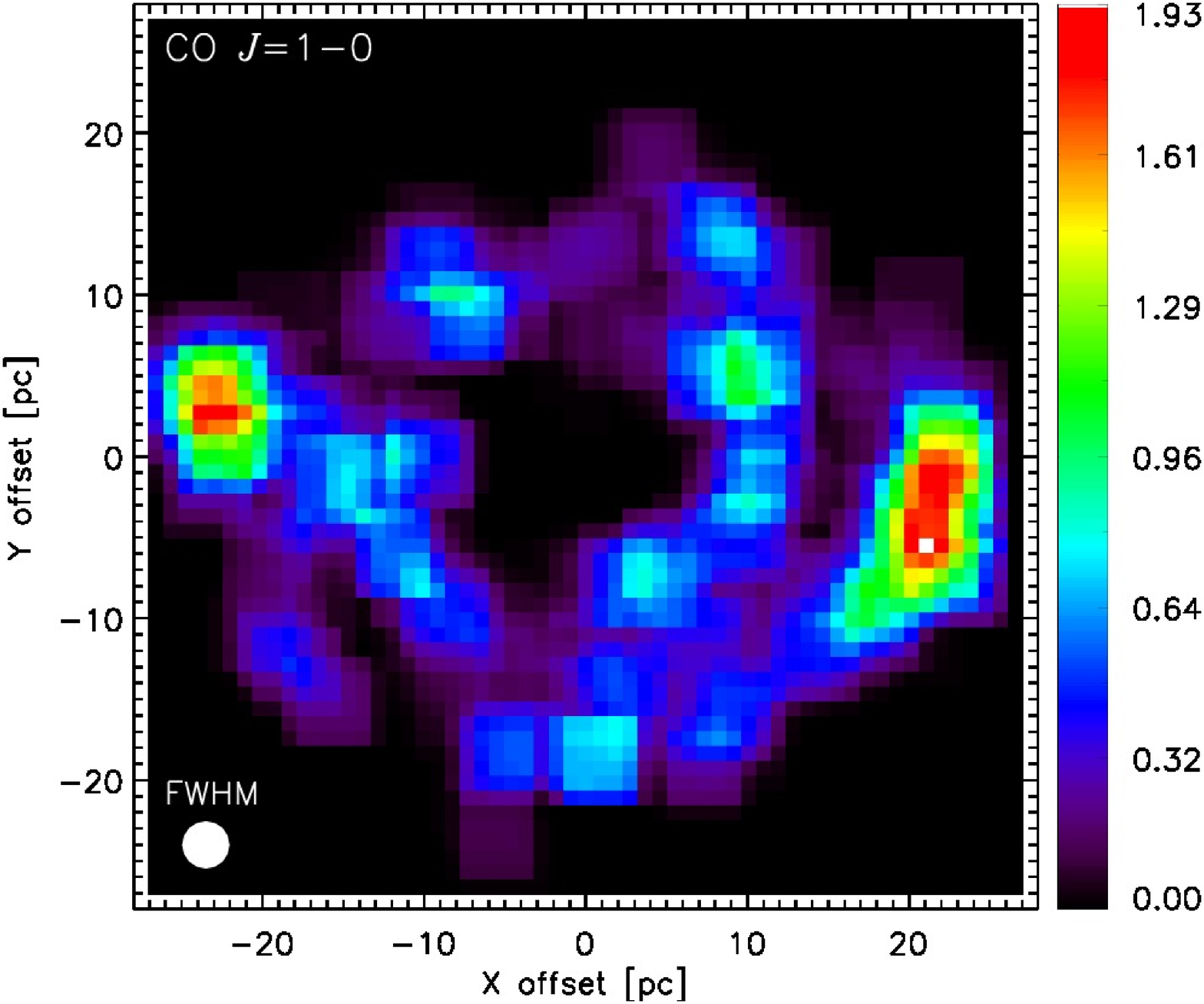}%
\hfill\includegraphics[angle=0,width=0.33\textwidth]{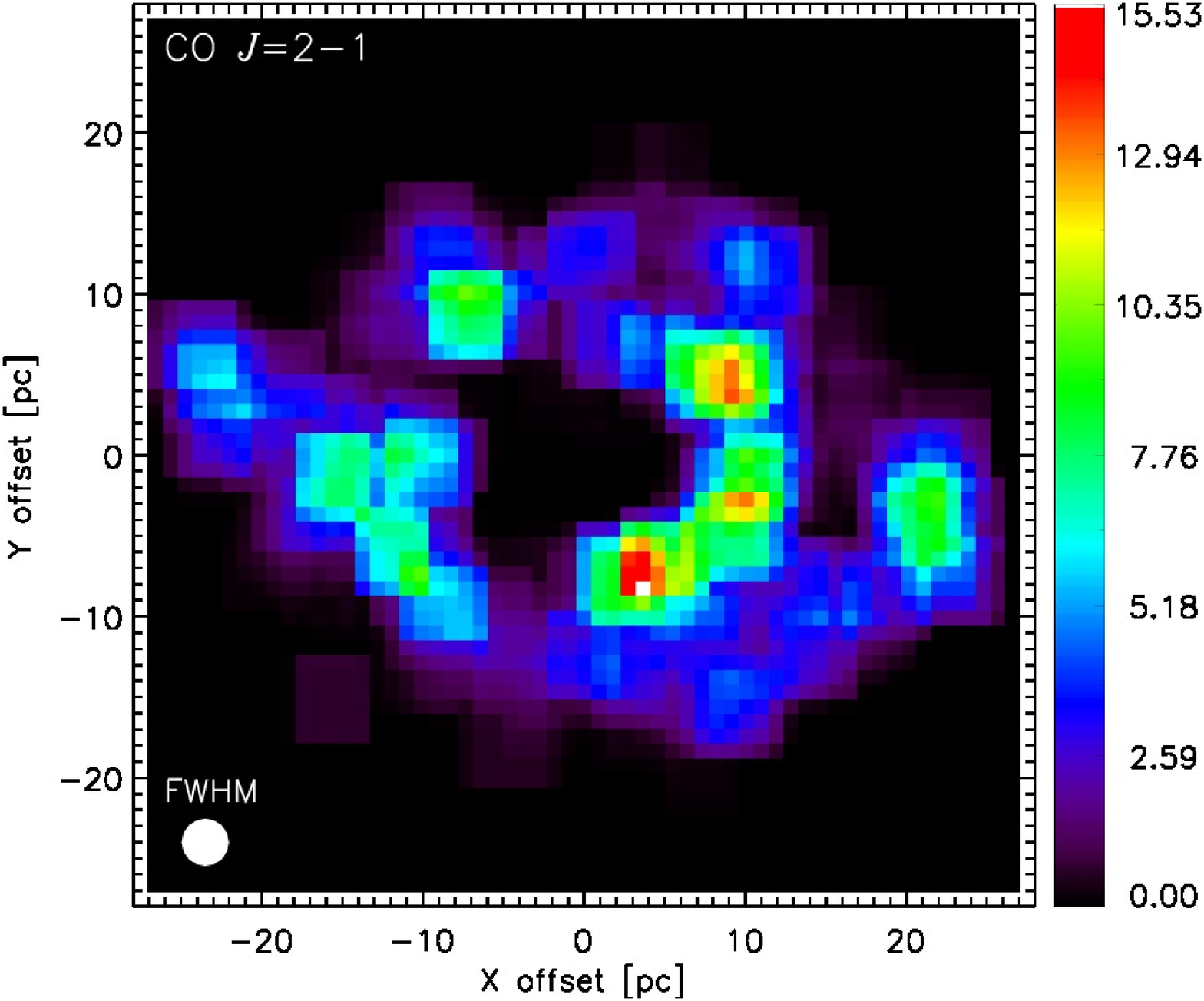}%
\hfill\includegraphics[angle=0,width=0.33\textwidth]{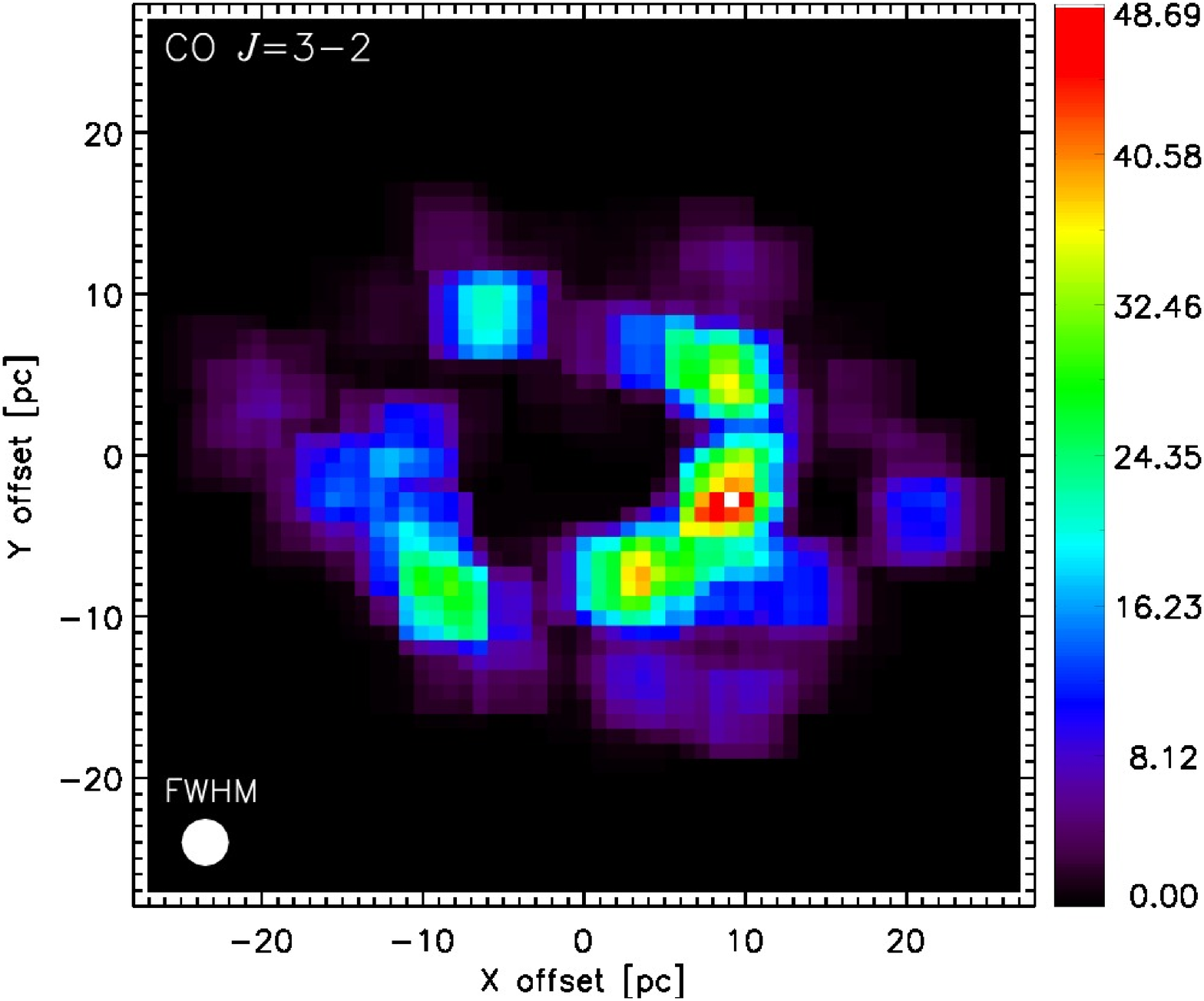}\hspace*{\fill}\\

\vspace{-0.5cm}

\hspace*{\fill}\includegraphics[angle=0,width=0.33\textwidth]{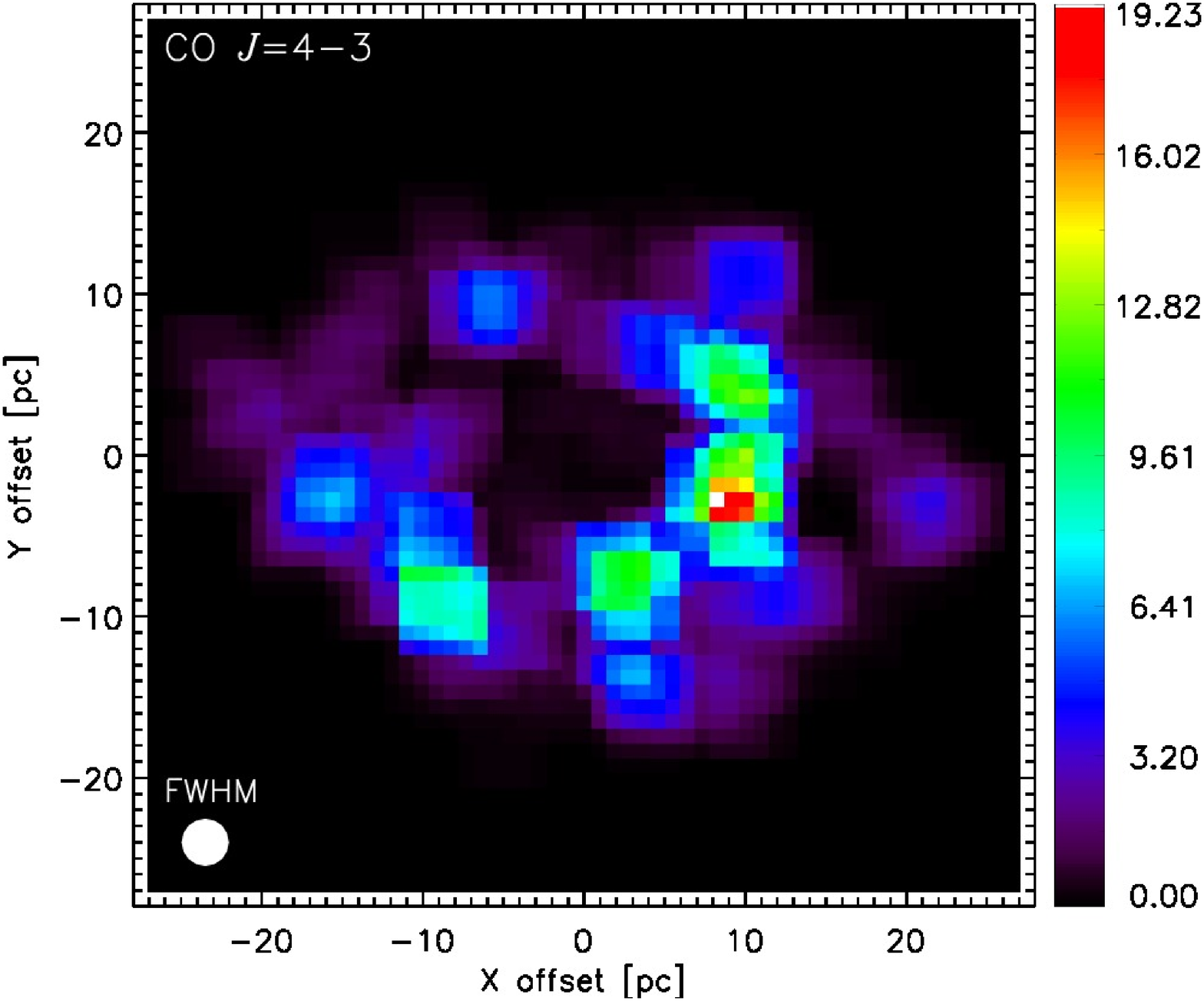}%
\hfill\includegraphics[angle=0,width=0.33\textwidth]{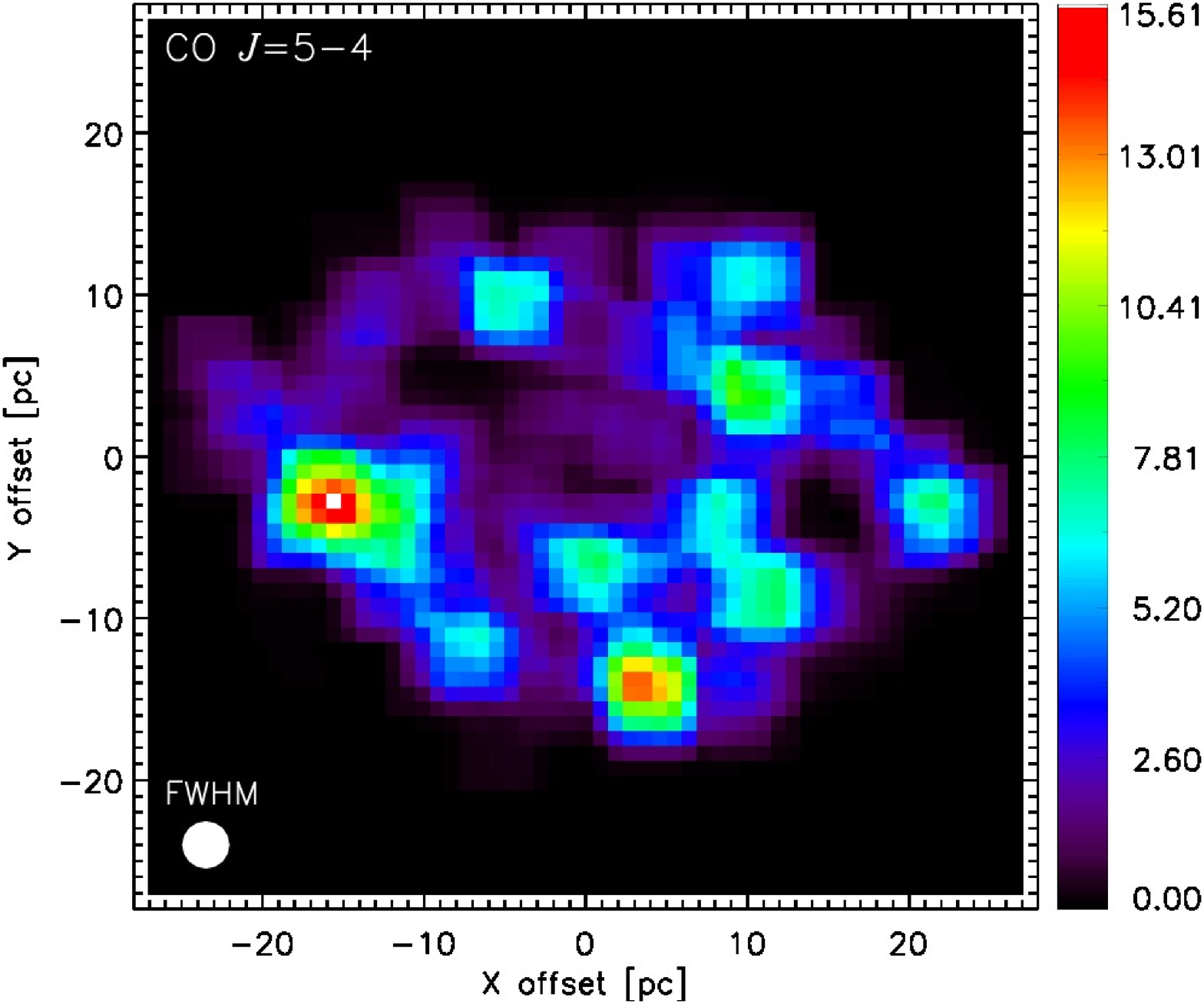}%
\hfill\includegraphics[angle=0,width=0.33\textwidth]{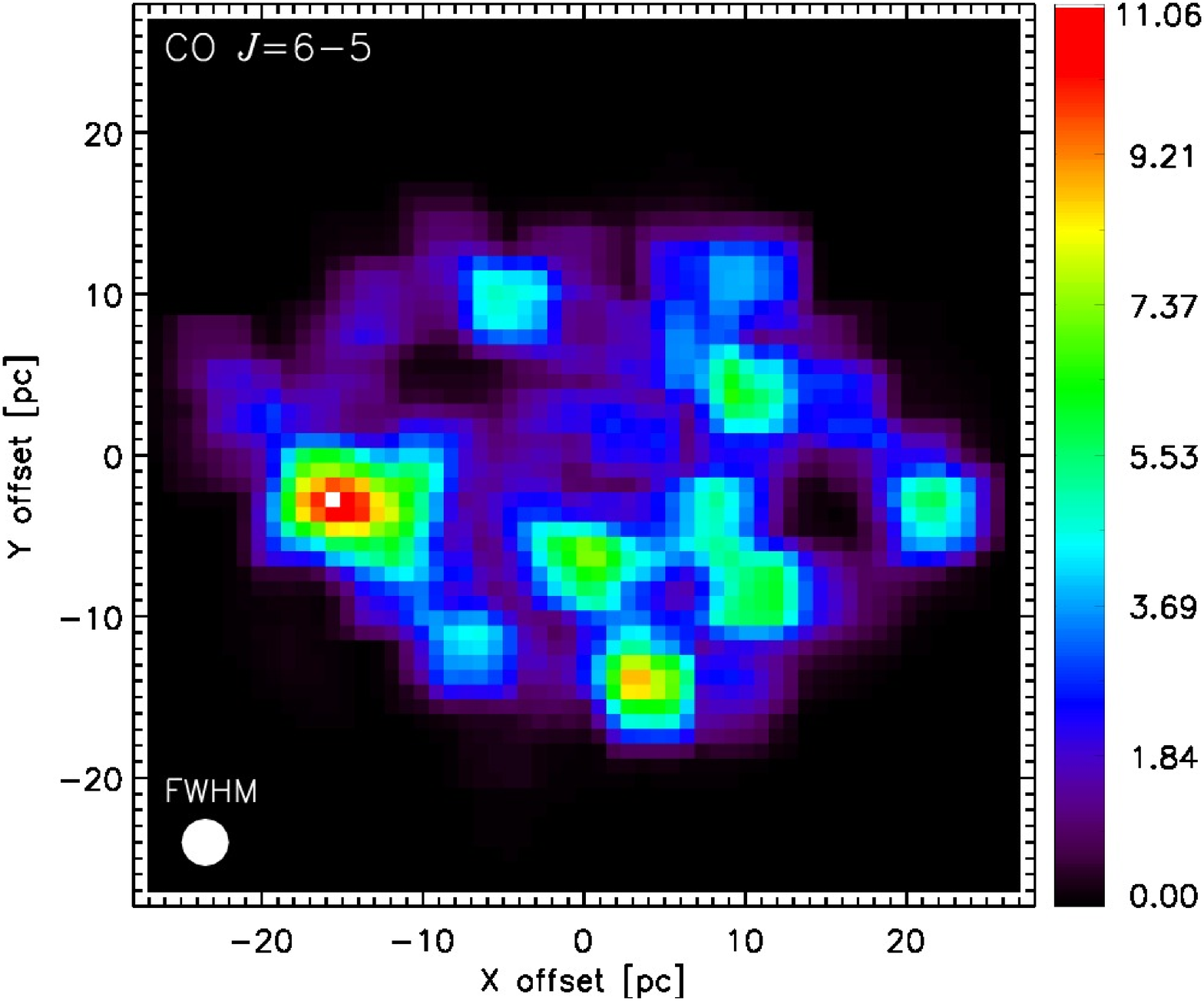}\hspace*{\fill}\\

\vspace{-0.5cm}

\hspace*{\fill}\includegraphics[angle=0,width=0.33\textwidth]{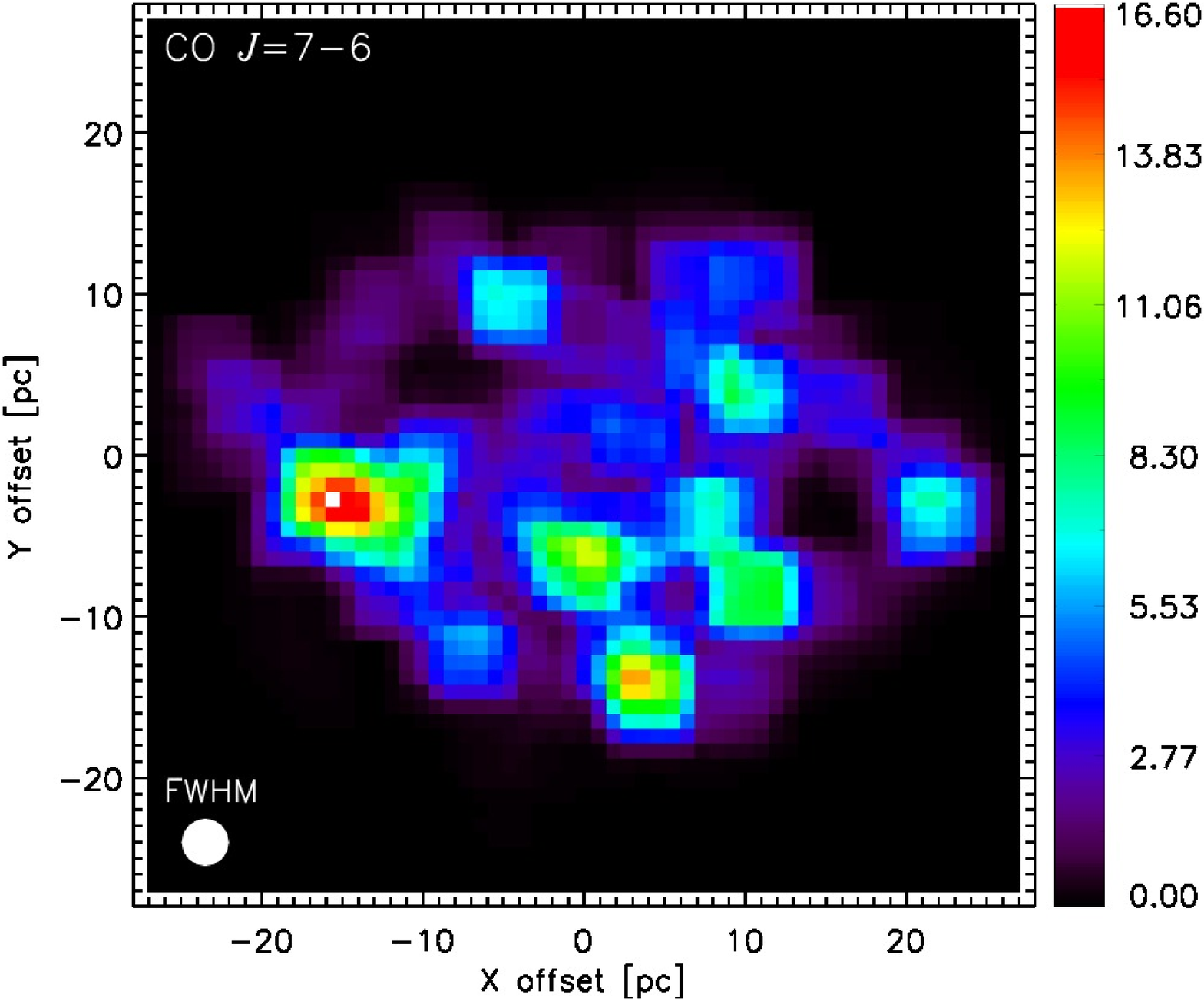}%
\hfill\includegraphics[angle=0,width=0.33\textwidth]{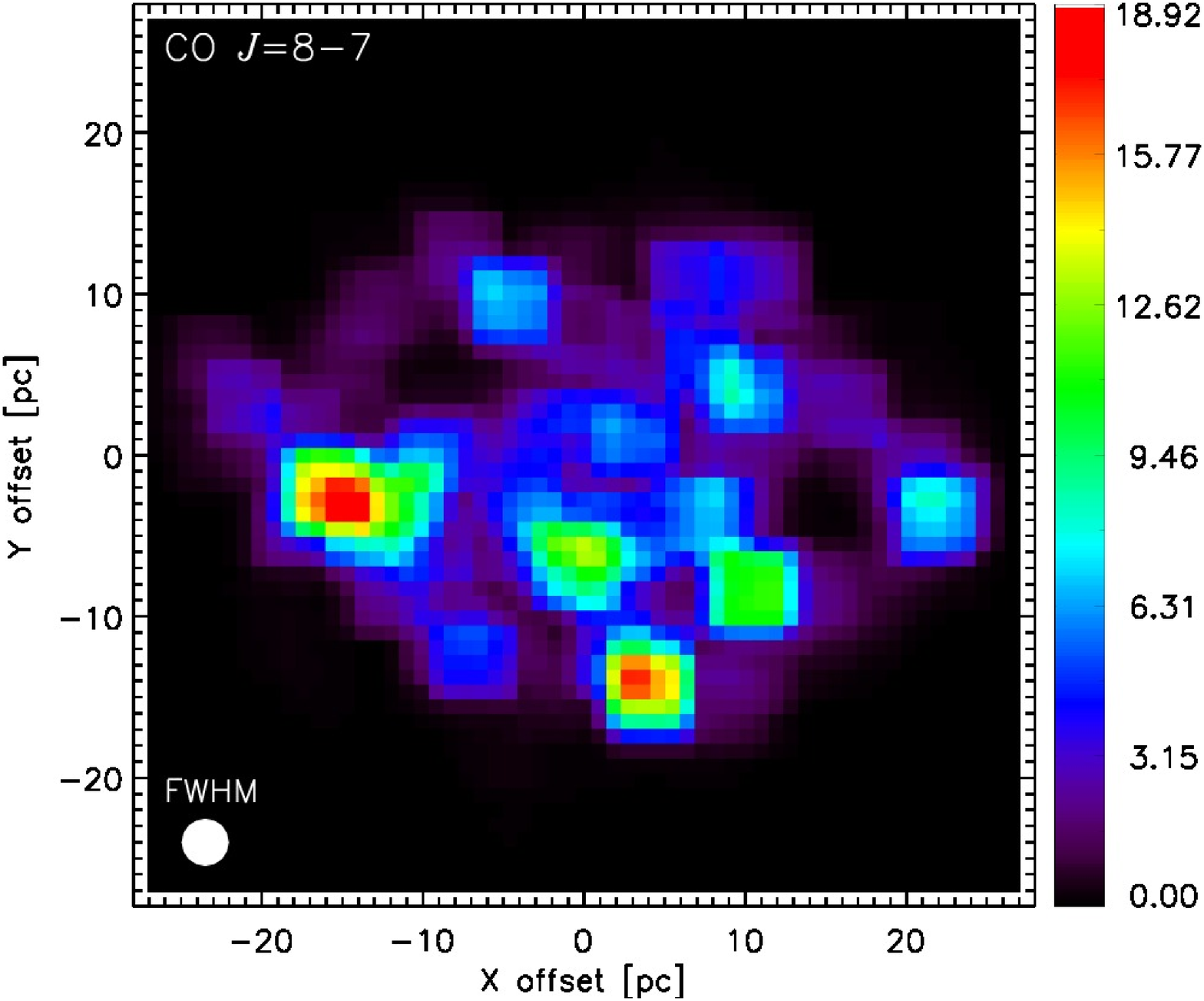}%
\hfill\includegraphics[angle=0,width=0.33\textwidth]{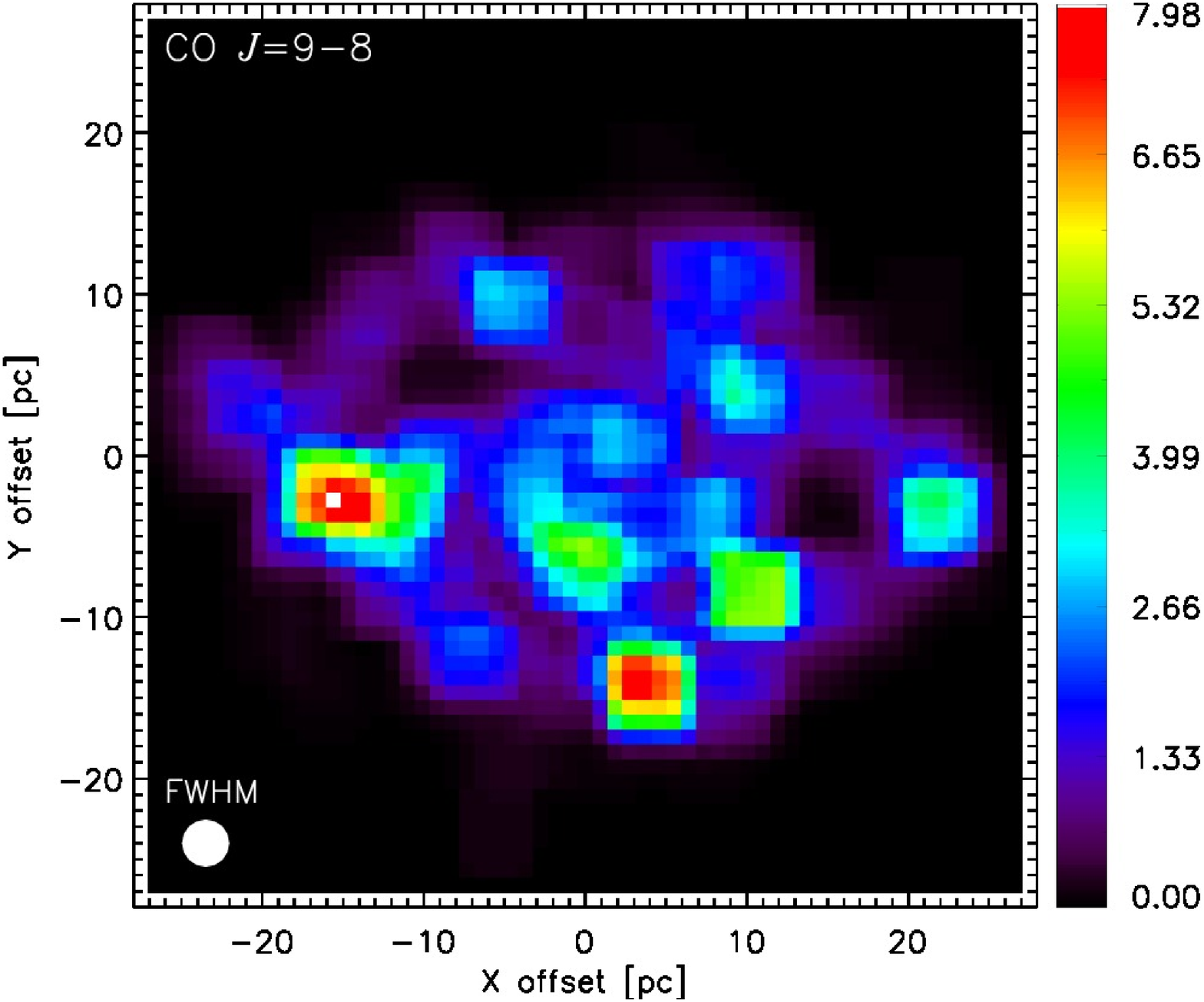}\hspace*{\fill}

\caption{{\footnotesize Maps of the flux ($\ergscm$) of the CO transitions (from $J=1\rightarrow0$ to $J=9\rightarrow8$) with a $45^o$ inclination about the X-axis and convolved with a single dish beam of FWHM=$0.15''$ ($\sim2.8~\rm pc$). We adopt a distance $D=3.82~\rm Mpc$ to the source. Although the $J=3\rightarrow2$ transition does not trace the warmer gas of the {inner NLR}, it is the brightest emission line of this configuration (see the colour scale).}}
\label{fig:maps-45x-fwhm}
\end{figure}

\newpage




\bibliographystyle{mn2e}
\setlength{\bibsep}{-2.1pt}
\bibliography{jp}

\begin{thebibliography}{78}
\expandafter\ifx\csname natexlab\endcsname\relax\def\natexlab#1{#1}\fi

\bibitem[{{Aalto} {et~al.}(2007){Aalto}, {Spaans}, {Wiedner} \&
  {H{\"u}ttemeister}}]{aalto07a}
{Aalto} S., {Spaans} M., {Wiedner} M.~C., {H{\"u}ttemeister} S., 2007, \aap,
  464, 193

\bibitem[{{Baan} {et~al.}(2010){Baan}, {Loenen} \& {Spaans}}]{baan10}
{Baan} W.~A., {Loenen} A.~F., {Spaans} M., 2010, \aap, 516, A40+

\bibitem[{{Ballantyne}(2008)}]{ballantyne08}
{Ballantyne} D.~R., 2008, \apj, 685, 787

\bibitem[{{Bieging} {et~al.}(1998){Bieging}, {Knee}, {Latter} \&
  {Olofsson}}]{bieging98}
{Bieging} J.~H., {Knee} L.~B.~G., {Latter} W.~B., {Olofsson} H., 1998, \aap,
  339, 811

\bibitem[{{Cecchi-Pestellini} {et~al.}(2002){Cecchi-Pestellini}, {Bodo},
  {Balakrishnan} \& {Dalgarno}}]{cecchi02}
{Cecchi-Pestellini} C., {Bodo} E., {Balakrishnan} N., {Dalgarno} A., 2002,
  \apj, 571, 1015

\bibitem[{{Dame} {et~al.}(2001){Dame}, {Hartmann} \& {Thaddeus}}]{dame01}
{Dame} T.~M., {Hartmann} D., {Thaddeus} P., 2001, \apj, 547, 792

\bibitem[{{de Jong} {et~al.}(1975){de Jong}, {Dalgarno} \& {Chu}}]{dejong75}
{de Jong} T., {Dalgarno} A., {Chu} S., 1975, \apj, 199, 69

\bibitem[{{Depristo} {et~al.}(1979){Depristo}, {Augustin}, {Ramaswamy} \&
  {Rabitz}}]{depristo79}
{Depristo} A.~E., {Augustin} S.~D., {Ramaswamy} R., {Rabitz} H., 1979, \jcp,
  71, 850

\bibitem[{{Dickman} {et~al.}(1986){Dickman}, {Snell} \& {Schloerb}}]{dickman86}
{Dickman} R.~L., {Snell} R.~L., {Schloerb} F.~P., 1986, \apj, 309, 326

\bibitem[{{Downes} \& {Solomon}(1998)}]{downes98}
{Downes} D., {Solomon} P.~M., 1998, \apj, 507, 615

\bibitem[{{Downes} {et~al.}(1993){Downes}, {Solomon} \& {Radford}}]{downes93}
{Downes} D., {Solomon} P.~M., {Radford} S.~J.~E., 1993, \apjl, 414, L13

\bibitem[{{Ferrarese} \& {Merritt}(2000)}]{ferrarese00}
{Ferrarese} L., {Merritt} D., 2000, \apjl, 539, L9

\bibitem[{{Flower}(2001)}]{flower01}
{Flower} D.~R., 2001, \mnras, 328, 147

\bibitem[{{Flower} \& {Launay}(1977)}]{flower77}
{Flower} D.~R., {Launay} J.~M., 1977, Journal of Physics B Atomic Molecular
  Physics, 10, 3673

\bibitem[{{Garc{\'{\i}}a-Burillo} {et~al.}(2007){Garc{\'{\i}}a-Burillo},
  {Combes}, {Usero} \& {Graci{\'a}-Carpio}}]{garcia07}
{Garc{\'{\i}}a-Burillo} S., {Combes} F., {Usero} A., {Graci{\'a}-Carpio} J.,
  2007, New Astronomy Review, 51, 160

\bibitem[{{Garc{\'{\i}}a-Burillo} {et~al.}(2008){Garc{\'{\i}}a-Burillo},
  {Combes}, {Usero} \& {Graci{\'a}-Carpio}}]{garcia08}
---, 2008, Journal of Physics Conference Series, 131, 012031

\bibitem[{{Goldflam} {et~al.}(1977){Goldflam}, {Kouri} \& {Green}}]{goldflam77}
{Goldflam} R., {Kouri} D.~J., {Green} S., 1977, \jcp, 67, 4149

\bibitem[{{Graham} {et~al.}(2001){Graham}, {Erwin}, {Caon} \&
  {Trujillo}}]{graham01}
{Graham} A.~W., {Erwin} P., {Caon} N., {Trujillo} I., 2001, \apjl, 563, L11

\bibitem[{{H{\"a}ring} \& {Rix}(2004)}]{haring04}
{H{\"a}ring} N., {Rix} H., 2004, \apjl, 604, L89

\bibitem[{{Henkel} {et~al.}(1987){Henkel}, {Jacq}, {Mauersberger}, {Menten} \&
  {Steppe}}]{henkel87}
{Henkel} C., {Jacq} T., {Mauersberger} R., {Menten} K.~M., {Steppe} H., 1987,
  \aap, 188, L1

\bibitem[{{Hocuk} \& {Spaans}(2010)}]{hocuk10}
{Hocuk} S., {Spaans} M., 2010, \aap, 510, A110+

\bibitem[{{Hogerheijde} \& {van der Tak}(2000)}]{hogerheijde00}
{Hogerheijde} M.~R., {van der Tak} F.~F.~S., 2000, \aap, 362, 697

\bibitem[{{Hollenbach} \& {Tielens}(1999)}]{hollenbach99}
{Hollenbach} D.~J., {Tielens} A.~G.~G.~M., 1999, Reviews of Modern Physics, 71,
  173

\bibitem[{{Hopkins} {et~al.}(2008){Hopkins}, {Hernquist}, {Cox} \& {Kere{\v
  s}}}]{hopkins08}
{Hopkins} P.~F., {Hernquist} L., {Cox} T.~J., {Kere{\v s}} D., 2008, \apjs,
  175, 356

\bibitem[{{Hopkins} {et~al.}(2006){Hopkins}, {Hernquist}, {Cox}, {Robertson} \&
  {Springel}}]{hopkins06}
{Hopkins} P.~F., {Hernquist} L., {Cox} T.~J., {Robertson} B., {Springel} V.,
  2006, \apjs, 163, 50

\bibitem[{{Imanishi} {et~al.}(2006{\natexlab{a}}){Imanishi}, {Dudley} \&
  {Maloney}}]{imanishi06a}
{Imanishi} M., {Dudley} C.~C., {Maloney} P.~R., 2006{\natexlab{a}}, \apj, 637,
  114

\bibitem[{{Imanishi} \& {Nakanishi}(2006)}]{imanishi06}
{Imanishi} M., {Nakanishi} K., 2006, \pasj, 58, 813

\bibitem[{{Imanishi} {et~al.}(2006{\natexlab{b}}){Imanishi}, {Nakanishi} \&
  {Kohno}}]{imanishi06b}
{Imanishi} M., {Nakanishi} K., {Kohno} K., 2006{\natexlab{b}}, \aj, 131, 2888

\bibitem[{{Imanishi} \& {Wada}(2004)}]{imanishi04}
{Imanishi} M., {Wada} K., 2004, \apj, 617, 214

\bibitem[{{Kaspi} {et~al.}(2000){Kaspi}, {Smith}, {Netzer}, {Maoz}, {Jannuzi}
  \& {Giveon}}]{kaspi00}
{Kaspi} S., {Smith} P.~S., {Netzer} H., {Maoz} D., {Jannuzi} B.~T., {Giveon}
  U., 2000, \apj, 533, 631

\bibitem[{{Kohno}(2005)}]{kohno05}
{Kohno} K., 2005, in American Institute of Physics Conf. Ser., Vol. 783, The
  Evolution of Starbursts, {S.~H{\"u}ttmeister, E.~Manthey, D.~Bomans, \&
  K.~Weis}, ed., pp. 203--208

\bibitem[{{Kohno} {et~al.}(2001){Kohno}, {Matsushita}, {Vila-Vilar{\'o}},
  {Okumura}, {Shibatsuka}, {Okiura}, {Ishizuki} \& {Kawabe}}]{kohno01}
{Kohno} K., {Matsushita} S., {Vila-Vilar{\'o}} B., {Okumura} S.~K.,
  {Shibatsuka} T., {Okiura} M., {Ishizuki} S., {Kawabe} R., 2001, in ASP Conf.
  Ser., Vol. 249, The Central Kiloparsec of Starbursts and AGN: The La Palma
  Connection, {J.~H.~Knapen, J.~E.~Beckman, I.~Shlosman, \& T.~J.~Mahoney},
  ed., pp. 672--+

\bibitem[{{Kohno} {et~al.}(2007){Kohno}, {Nakanishi} \& {Imanishi}}]{kohno07}
{Kohno} K., {Nakanishi} K., {Imanishi} M., 2007, in ASP Conf. Ser., Vol. 373,
  The Central Engine of Active Galactic Nuclei, {L.~C.~Ho \& J.-W.~Wang}, ed.,
  pp. 647--+

\bibitem[{{Launay} \& {Roueff}(1977)}]{launay77}
{Launay} J., {Roueff} E., 1977, Journal of Physics B Atomic Molecular Physics,
  10, 879

\bibitem[{{Levenson} {et~al.}(2007){Levenson}, {Sirocky}, {Hao}, {Spoon},
  {Marshall}, {Elitzur} \& {Houck}}]{levenson07}
{Levenson} N.~A., {Sirocky} M.~M., {Hao} L., {Spoon} H.~W.~W., {Marshall}
  J.~A., {Elitzur} M., {Houck} J.~R., 2007, \apjl, 654, L45

\bibitem[{{Levenson} {et~al.}(2001){Levenson}, {Weaver} \&
  {Heckman}}]{levenson01}
{Levenson} N.~A., {Weaver} K.~A., {Heckman} T.~M., 2001, \apj, 550, 230

\bibitem[{{Loenen} {et~al.}(2008){Loenen}, {Spaans}, {Baan} \&
  {Meijerink}}]{loenen08}
{Loenen} A.~F., {Spaans} M., {Baan} W.~A., {Meijerink} R., 2008, \aap, 488, L5

\bibitem[{{Madejski} \& {et al.}(1995)}]{madejski95}
{Madejski} G.~M., {et al.}, 1995, \apj, 438, 672

\bibitem[{{Magorrian} \& {et al.}(1998)}]{magorrian98}
{Magorrian} J., {et al.}, 1998, \aj, 115, 2285

\bibitem[{{Maloney} {et~al.}(1996){Maloney}, {Hollenbach} \&
  {Tielens}}]{maloney96}
{Maloney} P.~R., {Hollenbach} D.~J., {Tielens} A.~G.~G.~M., 1996, \apj, 466,
  561

\bibitem[{{Mart{\'{\i}}n} {et~al.}(2003){Mart{\'{\i}}n}, {Mauersberger},
  {Mart{\'{\i}}n-Pintado}, {Garc{\'{\i}}a-Burillo} \& {Henkel}}]{martin03}
{Mart{\'{\i}}n} S., {Mauersberger} R., {Mart{\'{\i}}n-Pintado} J.,
  {Garc{\'{\i}}a-Burillo} S., {Henkel} C., 2003, \aap, 411, L465

\bibitem[{{McKee} {et~al.}(1982){McKee}, {Storey}, {Watson} \&
  {Green}}]{mckee82}
{McKee} C.~F., {Storey} J.~W.~V., {Watson} D.~M., {Green} S., 1982, \apj, 259,
  647

\bibitem[{{Meijerink} \& {Spaans}(2005)}]{meijerink05}
{Meijerink} R., {Spaans} M., 2005, \aap, 436, 397

\bibitem[{{Meijerink} {et~al.}(2007){Meijerink}, {Spaans} \&
  {Israel}}]{meijerink07}
{Meijerink} R., {Spaans} M., {Israel} F.~P., 2007, \aap, 461, 793

\bibitem[{{Narayanan} {et~al.}(2009){Narayanan}, {Cox}, {Hayward}, {Younger} \&
  {Hernquist}}]{narayanan09}
{Narayanan} D., {Cox} T.~J., {Hayward} C.~C., {Younger} J.~D., {Hernquist} L.,
  2009, \mnras, 400, 1919

\bibitem[{{Narayanan} \& {et al.}(2008{\natexlab{a}})}]{narayanan08a}
{Narayanan} D., {et al.}, 2008{\natexlab{a}}, \apjs, 174, 13

\bibitem[{{Narayanan} \& {et al.}(2008{\natexlab{b}})}]{narayanan08b}
---, 2008{\natexlab{b}}, \apjs, 176, 331

\bibitem[{{Narayanan} {et~al.}(2010){Narayanan}, {Hayward}, {Cox}, {Hernquist},
  {Jonsson}, {Younger} \& {Groves}}]{narayanan10}
{Narayanan} D., {Hayward} C.~C., {Cox} T.~J., {Hernquist} L., {Jonsson} P.,
  {Younger} J.~D., {Groves} B., 2010, \mnras, 401, 1613

\bibitem[{{Nguyen-Q-Rieu} {et~al.}(1991){Nguyen-Q-Rieu}, {Henkel}, {Jackson} \&
  {Mauersberger}}]{nguyen91}
{Nguyen-Q-Rieu}, {Henkel} C., {Jackson} J.~M., {Mauersberger} R., 1991, \aap,
  241, L33

\bibitem[{{Ohsuga} \& {Umemura}(2001{\natexlab{a}})}]{ohsuga01a}
{Ohsuga} K., {Umemura} M., 2001{\natexlab{a}}, \aap, 371, 890

\bibitem[{{Ohsuga} \& {Umemura}(2001{\natexlab{b}})}]{ohsuga01b}
---, 2001{\natexlab{b}}, \apj, 559, 157

\bibitem[{{P{\'e}rez-Beaupuits} {et~al.}(2007){P{\'e}rez-Beaupuits}, {Aalto} \&
  {Gerebro}}]{pb07}
{P{\'e}rez-Beaupuits} J.~P., {Aalto} S., {Gerebro} H., 2007, \aap, 476, 177

\bibitem[{{P{\'e}rez-Beaupuits} {et~al.}(2010){P{\'e}rez-Beaupuits}, {Spaans},
  {Hogerheijde}, {G{\"u}sten}, {Baryshev} \& {Boland}}]{pb10}
{P{\'e}rez-Beaupuits} J.~P., {Spaans} M., {Hogerheijde} M.~R., {G{\"u}sten} R.,
  {Baryshev} A., {Boland} W., 2010, \aap, 510, A87+

\bibitem[{{P{\'e}rez-Beaupuits} {et~al.}(2009){P{\'e}rez-Beaupuits}, {Spaans},
  {van der Tak}, {Aalto}, {Garc{\'{\i}}a-Burillo}, {Fuente} \& {Usero}}]{pb09}
{P{\'e}rez-Beaupuits} J.~P., {Spaans} M., {van der Tak} F.~F.~S., {Aalto} S.,
  {Garc{\'{\i}}a-Burillo} S., {Fuente} A., {Usero} A., 2009, \aap, 503, 459

\bibitem[{{Poelman} \& {Spaans}(2005)}]{poelman05}
{Poelman} D.~R., {Spaans} M., 2005, \aap, 440, 559

\bibitem[{{Poelman} \& {Spaans}(2006)}]{poelman06}
---, 2006, \aap, 453, 615

\bibitem[{{Pounds} {et~al.}(1990){Pounds}, {Nandra}, {Stewart}, {George} \&
  {Fabian}}]{pounds90}
{Pounds} K.~A., {Nandra} K., {Stewart} G.~C., {George} I.~M., {Fabian} A.~C.,
  1990, \nat, 344, 132

\bibitem[{{Schleicher} {et~al.}(2010){Schleicher}, {Spaans} \&
  {Klessen}}]{schleicher10}
{Schleicher} D.~R.~G., {Spaans} M., {Klessen} R.~S., 2010, \aap, 513, A7+

\bibitem[{{Sch{\"o}ier} {et~al.}(2005){Sch{\"o}ier}, {van der Tak}, {van
  Dishoeck} \& {Black}}]{schoier05}
{Sch{\"o}ier} F.~L., {van der Tak} F.~F.~S., {van Dishoeck} E.~F., {Black}
  J.~H., 2005, \aap, 432, 369

\bibitem[{{Solomon} \& {Barrett}(1991)}]{solomon91}
{Solomon} P.~M., {Barrett} J.~W., 1991, in IAU Symposium, Vol. 146, Dynamics of
  Galaxies and Their Molecular Cloud Distributions, {F.~Combes \& F.~Casoli},
  ed., pp. 235--+

\bibitem[{{Solomon} {et~al.}(1997){Solomon}, {Downes}, {Radford} \&
  {Barrett}}]{solomon97}
{Solomon} P.~M., {Downes} D., {Radford} S.~J.~E., {Barrett} J.~W., 1997, \apj,
  478, 144

\bibitem[{{Solomon} {et~al.}(1987){Solomon}, {Rivolo}, {Barrett} \&
  {Yahil}}]{solomon87}
{Solomon} P.~M., {Rivolo} A.~R., {Barrett} J., {Yahil} A., 1987, \apj, 319, 730

\bibitem[{{Soltan}(1982)}]{soltan82}
{Soltan} A., 1982, \mnras, 200, 115

\bibitem[{{Spaans} \& {Meijerink}(2008)}]{spaans08}
{Spaans} M., {Meijerink} R., 2008, \apjl, 678, L5

\bibitem[{{Strong} \& {Mattox}(1996)}]{strong96}
{Strong} A.~W., {Mattox} J.~R., 1996, \aap, 308, L21

\bibitem[{{Tacconi} \& {et al.}(2008)}]{tacconi08}
{Tacconi} L.~J., {et al.}, 2008, \apj, 680, 246

\bibitem[{{Tuzun} {et~al.}(1998){Tuzun}, {Burkhardt} \& {Secrest}}]{tuzun98}
{Tuzun} R.~E., {Burkhardt} P., {Secrest} D., 1998, Computer Physics
  Communications, 112, 112

\bibitem[{{Usero} {et~al.}(2004){Usero}, {Garc{\'{\i}}a-Burillo}, {Fuente},
  {Mart{\'{\i}}n-Pintado} \& {Rodr{\'{\i}}guez-Fern{\'a}ndez}}]{usero04}
{Usero} A., {Garc{\'{\i}}a-Burillo} S., {Fuente} A., {Mart{\'{\i}}n-Pintado}
  J., {Rodr{\'{\i}}guez-Fern{\'a}ndez} N.~J., 2004, \aap, 419, 897

\bibitem[{{van der Tak} {et~al.}(2007){van der Tak}, {Black}, {Sch{\"o}ier},
  {Jansen} \& {van Dishoeck}}]{vdtak07}
{van der Tak} F.~F.~S., {Black} J.~H., {Sch{\"o}ier} F.~L., {Jansen} D.~J.,
  {van Dishoeck} E.~F., 2007, \aap, 468, 627

\bibitem[{{van der Werf} \& {et al.}(2010)}]{vdwerf10}
{van der Werf} P.~P., {et al.}, 2010, \aap, 518, L42+

\bibitem[{{Verner} \& {Yakovlev}(1995)}]{verner95}
{Verner} D.~A., {Yakovlev} D.~G., 1995, \aaps, 109, 125

\bibitem[{{Wada} \& {Norman}(2002)}]{wada02}
{Wada} K., {Norman} C.~A., 2002, \apjl, 566, L21

\bibitem[{{Wada} {et~al.}(2009){Wada}, {Papadopoulos} \& {Spaans}}]{wada09}
{Wada} K., {Papadopoulos} P.~P., {Spaans} M., 2009, \apj, 702, 63

\bibitem[{{Wada} \& {Tomisaka}(2005)}]{wada05}
{Wada} K., {Tomisaka} K., 2005, \apj, 619, 93

\bibitem[{{Wernli} {et~al.}(2006){Wernli}, {Valiron}, {Faure}, {Wiesenfeld},
  {Jankowski} \& {Szalewicz}}]{wernli06}
{Wernli} M., {Valiron} P., {Faure} A., {Wiesenfeld} L., {Jankowski} P.,
  {Szalewicz} K., 2006, \aap, 446, 367

\bibitem[{{Wilson} \& {Bell}(2002)}]{wilson02}
{Wilson} N.~J., {Bell} K.~L., 2002, \mnras, 337, 1027

\bibitem[{{Yamada} {et~al.}(2007){Yamada}, {Wada} \& {Tomisaka}}]{yamada07}
{Yamada} M., {Wada} K., {Tomisaka} K., 2007, \apj, 671, 73

\bibitem[{{Zdziarski} {et~al.}(1995){Zdziarski}, {Johnson}, {Done}, {Smith} \&
  {McNaron-Brown}}]{zdziarski95}
{Zdziarski} A.~A., {Johnson} W.~N., {Done} C., {Smith} D., {McNaron-Brown} K.,
  1995, \apjl, 438, L63

\end{thebibliography}

\end{document}